# The structural-size effect, aging time, and pressure-dependent functional properties of Mn-containing perovskite nanoparticles

Danyang Su[a], N.A. Liedienov[a,b,*], V.M. Kalita[c,d,e], I.V. Fesych[f], Wei Xu[g], A.V. Bodnaruk[e], Yu.I. Dzhezherya[c,d,e], Quanjun Li[a], Bingbing Liu[a], G.G. Levchenko[a,b,h,*]

[a]*State Key Laboratory of Superhard Material, Jilin University, 130012 Changchun, China*
[b]*Donetsk Institute for Physics and Engineering named after O.O. Galkin, NAS of Ukraine, 03028 Kyiv, Ukraine*
[c]*National Technical University of Ukraine "Igor Sikorsky Kyiv Polytechnic Institute", 03056 Kyiv, Ukraine*
[d]*Institute of Magnetism, NAS of Ukraine and MES of Ukraine, 03142 Kyiv, Ukraine*
[e]*Institute of Physics, NAS of Ukraine, 03028 Kyiv, Ukraine*
[f]*Taras Shevchenko National University of Kyiv, 01030 Kyiv, Ukraine*
[g]*State Key Laboratory of Inorganic Synthesis and Preparative Chemistry, College of Chemistry, Jilin University, 130012 Changchun, China*
[h]*International Center of Future Science, Jilin University, 130012 Changchun, China*

Corresponding author
E-mail addresses:    nikita.ledenev.ssp@gmail.com (N.A. Liedienov)
                    g-levch@ukr.net (G.G. Levchenko)

**Abstract**
The properties of nanoparticles are determined by their size and structure. When exposed to external pressure $P$, their structural properties can change. The improvement or degradation of samples' properties depending on time is particularly interesting. The knowledge of the influence of structural-size effect, aging time, and pressure on compounds' behavior is essential for fundamental and applied purposes. Therefore, the first attempts have been made to shed light on how the functional properties of the Mn-containing perovskites change depending on them. The nanoparticles of different sizes, 20–70 nm, have been obtained. After 3 years, their structural properties underwent significant changes, including an increase in particle size, bandwidth, and microstrains, as well as a reduction in dislocation density. The greatest change in the structure is observed for the smallest nanoparticles. The phase transition temperatures increase with nanoparticle size, time, and pressure. The aging time has the strongest influence on the changing Curie temperature $T_C$ for the smallest and most magnetically inhomogeneous nanoparticles with $dT_C/dP \approx 100$ K/GPa. Conversely, the structural-size effect and external pressure have the greatest influence on the biggest and most magnetically uniform nanoparticles with $dT_C/dP \approx 91$ and 16 K/GPa, respectively. After 3 years, the biggest nanoparticles demonstrate the most stable phase transition temperatures with improved magnetocaloric parameters near room temperature. These structural-size effect, aging time, and pressure are powerful instruments to tune the nanoparticles' phase transition temperatures and magnetocaloric effect. These outcomes may have implications for the whole class of perovskites and could initiate a new mainstream.

# 1. Introduction

Mn-containing $R_{1-x}A_x$MnO$_3$ inorganic perovskites, where $R$ is the rare-earth cation and $A$ is the divalent and/or monovalent doped cation, still never cease to amaze with their abundance of unique physical properties: the diversity of the phase transitions, starting from paramagnetic (PM) through the phase separation (PS), ferromagnetic (FM), superparamagnetic (SPM), and ending to ferri- and/or antiferromagnetic transitions and accompanied simultaneously by the insulator-metal one [1-4]; the presence of the colossal magnetoresistance effect [5-7], large low-field magnetoresistance effect [8], and giant magnetocaloric effect (MCE) [9, 10]. Moreover, their low price, ease of manufacture, high energy efficiency, non-polluting, eco-friendly, adjustable Curie temperature ($T_C$), and high chemical stability [11, 12] make them "hot" materials for applied applications such as magnetic refrigerators, sensitive magnetic field sensors, magnetic memory devices, microwave coatings, drug delivery, magnetic hyperthermia therapy, batteries, fuel cells, etc. [13-17]. However, knowledge of the influence of the internal reversible (structural-size effect and aging time of the sample $t$) and external irreversible (hydrostatic $P$) pressures on the functional properties of Mn-containing perovskites from both fundamental and applied points of view is currently practically lacking [18-20].

Manganite perovskites' physical and chemical properties strongly depend on their structure, morphology, type and concentration of ions, as well as particle size [21]. When the particle size is reduced to the nanometer scale, such basic magnetic parameters as transition ordering temperatures, spontaneous magnetization $M_S$, coercivity $H_C$, residual magnetization $M_R$, magnetic entropy change $\Delta S_M$, etc., differ significantly from those of bulk [22-24]. As the particle size decreases, the specific surface area increases, and the size role becomes more and more prominent [25, 26]. A "core-shell" structure appears to have various structural (defects, imperfections, distortions, disorders, etc.) and magnetic (FM, PM, SPM, etc.) natures [27, 28]. Each nanoparticle (NP) has its own Curie temperature $T_C$. As the NP's size reduces, its $T_C$ also decreases due to the influence of structural-size, magnetic, and thermal fluctuations on the magnetic ordering. The $T_C$ can also be adjusted by changing the doping level. Moreover, the NPs' Curie temperature is affected by their structure. With varying particle sizes, the interionic distances are changed and, consequently, the exchange interactions and magnetic ordering temperatures [20]. Such parameters as bond angle $\theta_{<Mn-O-Mn>}$ and bond length $<d_{Mn-O}>$ from the structural side effect play an important role in the formation of magneto-transport properties of the manganites [29]. Additionally, an essential parameter as the Néel relaxation and attributed blocking temperature $T_B$ for the particles' magnetic moment vector orientation depends on the size of the NPs, which affects the type and character of the magnetization curves of the ensemble of NPs [30]. The blocked and/or unblocked SPM states arise [31-33]. Thus, the perovskite NPs' magnetic ordering temperature should depend not only on the chemical composition and structural parameters but also on the finite size of particles. These parameters should strongly depend on each other for NPs. For instance, after additional annealing at higher temperatures, an ensemble of larger particles can be obtained, and the distance between the ions also has to be different, which influences the Curie temperature. In other words, there is a structural-size effect when the structure and size of NPs interconnectedly influence their functional properties [34]. It is important to know what affects the magnetic ordering temperatures: fluctuations due to the small particles and/or structure, which is also sensitive to the particle size since it is difficult to distinguish in a real experiment. Usually, an annealing process is used to obtain NPs, which affects their structure, size, and functional properties. However, there is no certainty that after annealing and cooling the samples, all relaxation processes, including diffuse ones, will be completed even during the long annealing time. This means that over a long period, at least several years of sample exposure under normal conditions, slow changes in the structure and size of NPs are possible, and it is exiting how. If slow relaxation changes occur after obtaining NPs, they are expected to be detected by monitoring their structure and analyzing changes in their functional properties.

Another crucial and underestimated physical parameter is aging time $t$. It does not mean the measurement/experiment time. Here, it is the waiting time after obtaining the sample or its aging time. This time of several years significantly exceeds the time required for measurements. For instance, for NPs with the Néel type of relaxation of their magnetization, the measurement time is about 100 s [35] or comparable to this time, which is much less than the influence of waiting time on the structural and functional properties of the samples, considering here. Only several papers were dedicated to the bulk manganite samples [18, 19, 36, 37], where the influence of measurement/experiment time on the value resistivity and magnetization was considered. This time usually varied within $10^0$-$10^4$ sec and achieved its maximum value of 20 hours. The influence of aging time $t$ after annealing on the phase transition temperatures: paramagnetic Curie temperature $\theta$, Curie temperature $T_C$, and blocking temperature $T_B$, as well as the fraction of magnetic phase, magnetic entropy change $\Delta S_M$, and relative cooling power $RCP$, is fully absent. Moreover, despite the great importance of having data about the influence of this time $t$ on the functional properties of materials for their practical application, there are currently no studies elucidating the effects of long-term waiting time of several years for both bulk and nanopowder materials.

The external pressure $P$ is another valuable tool for studying and tuning the functional properties of the compounds. It can modulate the bandwidth $W$ through the bond angle $\theta_{<Mn-O-Mn>}$ and bond length $<d_{Mn-O}>$ and, consequently, define the influence of only structural effect without changing particle size on the exchange interactions and phase transition temperatures [38]. Studying the effect of compression of NPs under external pressure is topical since this makes it possible to separate the influence of different factors on magnetic ordering: the manifestation of the fluctuation mechanism and the influence of changes in interionic distances on exchange, which is a challenge to implement under normal conditions. Moreover, the $P$ can change the magnetization $M$ and, consequently, MCE $\Delta S_M$ *via* the spin-lattice coupling [24]. However, again, there is no data on how the magnetic and magnetocaloric properties of the Mn-containing perovskite NPs will be changed after several years under high pressure.

Therefore, realizing the importance and necessity of knowing how the functional properties of the Mn-containing perovskite NPs depend on the profoundly interrelated structural-size effect, aging time $t$, and pressure $P$, we have tried to pull back the curtain and shed light on that. For this purpose, nine in total different $La_{0.8-x}Cd_xNa_{0.2}MnO_3$ ($x = 0$ and $0.05$) NPs under various annealing temperatures $t_{ann}$ = 500–900 °C and additionally the $La_{0.7}A_{0.2}Mn_{1.1}O_3$ ($A$ = Na$^+$, Ag$^+$, K$^+$) NPs upon the $t_{ann}$ = 900 °C have been obtained. Their structure, morphology, particle size, phase transition temperatures, and MCE immediately after obtaining samples and after a long-term waiting time of 3 years have been thoroughly and systematically studied to generalize and expand the obtained results for a whole class of manganites. Moreover, the studied samples were also tested under high external pressure. It has allowed us to plot the phase diagrams of states and establish the structure-functional relationships of Mn-containing perovskite NPs modulated by doping level, type of ions, annealing temperature $t_{ann}$, aging time $t$, and external high hydrostatic pressure $P$, and, as a consequence, define their role in the formation of the NPs' functional properties, which provides the relevance, significance, and worth of this research.

## 2. Experimental section
## 2.1. Sample preparation

The $La_{0.8-x}Cd_xNa_{0.2}MnO_3$ ($x = 0$ and $0.05$) nanopowder was synthesized using the citrate-nitrate combustion method [15, 31, 39]. The weighed stoichiometric amounts of lanthanum nitrate $La(NO_3)_3 \cdot 6H_2O$ (99.9% metals basis, Sigma-Aldrich), cadmium nitrate $Cd(NO_3)_3 \cdot 6H_2O$ (99.99% trace metals basis, Sigma-Aldrich), sodium nitrate $NaNO_3$ (ACS, $\geq$ 99.0%, Sigma-Aldrich), and manganese nitrate $Mn(NO_3)_2 \cdot 4H_2O$ (98%, Sigma-Aldrich) were first dissolved in 100 mL deionized water and then added citric acid $C_6H_8O_7 \cdot H_2O$ (ACS reagent, $\geq$ 99.0%, Sigma-Aldrich) to the above metal nitrate solution. The molar amount of citric acid was equal to the total molar amount of metal nitrates in the solution. pH of the obtained well-mixed citrate-nitrate solution was adjusted to 6–7 with the addition of 25 % ammonia water solution under continuous stirring at ambient temperature. The homogeneous solution mixture was heated at 85–90 °C on a hot plate and evaporated to form a highly viscous and dried gel. Then, the dried viscous gel is heated at 200 °C to initiate the spontaneous combustion. The self-ignited final product was ground well and calcined at different temperatures $t_{ann}$ = 500, 700, and 900 °C (20 h for each temperature). All samples annealed at 500, 700, and 900 °C were noted as LNMO-500, LNMO-700, LNMO-900 for the $La_{0.8}Na_{0.2}MnO_3$ and LCNMO-500, LCNMO-700, LCNMO-900 for the $La_{0.75}Cd_{0.05}Na_{0.2}MnO_3$, respectively.

## 2.2. Characterization

The phase formation in the L(C)NMO system was studied by simultaneous thermogravimetry/differential thermal analysis (TG/DTA) using DTG-60H (Shimadzu, Japan). The TG/DTA curves were recorded in the 16–900 °C range with a heating rate of 10 °C min$^{-1}$. The initial precursor mixture mass was about 1–2 mg and weighed into alumina crucibles. The experiments were conducted in air at a flow rate of 100 mL min$^{-1}$. The reference substance was pure α-$Al_2O_3$. The structure, phase composition, and crystallite size were examined using X-ray diffraction (XRD) on a Shimadzu LabX XRD-6000 in Cu$K_α$-radiation (λ = 1.5406 Å) and a Rigaku Smartlab SE in Cu$K_α$-radiation (λ = 1.5406 Å) at room temperature. Additionally, an X-ray diffractometer MicroMax-007 HF (Rigaku, Japan) in Mo$K_α$-radiation (λ = 0.71146 Å) was used to study structure under high pressure up to $P ≈ 5$ GPa at room temperature. The lattice parameters were obtained by Rietveld refinement [40] using the Full Prof software [41]. The morphology and size of the nanoparticles, as well as their chemical composition were studied using transmitting electron microscopy (TEM) on JEM-2200FS, scanning electron microscopy (SEM) on FEI Magellan 400 with an energy-dispersive X-ray spectroscopy (EDS) module. The distribution function of particle size was obtained from the analysis of TEM and SEM images using Nano Measure 1.2.5 software [42]. X-ray photoelectron spectroscopy (XPS) measurements were performed on a Thermo ESCALAB 250XI X-ray photoelectron spectrometer. The XPS spectra were fitted and analyzed using XPSPEAK41. The background of the X-ray photoelectronic lines was cut using the Shirley method. The spectra were excited using monochromatized Al$K_α$-radiation. The state of the surfaces was monitored through the C1s line, allowing calibration of the energy scales for all spectra. The C1s line binding energy was approximately 285 eV. Magnetic properties of the nanopowders were tested using LDJ 9500 and Quantum Design SQUID MPMS 3 magnetometers. The temperature dependences of the magnetization $M(T)$ under different fields of $H$ = 50 Oe, 2 kOe, and 10 kOe were carried out within the temperature range from 100 to 350 K. The magnetization isotherms $M(H)$ were measured up to 10 kOe near Curie temperature with a step of $ΔT$ = 10 K and $ΔH$ = 100 Oe to determine MCE. Additionally, the magnetic measurements under high pressure up to $P ≈ 0.8$ GPa were conducted using a piston-type pressure cell made of Ni–Cr–Al alloy [43, 44]. Silicon oil of low viscosity DC 200 was used as a pressure-transmitting medium. The pressure inside was calibrated using the pressure dependence of the superconducting transition temperature of high-purity lead.

## 3. Results and discussion
### 3.1. Structural properties

To determine the heat treatment and annealing conditions for obtaining single-phase L(C)NMO manganites, it is necessary to carry out the TG/DTA studies (see Fig. S1 in Supplementary Material (SM) 1). As turned out, the first attempts of a perovskite structure formation for the LNMO compound occurred after self-ignition at 200 °C with an appearance of the lanthanum oxycarbonate $La_2O_2CO_3$ (JCPDS Card No. 00-048-1113), sodium nitrite $NaNO_2$ (JCPDS Card No. 00-006-0392), and manganese oxide $Mn_3O_4$ (JCPDS Card No. 00-024-0734) (see Fig. S2). At 500, 700, and 900 °C, the single-phase L(C)NMO samples with a disordered rhombohedral $R\bar{3}c$ perovskite structure (JCPDS Card No. 96-152-5859) are already formed (see Fig. 1). The main lattice parameters are listed in Table 1.

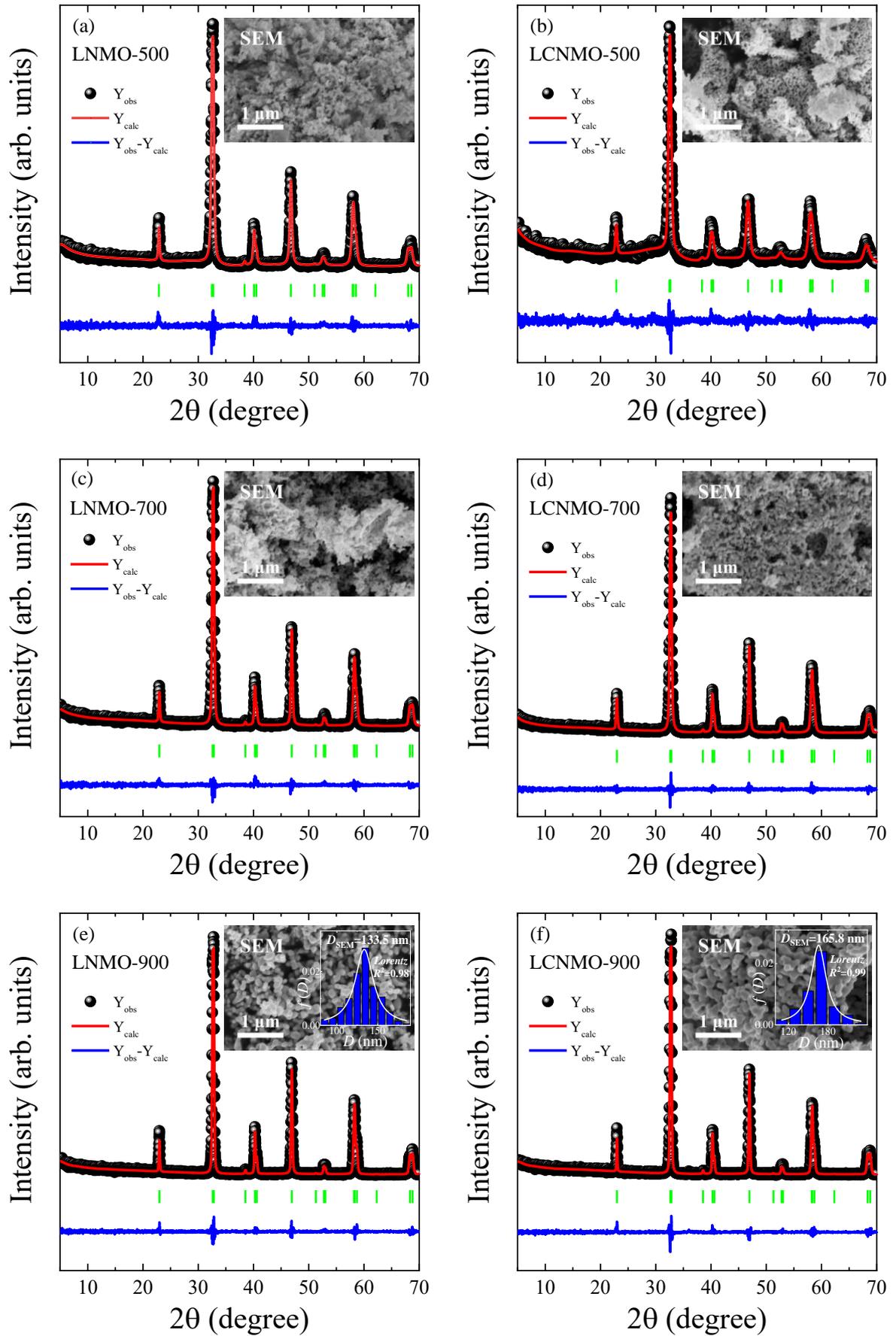

**Fig. 1**. Rietveld refinement XRD patterns of the LNMO (a, c, e) and LCNMO (b, d, f) samples obtained under different annealing temperatures $t_{ann}$ = 500 (a, b), 700 (c, d), and 900 °C (e, f). The experimental (black circles) and calculated (red line) values with a difference (blue line) are plotted. The vertical (green) bars indicate the angular positions of the allowed Bragg reflections (JCPDS Card

No. 96-152-5859). The insets indicate SEM images of the studied nanopowders with a particle-size distribution for the highest $t_{ann}$.

**Table 1**
Rietveld refinement crystallographic parameters, as well as SEM and TEM data for the L(C)NMO nanopowders under different $t_{ann}$.

| Sample | LNMO | | | LCNMO | | |
|---|---|---|---|---|---|---|
| $t_{ann}$ (°C) | 500 | 700 | 900 | 500 | 700 | 900 |
| $D_{XRD}$ (nm) | 27±1 | 38±1 | 62±2 | 20±1 | 38±1 | 71±3 |
| $D_{SEM}$ (nm) | – | – | 134±1 | – | – | 166±1 |
| $D_{TEM}$ (nm) | 38±1 | 62±1 | 132±2 | 25±1 | 60±1 | 164±1 |
| $t_{shell}^{TEM}$ (nm) | 1.1 | – | – | 2.7 | – | – |
| $t_{shell}^{M(H)}$ (nm) | 3.9 | 0.8 | – | 3.1 | 1.5 | – |
| $\delta \cdot 10^{-4}$ (nm$^{-2}$) | 14.0 | 7.0 | 2.6 | 25.8 | 7.0 | 2.0 |
| $\varepsilon$ (%) | 0.278(5) | 0.208(5) | 0.194(11) | 0.651(15) | 0.233(17) | 0.205(1) |
| Space group | $R\bar{3}c$ (No. 167) | $R\bar{3}c$ (No. 167) | $R\bar{3}c$ (No. 167) | $R\bar{3}c$ (No. 167) | $R\bar{3}c$ (No. 167) | $R\bar{3}c$ (No. 167) |
| $a$ (Å) | 5.51036(49) | 5.48910(120) | 5.48781(12) | 5.51025(150) | 5.48757(17) | 5.48491(32) |
| $b$ (Å) | 5.51036(49) | 5.48910(120) | 5.48781(12) | 5.50855(146) | 5.48757(17) | 5.48491(32) |
| $c$ (Å) | 13.34804(166) | 13.32137(296) | 13.31984(50) | 13.38981(734) | 13.32232(77) | 13.31780(92) |
| $V$ (Å$^3$) | 351.001(62) | 347.602(132) | 347.398(17) | 352.070(229) | 347.432(25) | 346.978(38) |
| $Z$ | 6 | 6 | 6 | 6 | 6 | 6 |
| $<d_{Mn-O}>$ (Å) | 1.9642(21) | 1.9523(20) | 1.9524(17) | 1.9639(39) | 1.9504(21) | 1.9500(18) |
| $\theta_{<Mn-O-Mn>}$ (°) | 162.3(3) | 164.6(3) | 164.3(2) | 163.2(5) | 165.2(3) | 165.1(2) |
| $W$ | 0.0930 | 0.0953 | 0.0953 | 0.0932 | 0.0957 | 0.0958 |
| $\rho$ (g/cm$^3$) | 6.207 | 6.268 | 6.271 | 6.150 | 6.233 | 6.241 |
| $R_p$ (%) | 10.1 | 9.2 | 9.8 | 10.8 | 8.6 | 9.4 |
| $R_{wp}$ (%) | 13.7 | 13.8 | 14.1 | 13.8 | 12.8 | 14.3 |
| $R_{exp}$ (%) | 11.9 | 11.9 | 11.9 | 11.8 | 11.9 | 11.9 |
| $\chi^2$ (%) | 1.3 | 1.3 | 1.4 | 1.4 | 1.1 | 1.4 |

$R_p = 100 \cdot \Sigma|y_{obs} - y_{calc}| / \Sigma|y_{obs}|$; $R_{wp} = 100 \cdot [\Sigma w|y_{obs} - y_{calc}|^2 / \Sigma w|y_{obs}|^2]^{1/2}$; $R_{exp} = 100 \cdot [(N-P+C) / \Sigma w(y_{obs}^2)]^{1/2}$; N-P+C = degrees of freedom; $\chi^2 = (R_{wp}/R_{exp})^2$

With increasing the $t_{ann}$ and concentration of the cadmium $x$, the lattice parameters ($a$, $c$) and unit cell volume ($V$) tend to decrease. It can be associated with reducing the number of dislocations $\delta$ (see Fig. S4 in SM2), structural defects and distortions, as well as broken chemical bonds with increasing the $t_{ann}$ [14, 31, 44-46] and replacement of the La$^{3+}$ ($r_A^{XII}$ = 1.36 Å) ions by smaller Cd$^{2+}$ ($r_A^{XII}$ = 1.31 Å) ones in the $A$-position for the 12-fold coordination number [47]. At the same time, the bandwidth $W$, characterized by the overlap between the Mn3d orbital and O2p orbital in the MnO$_6$ oxygen octahedra, increases. This structural parameter plays an important role in the formation of magneto-transport properties of the manganites and defined as $W \sim \cos(1/2(\pi-\theta_{<Mn-O-Mn>}))/(<d_{Mn-O}>)^{3.5}$ [48], where bond angle $\theta_{<Mn-O-Mn>}$ is an average angle between manganese ions Mn-O-Mn and bond length $<d_{Mn-O}>$ is an average distance between Mn-O. With increasing $t_{ann}$, the bond angle $\theta_{<Mn-O-Mn>}$ and length $<d_{Mn-O}>$ increases and decreases, respectively, leading to the strengthening DE and, as a result, increasing phase transition temperatures (see below). On the other side, besides the structural effect, the effect of particle size also works and is defined as a coherent scattering region $D_{XRD}$ or, in the first approximation, an average crystallite size. The $D_{XRD}$ values were obtained using Scherrer equation (see SM2). All L(C)NMO samples are NPs. Their average particle size tends to increase as both the $t_{ann}$ and $x$ increase (see Table 1). It can be associated with increasing the internal energy of the crystal structure, promoting atomic diffusion, and forming larger particles [49]. Moreover, the reasons for the broadening of XRD peaks and, consequently, changing effect of particle size were found out (see SM2). For the LNMO system, with increasing $t_{ann}$, the role of crystallite size decreases, whereas the lattice microstrains become dominant, respectively. At the same time, the prevalent lattice microstrains are maintained for almost all $t_{ann}$ for the LCNMO system. However, it should not be forgotten that with increasing $t_{ann}$, the peak sharpness increases, clearly denoting the reduction of lattice microstrains $\varepsilon$ and increase in crystallinity $D_{XRD}$ of the L(C)NMO samples (see Table 1, Fig. S3, and Table S1).

Additionally, the average particle size $D_{SEM}$ was determined from the SEM data (see the insets in Fig. 1). All L(C)NMO samples consist of well-formed spherical-like NPs with clear boundaries, the values of which are higher than $D_{XRD}$. Furthermore, the TEM data confirms the spherical nature of the samples with the same tendency of growing $D_{TEM}$ as $D_{XRD}$ with increasing $t_{ann}$ and $x$ (see Fig. 2 and Table 1). The reasons for the discrepancies in the average particle size obtained from the SEM, TEM, and XRD data can be due to insufficient particle size data sets with relatively high scale resolution (50-1000 nm) for their determination and, as a result, reduced coefficient of determination $R^2$, as well as their magnetic adhesion at room temperature. The EDS data indicates the homogeneous distribution of the La, Cd, Na, Mn, and O elements on the surface of the L(C)NMO samples with their approximate stoichiometric atomic percentage 16:4:20:60 and 15:1:4:20:60 ratio (see SM3). The SAED patterns and HRTEM data additionally confirm the rhombohedral structure of the L(C)NMO samples with determined (hkl) Miller's indexes (see Fig. 2). The size effect attributed to a core-shell structure was also detected using the HRTEM images and FFT patterns. With increasing $t_{ann}$, all L(C)NMO samples demonstrate a transformation of their surface from the defect (dislocations, distortions, broken chemical bonds, etc.) core-shell structure with dispersed diffraction spots to the more structurally homogeneous core [50, 51] that is also consistent with the reducing dislocation density $\delta$ (see Table 1). The most apparent core-shell structure can be observed for the L(C)NMO-500 samples with the shell thickness $t_{shell}^{TEM} \approx 1–2$ nm (see Fig. 2 (c) and (f), and Figs. S7-S8), which is additionally confirmed by $t_{shell}^{M(H)}$ values obtained from the magnetic data (see Table S3 in SM4). It can be conditionally noted that the L(C)NMO-700 samples are in the transitional stage from the core-shell structure to the pure core one.

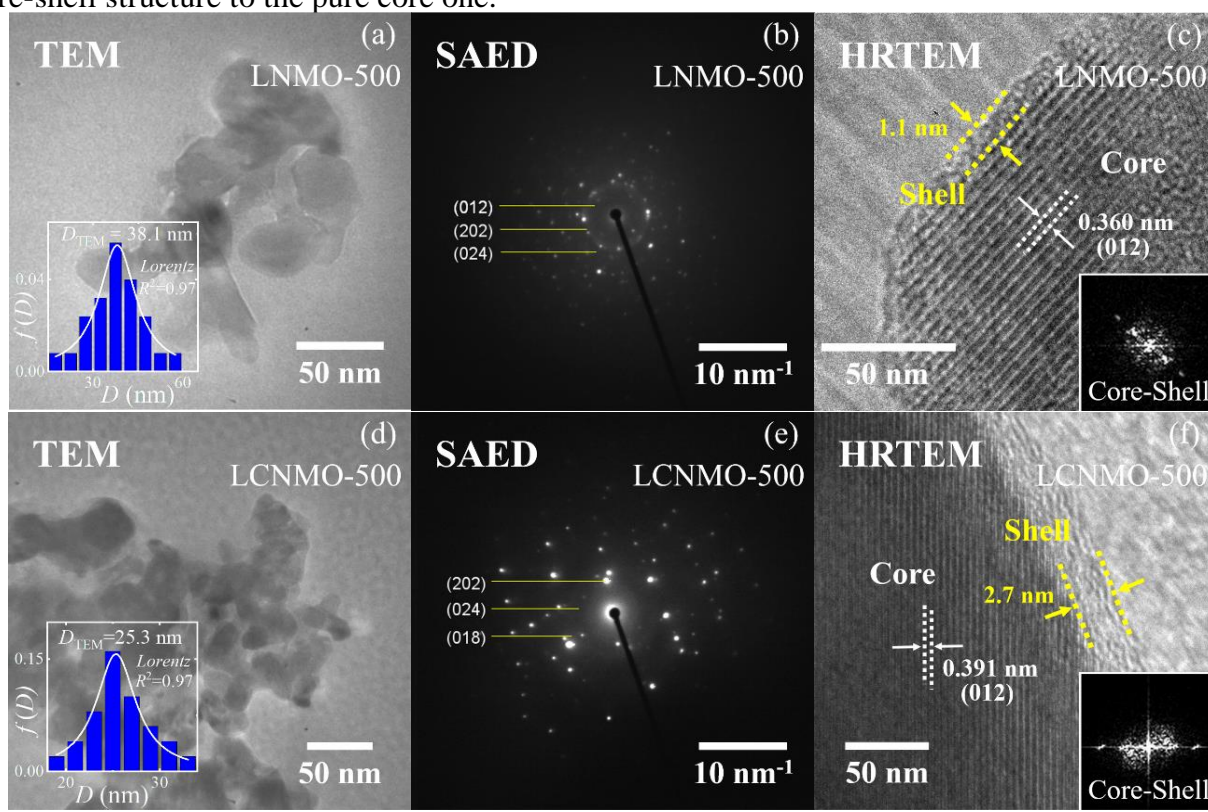

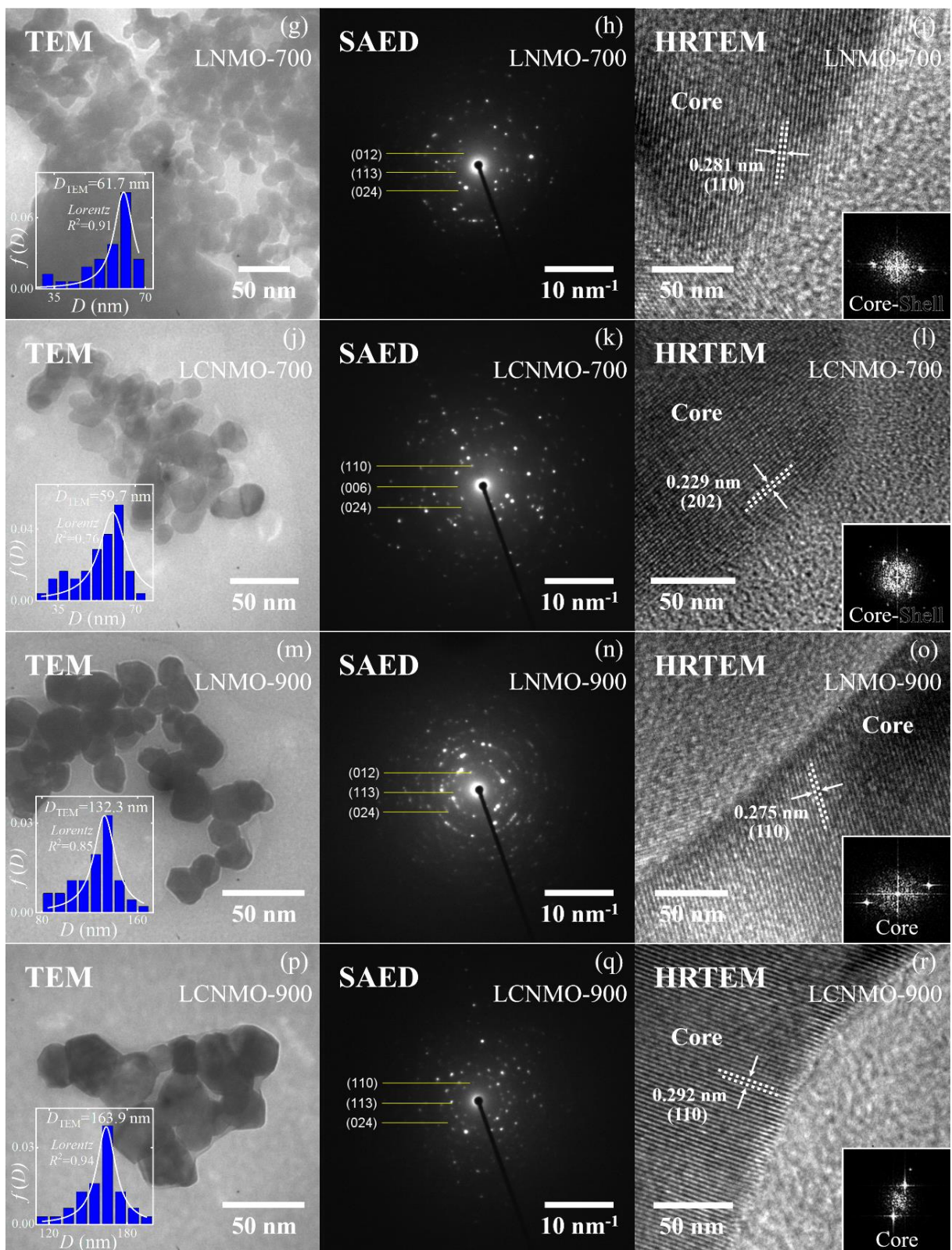

**Fig. 2**. TEM images with a particle size distribution (a, g, m) and (d, j, p), SAED patterns (b, h, n) and (e, k, q), HRTEM images with FFT patterns (c, i, o) and (f, l, r) for the LNMO and LCNMO samples, respectively.

XPS method was employed to study the surface of the L(C)NMO samples and determine the valence state of the included ions. As an example, the following XPS curves were presented for the LNMO-500 and LCNMO-500 samples only (Fig. 3). Survey spectra clearly demonstrate the presence of the La, Na, Cd, Mn, and O elements (Fig. 3(a, b)). All peak positions are presented in Table 2. The La3d spectra indicate two major peaks (Fig. 3(c, d)), corresponding spin doublet La3$d_{5/2}$ (A) and La3$d_{3/2}$ (C) with a difference of 16.8 eV, which agrees with the standard deviation [52]. Moreover, these spectra show charge transfer satellites from B and D features, as well as plasmon lines and MNN Auger lines from B' and D' features. Similar results were also observed for the XPS spectra of $La_{0.8-x}Na_{0.2}Mn_{1+x}O_3$ [31], $La_{0.9}Mn_{1.1}O_{3-\delta}$ [53], and $La_{0.7}Sr_{0.3}CoO_3$ [54], where La ions were in a trivalent state. The Na1s spectra (Fig. 3(e, f)) indicate a weak single symmetric peak at around 1071.4–1071.5 eV from $Na^+$ [31]. The Cd3d spectrum (Fig. 3(g)) shows spin-orbit splitting at $Cd3d_{5/2}$ and $Cd3d_{3/2}$ with a difference of 6.74 eV, which is in good agreement with literature data [55], indicating the presence of $Cd^{2+}$ ions. The Mn2p spectra have two major peaks at around 641 and 653 eV (Fig. 3(h)), corresponding to the $Mn2p_{3/2}$ and $Mn2p_{1/2}$ states, respectively. Their deconvolution into two components exhibits the availability of mixed valence state of manganese ions, i.e. $Mn^{3+}$ and $Mn^{4+}$. While $Cd^{2+}$ substitution, the ratio of $Mn^{3+}/Mn^{4+}$ reduces from 46.3/53.7 for the LNMO-500 to 42.0/58.0 for the LCNMO-500, indicating the increase in the $Mn^{4+}$ concentration that should influence the magnetic properties of the L(C)NMO samples. Moreover, the LCNMO-500 spectrum is a bit shifted towards higher energy relative to the LNMO-500 one, confirming additionally the rise in the $Mn^{4+}$ concentration and successful embedding of $Cd^{2+}$ ions in the perovskite structure. Regarding O1s spectra (Fig. 3(i, j)), three components with energy positions of about 529.1–529.3 (O1), 531.0–531.1 (O2), and 532.8–533.1 (O3) eV can be distinguished. They can be associated with the lattice oxygen species O1, the adsorbed oxygen O2, and hydroxyl groups or water on the surface O3 [56].

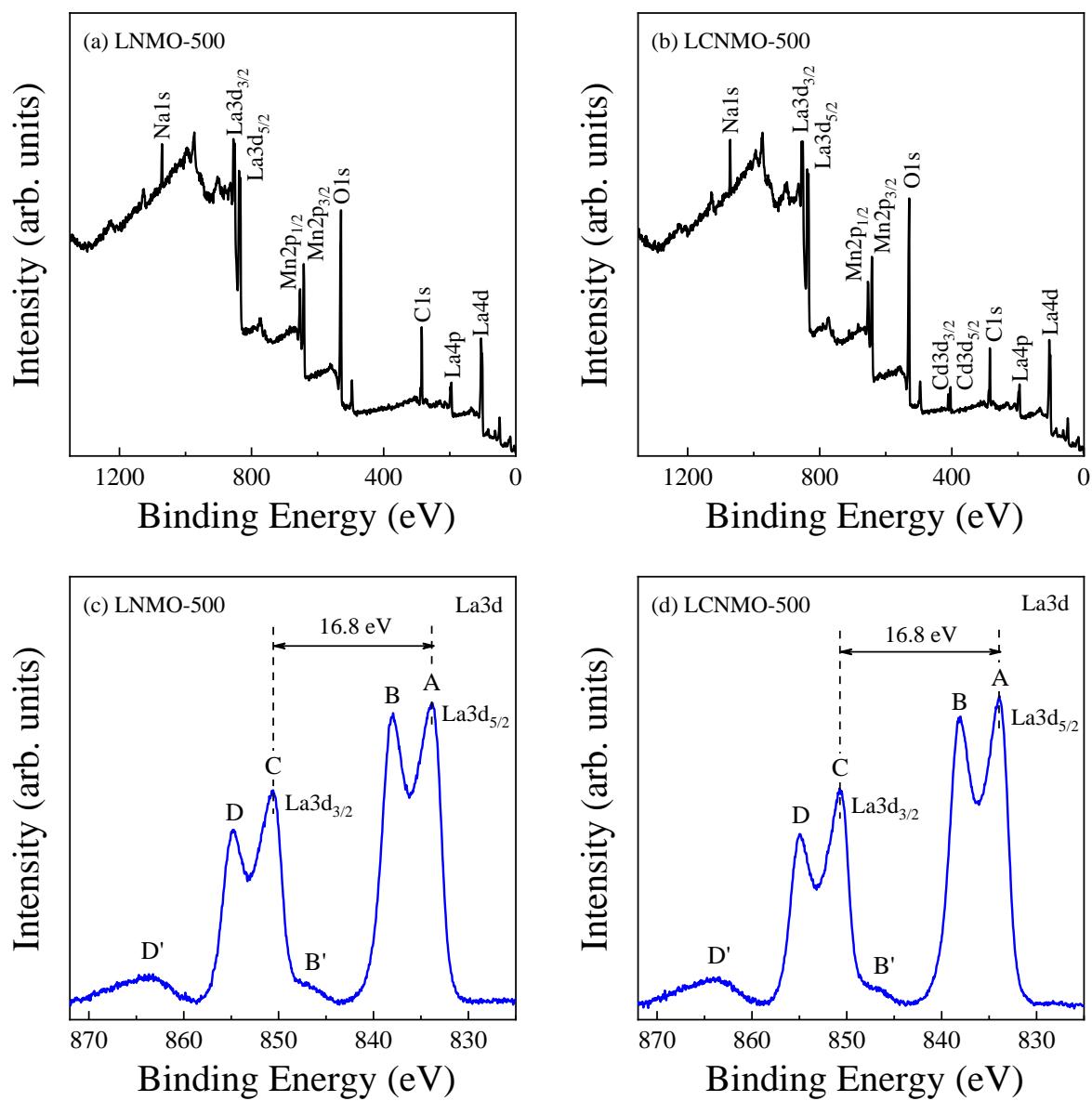

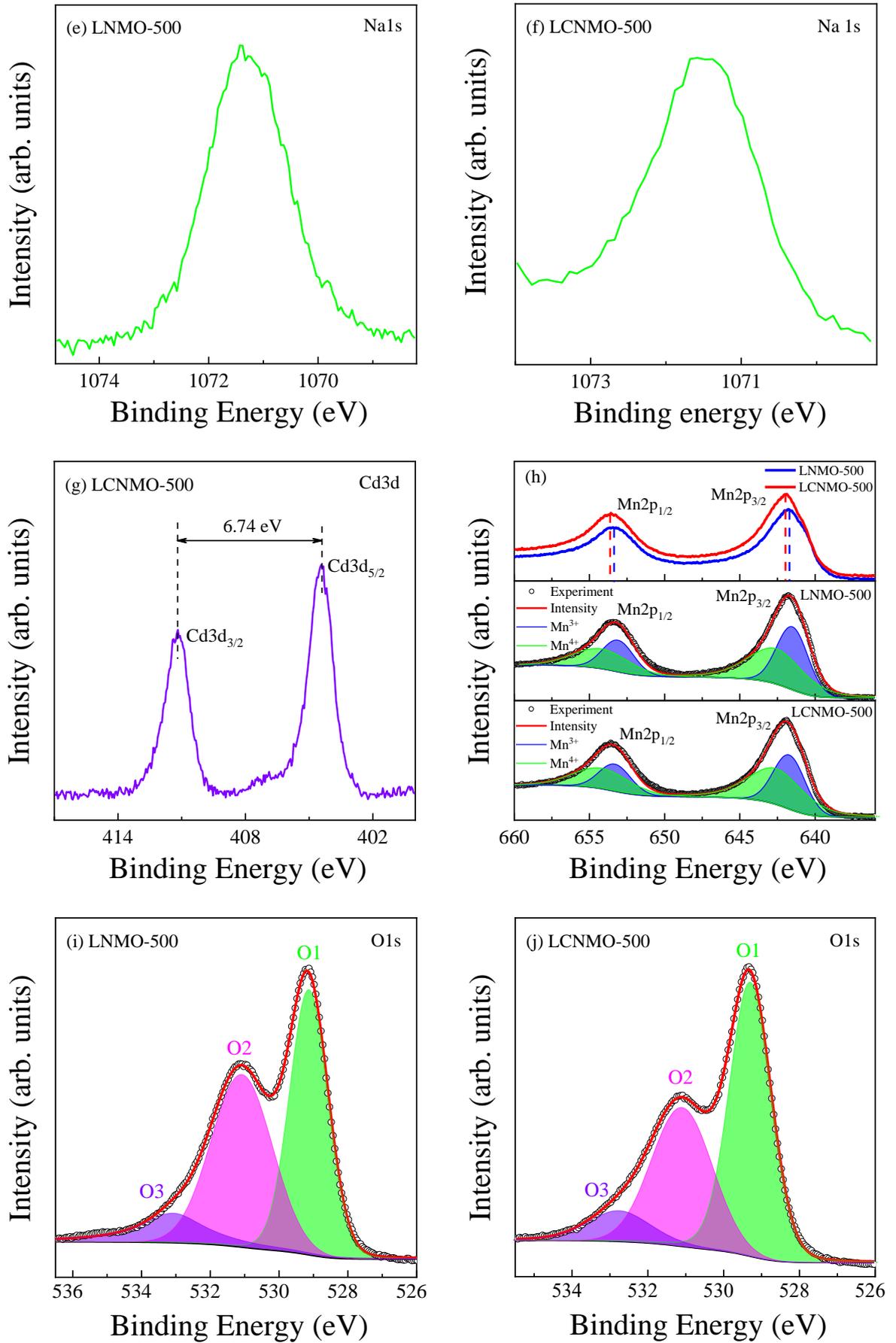

**Fig. 3**. XPS curves for the LNMO-500 and LCNMO-500 samples, respectively: (a) and (b) survey spectra; (c) and (d) La3d spectra with spin doublet (A and C), charge transfer satellites (B and D), as well as plasmon lines and MNN Auger lines (B' and D'); (e) and (f) Na1s spectrum; (g) Cd3d spectrum with spin-orbit splitting; (h) Mn2p spectra approximated by red envelope and deconvoluted by $Mn^{3+}$

(blue) and $Mn^{4+}$ (green) components; (i) and (j) O1s spectra deconvoluted by O1 lattice oxygen (green), O2 adsorbed oxygen (magenta) and O3 hydroxyl groups and/or water (violet) on the surface.

**Table 2**
Energy positions of the La3d, Na1s, Cd3d, Mn2p, and O1s XPS lines for the L(C)NMO-500 samples

| Sample | Binding energy (eV) | | | | | | | | | | | | |
|---|---|---|---|---|---|---|---|---|---|---|---|---|---|
| | La3d | | | | Na1s | Cd3d | | Mn2p | | | | O1s | | |
| | $La3d_{5/2}$ | Sat | $La3d_{3/2}$ | Sat | | $Cd3d_{5/2}$ | $Cd3d_{3/2}$ | $Mn2p_{3/2}$ | | $Mn2p_{1/2}$ | | O1 | O2 | O3 |
| | | | | | | | | $Mn^{3+}$ | $Mn^{4+}$ | $Mn^{3+}$ | $Mn^{4+}$ | | | |
| LNMO | 833.8 | 838.0 | 850.6 | 854.78 | 1071.4 | – | – | 641.5 | 642.6 | 653.1 | 654.2 | 529.1 | 531.0 | 533.1 |
| LCNMO | 833.9 | 838.1 | 850.7 | 854.98 | 1071.5 | 404.4 | 411.2 | 641.7 | 642.6 | 653.3 | 654.2 | 529.3 | 531.1 | 532.8 |

Thus, all L(C)NMO samples have the same rhombohedral crystal structure but differ only in chemical composition, particle size, interionic distance, microstrains, and defects. Their average particle size is within the 20–70 nm range with a small dispersion, making them technologically reliable and suitable samples for studying their functional properties. Moreover, the comprehensively defined structural and size parameters by different experimental methods have to affect the spontaneous magnetization $M_S$, coercivity $H_C$, residual magnetization $M_R$, Curie temperature $T_C$, blocking temperature $T_B$, paramagnetic Curie temperature $\theta$, magnetic phase content, magnetic entropy change $-\Delta S_M$, and $RCP$ of the studied L(C)NMO manganite NPs and also to be dependent from such critical factors as aging time $t$ and external high hydrostatic pressure $P$.

### 3.2. Magnetic properties

The influence of internal irreversible (structural-size effect and aging time $t$) and external reversible (hydrostatic $P$) pressures on the L(C)NMO NPs' magnetic properties is presented as the phase diagrams of states in Fig. 4. Here, there are two utmost states of particles: PM and FM. In an ensemble of NPs, due to the Boltzmann type of thermal fluctuations, the direction of the NPs' magnetic moments in the FM phase will not be fixed. Therefore, FM NPs are magnetized SPM with practically no coercivity. If thermal excitations are insufficient, then the magnetic orientation of the NPs' magnetic moments turns out to be partially blocked by their magnetic anisotropy fields. In this case, when NPs are magnetized, a coercivity arises. It has a relaxation type and is associated with the fact that the time of reversal of the particles' magnetic moments is comparable to the time of their demagnetization by the magnetic field. The effect of blocking magnetic moments is characterized by the blocking temperature $T_B$. The phase diagrams show: (i) paramagnetic Curie temperature $\theta$, which is a characteristic of the high-temperature PM phase of particles; (ii) the Curie temperature $T_C$, which is the critical temperature for the onset of magnetic ordering in NPs; (iii) blocking temperature $T_B$, below which the blocked SPM state of NPs is realized. The difference between $\theta$ and $T_C$ temperatures is associated with the influence of fluctuations of the particle ions' magnetic moments on the magnetic ordering and the appearance of the FM phase in particles. The $T_B$ is connected with fluctuations of the magnetic moment vector of the entire NP. Thus, phase diagrams also make it possible to determine the influence of fluctuations on the magnetic properties of the samples. However, if the particles of the ensemble are not the same, both structurally and magnetically, then this can also affect the behavior of the curves in the phase diagram.

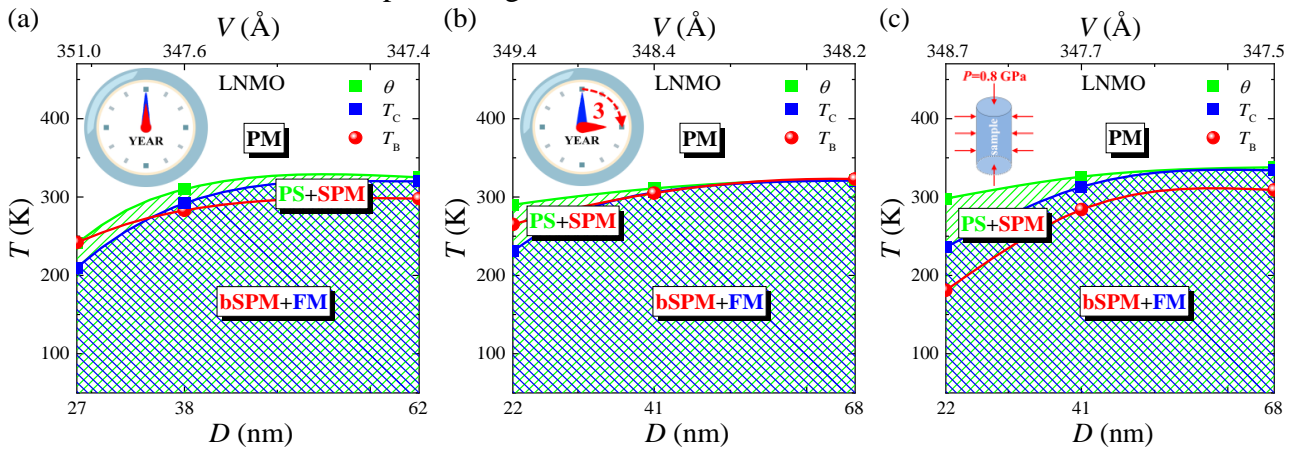

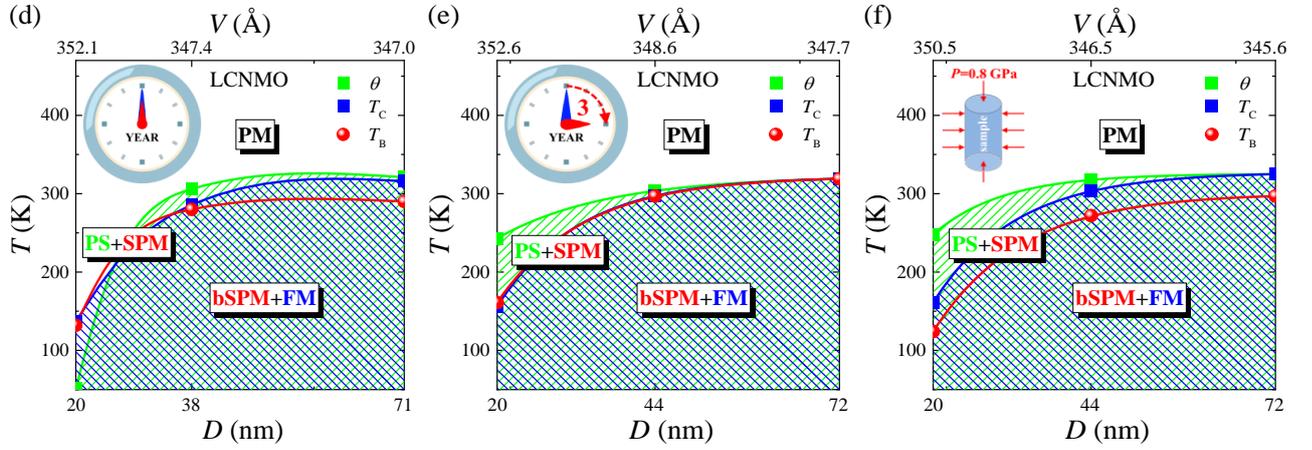

**Fig. 4**. Evolution of phase diagrams of states depending on the structural-size effect (a) and (d), aging time $t = 3$ years (b) and (e), and external high hydrostatic pressure $P = 0.8$ GPa (c) and (f) for the LNMO and LCNMO samples, respectively. $\theta$ is a paramagnetic Curie temperature, obtained from the $\chi^{-1}(T) = H/M(T)$ dependences; $T_C$ is a Curie temperature, obtained from the minimum derivative of $dM/dT(T)$; and $T_B$ is a blocking temperature, obtained from the $H_C(T^2)$ or $M_{ZFC}(T)$ peak dependences. PM is a paramagnetic state; PS is a phase separation; SPM is a superparamagnetic state; bSPM is a blocked superparamagnetic state; and FM is a ferromagnetic state.

The phase diagrams are the final result ("quintessence") of the deep analysis of many experimental data and approaches. They have considerable significance, allowing us to create new materials based on them and predict their magnetic behavior depending on these $t_{ann}$, $t$, and $P$ characteristics. The structural-size effect dependent phase diagrams (Fig. 4 (a) and (d)) of the L(C)NMO NPs were obtained based on the following procedure (see SM 4). First of all, the field dependences of the magnetization $M(H)$ upon different temperatures were measured in the magnetic field up to 1 T, then the temperature dependence of the coercivity $H_C(T)$ and residual magnetization $M_R(T)$ were defined (see Fig. S9). It should be noted that small values of coercivity indicate a magnetically soft nature of the studied L(C)NMO NPs. The blocking temperature $T_B$ was determined from the $H_C(T)$ curves, indicating the SPM state at $T > T_B$ and the blocked SPM state at $T < T_B$. Moreover, it was found that the coercive force is proportional to the square of the temperature $H_C \sim T^2$, allowing us to define the $T_B$ more accurately (see Fig. S10).

The Curie temperature $T_C$ was clarified from the temperature dependences of magnetization $M(T)$ under 2 and 10 kOe (see Fig. S11). The phase transition between PM and FM states is associated with the spontaneous appearance of an exchange field at the $T_C$, which orients the magnetic moments of particle ions, counteracting thermal fluctuations. The values of the $T_C$ were obtained from the derivative of $dM/dT(T) = \min$, indicating the PM state at $T > T_C$ and FM state at $T < T_C$. The $dM/dT(T)$ curve was narrower upon a smaller magnetic field, typical for the second-order phase transitions. Additionally, using the critical phenomena theory, the $T_C$ and critical exponent $\beta$ values were clarified (see Fig. S12), showing good agreement with the aforementioned determined Curie temperature and indicating the existence of the 3D-Ising and/or 3D-Heisenberg models with the short-range FM interactions caused by inhomogeneities, magnetic disorders, structural defects, charge orderings, and/or mixed magnetic states [31, 57].

Finally, the paramagnetic Curie temperature $\theta$ was determined from the $\chi^{-1}(T) = H/M(T)$ dependences during the intersection of the straight fitting line with the temperature axis (see Fig. S13), indicating the presence of PS from $\theta$ to $T_C$ [58]. The difference between these temperatures can also be caused by the fluctuations in the ions' magnetic moments (order parameter fluctuations), which are the most intense in the critical temperature region. Additionally, for the L(C)NMO particles, when the magnetic order changes, a change in ionic states can be observed due to the localization/delocalization of conduction $e_g$-electrons on magnetic ions [31]. The L(C)NMO-500 samples may have a significant dispersion of exchange parameters, so the particles may have different individual Curie temperatures. In this case, the experimentally measured $T_C$ of the sample is an effective quantity for a heterogeneous ensemble of particles with unequal magnetic properties [32] but with the same rhombohedral structure. It is clear from the phase diagrams of states that as particle size grows, the difference between the $\theta$ and $T_C$ disappears, indicating the forming of a more uniform ensemble of particles.

A similar procedure was carried out for plotting the time $t$ (see details in SM 5) and pressure $P$ (see details in SM 6) dependent phase diagrams except for the approach in determining the blocking temperature $T_B$ from the peak of the $M_{ZFC}(T)$ curves in the field of $H = 50$ Oe (see Figs. S14 and S16). All critical temperatures ($\theta$, $T_C$, and $T_B$) depending on the structural-size effect, aging time $t$, and pressure $P$ are collected in Table S4.

Considering magnetism of the initial L(C)NMO nanopowders in the structural-size effect plane (Fig. 4 (a) and (d)), it can be noted that with increasing $t_{ann}$ and, consequently, particle size $D$, the magnification of the phase transition temperatures ($\theta$, $T_C$, and $T_B$) with their saturation is observed. At the same time, the LCNMO compounds, compared with LNMO ones, have lower critical temperatures. All this can be explained within the structural-size effect framework [20, 59]. At the lowest $t_{ann} = 500$ °C, the particle size effect $D$ should be major in forming the L(C)NMO's magnetic properties. The particles are small, the role of interionic fluctuations is large, and there is a big difference between $T_C$ and $\theta$. However, the $T_B$ turned out to be close to the $\theta$, which is not associated with fluctuations. Indeed, the particle size effect $D$ and the presence of the structural defects $\delta$ are up to ~ 3.5 times smaller and up to ~ 14 times higher compared with the samples at $t_{ann} = 900$ °C (Table 1), respectively. Moreover, among the L(C)NMO samples at the same $t_{ann} = 500$ °C, the LCNMO has the smallest particle size $D = 20$ nm and highest microstrains $\varepsilon = 0.651\%$, indicating the most structurally inhomogeneous system and, consequently, the lowest phase transition temperatures [60]. With increasing $t_{ann}$, the particle size $D$ increases with growing core, decreasing shell $t_{shell}$, and reducing structural defects $\delta$ (Table 1). The structural effect is becoming dominant with the relative increasing bandwidth $\Delta W/W$ up to ~ 3%, responsible for the FM DE $Mn^{3+}\leftrightarrow O^{2-}\leftrightarrow Mn^{4+}$ [61]. Additionally, with increasing $t_{ann}$, the rate of changing $W$ slows down (Table 1), which is consistent with a trend of the critical temperature changes (Fig. 4 (a) and (d)).

After $t = 3$ years, the phase diagrams of the L(C)NMO samples have changed significantly (Fig. 4 (b) and (e)). The phase transition temperatures have increased noticeably, and the magnetic system has become more FM homogeneous with a reduced $\theta$-$T_C$ PS range. Remarkably, the most change in the phase transition temperatures is observed for the smallest L(C)NMO-500 nanoparticles. In general, all samples have become more structurally uniform, which can be seen from the magnetic data: the increased Curie temperature and blocking temperature, which are closer to the $\theta$ and $T_C$, respectively. It can be explained again by the structural-size effect (see SM7). However, here, the structural effect plays a crucial role with increased bandwidth $W$ and microstrains $\varepsilon$ (Table S5) compared with the initial L(C)NMO samples (Table 1) since the particle size of the L(C)NMO samples after 3 years has increased slightly, which may be due to the coalescence and/or diffusion process [42, 62, 63]. In other words, it may be supposed that time $t$ increases internal chemical pressure. The procedure for defining microstrains $\varepsilon$ of the L(C)NMO after 3 years was the same as for the initial samples in SM 2. The nature of the structural distortions is very similar, and the values of $\varepsilon$ decrease with increasing $t_{ann}$ (Tables S5 and S6).

Additionally, the studies of the time influence on the structural and magnetic properties of other Mn-containing $La_{0.7}A_{0.2}Mn_{1.1}O_3$ (L$A$MO) nanoparticles obtained at the $t_{ann} = 900$ °C, where $A$ = Na$^+$, Ag$^+$, K$^+$, have been carried out (see SM 8). Based on our previously published works [15, 31, 44, 64] and the data obtained after 3 years (Table S9), the phase diagrams of the L$A$MO samples have been plotted (Fig. 5). The procedure of defining structural and magnetic parameters was the same as aforementioned in SM2 and 5 and described in details in SM 8. Intriguing, for the biggest particles of the L(C)NMO and L$A$MO samples obtained at the same $t_{ann} = 900$ °C before and after 3 years, the change in phase transition temperatures is the smallest: the PM Curie temperature and Curie temperature are almost the same without PS region, but blocking temperature has become a little bit lower (see Figs. 4 (a, b, d, e) and 5). In general, it indicates that Mn-containing NPs annealed at $t_{ann} = 900$ °C are the most magnetically homogeneous and stable system regardless of time, and the obtained threshold $t_{ann}$ is optimal, which is valuable information for designing new materials with their further possible applied application. Moreover, the microstrains $\varepsilon$ of the L$A$MO after 3 years have increased, the nature of the structural distortions has remained the same, and the particle size has grown slightly (Tables S7 and S8).

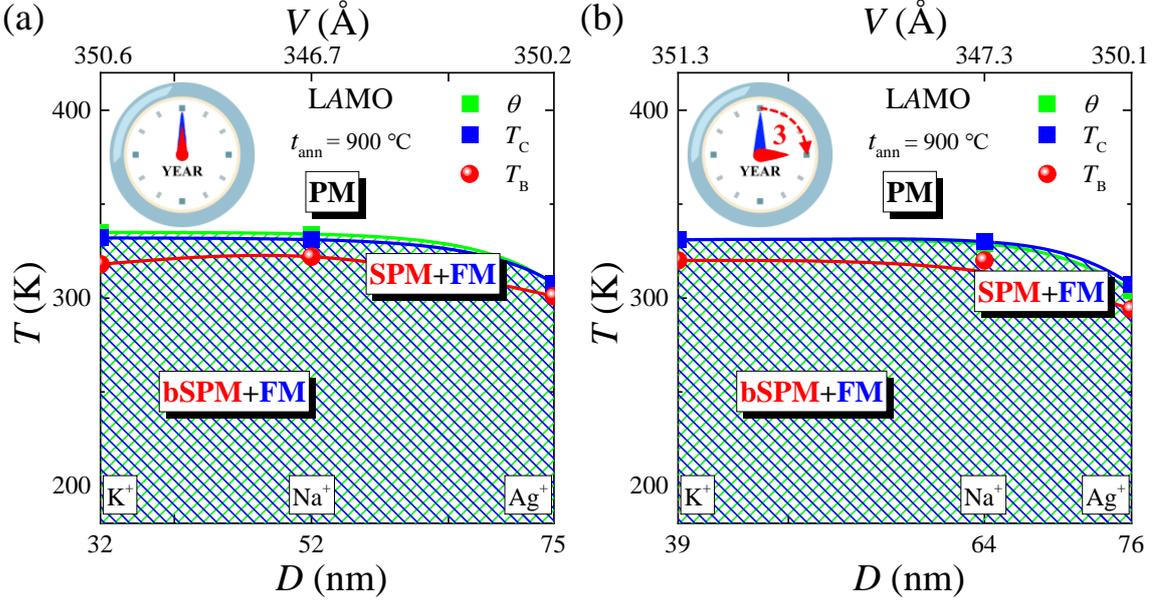

**Fig. 5**. Evolution of phase diagrams of states depending on aging time $t$ before (a) and after 3 years (b) for the L$A$MO ($A$ = Na$^+$, Ag$^+$, K$^+$) samples annealed at $t_{ann}$ = 900 °C, respectively. $\theta$ is a paramagnetic Curie temperature, obtained from the $\chi^{-1}(T) = H/M(T)$ dependences; $T_C$ is a Curie temperature, obtained from the minimum derivative of $dM/dT(T)$; and $T_B$ is a blocking temperature, obtained from the $M_{ZFC}(T)$ peak dependences. PM is a paramagnetic state; SPM is a superparamagnetic state; bSPM is a blocked superparamagnetic state; and FM is a ferromagnetic state.

Regarding the pressure $P$ effect on the L(C)NMO nanopowders' magnetic properties after 3 years (Fig. 4 (c) and (f)), it can be noted that $P$ leads to increasing both $\theta$ and $T_C$ and decreasing $T_B$ with their saturation at the highest $t_{ann}$. The most change in $T_C$ is observed for the biggest and the most magnetically uniform LNMO-900 NPs, up to ~ 13 K. In contrast, for the smallest and the most magnetically inhomogeneous LCNMO-500 NPs, it is up to ~ 5 K. A similar situation was observed in Co-containing nanoparticles [54]. Unexpectedly, the $T_B$ turned out to be supersensitive to external pressure. For the smallest L(C)NMO-500 NPs, the $T_B$ can change down to ~ 84 K. For the small particle samples, an external pressure preserved the manifestation of magnetic fluctuations and eliminated the manifestations of particle structure inhomogeneity in magnetization. In contrast, for the biggest L(C)NMO-900 NPs, the $T_B$ can vary down to ~ 22 K (Table S4). For all cases, the $T_B < T_C$ is fulfilled. A high external hydrostatic pressure makes the phase transitions more prominent and strengthens the system's magnetic nature (PM, FM, AFM, etc.). Besides the structural-size effect and aging time $t$, it is another tool and degree of freedom to control the magnetic properties of Mn-containing nanoparticles. Here, the change in particle size can be neglected during compression since the Curie temperature $T_C$ is the same as before and after the influence of high pressure (see Fig. S18). It indicates that the external reversible pressure works directly only with a structure. Therefore, the major factor for increasing $T_C$ and, consequently, strengthening FM DE is a modification of structural parameters (see SM 9), particularly bond angle $\theta_{<Mn-O-Mn>}$ and bond length $<d_{Mn-O}>$, straightening Mn-O-Mn and contracting Mn-O, respectively [20]. Compared with the $t_{ann}$ and $t$, the $P$ is the single way to reduce $T_B$. In other words, the pressure $P$ promotes the SPM of the NPs, decreases the magnetocrystalline anisotropy, and lowers the energy barrier, which is an interesting direction and requires further research.

Another valuable characteristic of the studied L(C)NMO nanopowders from both fundamental and applied points of view is MCE depending on the structural-size effect and aging time $t$ (Fig. 6). MCE is associated with a change in magnetic entropy when a sample is magnetized. The higher the magnetic field, the higher the degree of ordering and, accordingly, the more significant the change in entropy. Ordering is associated with overcoming thermal fluctuations, the intensity of which depends on temperature. The higher the temperature, the higher the disorder. The highest MCE can be expected in magnets with an order-disorder phase transition, and the MCE is most pronounced near $T_C$. Since the $T_C$ of NPs correlates with their sizes, it is interesting to compare the MCE values with the particle sizes before and after 3 years. MCE was measured indirectly through the magnetic entropy change $\Delta S_M$ upon isothermal conditions using the numerical integration method of the Maxwell relation [16]:

$$\Delta S_M(T, \mu_0 H) = S_M(T, \mu_0 H) - S_M(T, 0) = \int_0^{\mu_0 H} \left(\frac{\partial M}{\partial T}\right)_{\mu_0 H} d(\mu_0 H). \qquad (1)$$

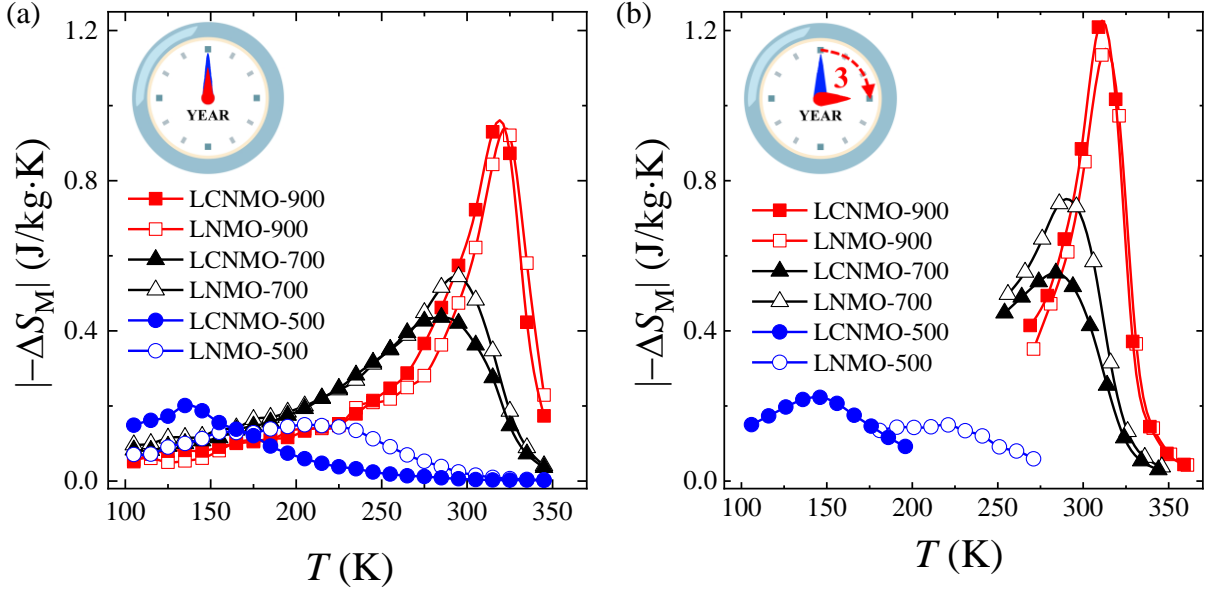

**Fig. 6**. Evolution of temperature-dependent magnetic entropy change $\Delta S_M(T)$ depending on the structural-size effect (a) and aging time $t = 3$ years (b) for the L(C)NMO samples.

All magnetic field dependences of magnetization $M(H)$ before and after 3 years are collected in SM 10. For all L(C)NMO samples with increasing $t_{ann}$, a peak position $T_{MCE}$ shifts to the higher temperature region, and a maximum of MCE $\Delta S_M^{max}$ increases with a sharper peak (Fig. 6 (a)). There is a trend of increasing both $\Delta S_M^{max}$ and $T_{MCE}$ for the LCNMO with increasing $t_{ann}$ compared with LNMO (Table 3). Moreover, with increasing $t_{ann}$, the $RCP$, defined as $RCP = \Delta S_M^{max} \cdot \delta T_{FWHM}$, increases and is higher for the LCNMO samples compared with LNMO at the same conditions $t_{ann}$. It may be said that in the LCNMO samples, there is an optimal concentration of structural disorders, imperfections, distortions, etc. ($t_{shell}, \delta, \varepsilon$ in Table 1), which helps to obtain higher values of the $\Delta S_M^{max}$ and $RCP$. The bigger particles, the higher values of the magnetization $M$ and, consequently, $\Delta S_M^{max}$, which has a linear behavior $\Delta S_M^{max}(D^{-1})$ (see Fig. S27) and is also observed in Ca-manganites [23]. It should be noted that MCE strongly depends on the exchange interaction constant, which defines phase transition temperature $T_C$ and magnetization $M$ values, which are higher for the bigger particles. Moreover, the $\Delta S_M$ also depends on the magnetization rate of the sample. At the Curie point, the magnetization has a power-law field dependence $M(H, T = T_C) \sim H^\Delta$ [15]. The greater the degree $\Delta$, the easier the sample is magnetized and the higher the MCE. Therefore, the MCE achieves the maximum values with the sharper peak for the bigger L(C)NMO-700 and L(C)NMO-900 NPs due to higher both the exchange interaction constant and magnetization rate because of the nonlinearity of spin-spin interactions in manganites [31]. In contrast, for the smaller L(C)NMO-500 NPs, the $\Delta S_M(T)$ shows broader peaks without sharpness, which indicates higher magnetic inhomogeneity.

**Table 3**

Magnetocaloric parameters (-$\Delta S_M$, MCE peak temperature $T_{MCE}$, and $RCP$) for the L(C)NMO nanopowders depending on the structural-size effect and time $t$ parameters.

| Sample | $t_{ann}$ (°C) | D (nm) | V (Å³) | t (year) | $\mu_0\Delta H$ (T) | $-\Delta S_M^{max}$ (J/kg·K) | RCP (J/kg) | $T_{MCE}$ (K) | $T_C$ (K) |
|---|---|---|---|---|---|---|---|---|---|
| LNMO | 500 | 27 | 351.001 | 0 | 1 | 0.15 | 17.2 | 205 | 210 |
| | | 22 | 349.411 | 3 | 1 | 0.15 | 12.6 | 221 | 231 |
| | 700 | 38 | 347.602 | 0 | 1 | 0.54 | 27.1 | 295 | 292 |
| | | 41 | 348.374 | 3 | 1 | 0.73 | 42.1 | 291 | 306 |
| | 900 | 62 | 347.398 | 0 | 1 | 0.92 | 40.8 | 325 | 320 |
| | | 68 | 348.185 | 3 | 1 | 1.14 | 41.8 | 311 | 321 |
| LCNMO | 500 | 20 | 352.070 | 0 | 1 | 0.20 | 18.4 | 135 | 137 |
| | | 20 | 352.566 | 3 | 1 | 0.22 | 18.8 | 146 | 156 |
| | 700 | 38 | 347.432 | 0 | 1 | 0.44 | 30.1 | 285 | 286 |
| | | 44 | 348.562 | 3 | 1 | 0.55 | 31.6 | 284 | 298 |
| | 900 | 71 | 346.978 | 0 | 1 | 0.93 | 44.3 | 315 | 316 |
| | | 72 | 347.734 | 3 | 1 | 1.21 | 46.5 | 309 | 319 |

Incredibly, the MCE of all L(C)NMO samples after 3 years has increased significantly up to 26% for the $\Delta S_M^{max}$ and 36% for the $RCP$, whereas the peak appearance of $T_{MCE}$ has decreased slightly (Fig. 6 (b)). The highest change in the $\Delta S_M^{max}$ and $RCP$ is observed for the biggest L(C)NMO-900 NPs with the most magnetically homogeneous and structurally perfect system before and after 3 years. In contrast, the lowest change occurs for the smallest L(C)NMO-500 NPs with the most magnetically and structurally inhomogeneous system. On the other hand, there is a change in the $T_{MCE}$ for the smallest and biggest NPs (Table 3). Additionally, a large difference between $T_{MCE}$ and $T_C$ up to 15 K is observed, which may be connected with the big step in $\Delta T = 10$ K during $M(H)$ measurements and demands additional research. A similar situation was also observed in other manganites [15, 31, 44]. The best sample with the most improved MCE parameters near room temperature phase transition after 3 years is the LNMO-700. In general, such enhancement of the MCE characteristics of the L(C)NMO samples after 3 years can be explained by changes in structural-size effect: growing particle size $D$, increasing microstrains $\varepsilon$, and strengthening DE $W$ (see Tables 1 and S5).

Summarizing, the following important fundamental and practical statements can be made: (i) the bigger NP size with less structural defects, the wider bandwidth $W$ and, consequently, phase transition temperature $T_C$; (ii) after $t = 3$ years, the largest change in the phase transition temperatures occur for the smallest NPs with the minimal change in MCE, whereas for the biggest NPs, it is oppositely, the maximum change in MCE and the lowest change in the phase transition temperatures; (iii) time $t$ makes the magnetic system more homogeneous, especially for the smallest "sensitive" NPs; (iv) an external high pressure has the greatest impact on the biggest and the most magnetically uniform NPs, whereas for the smallest one, it has the lowest effect; (v) the most stable magnetic system with improved MCE and the lowest deviation in the phase transition temperature after $t = 3$ years is the biggest NPs; (vi) for the smallest NPs, the fluctuations play an important role, manifesting in reduced $T_C$, increased PS, and broadened $\Delta S_M(T)$ compared with the bigger NPs, for which fluctuations weaken with increased $T_C$, reduced PS, and narrowed $\Delta S_M(T)$; and (vii) additionally, the contribution of the internal (structural-size effect and aging time) and external (hydrostatic) pressures on the Curie temperature as $dT_C/dP$ has been determined (see SM9): 1) the most "sensitive" NPs to the internal pressure have turned out LCNMO ones, and among them, the aging time $t$ has the greatest influence on the phase transition temperature with $dT_C/dP \approx 100$ K/GPa for the smallest LCNMO-500 NPs, whereas the structural-size effect significantly changes the Curie temperature with $dT_C/dP = 91$ K/GPa for the biggest LCNMO-900 NPs (Table S11); (2) in contrast, the external hydrostatic pressure has the highest impact on the biggest and more magnetically homogeneous LNMO-900 NPs with $dT_C/dP = 16$ K/GPa; (3) amazingly, at comparing absolute values of the internal irreversible and external reversible pressures, the aging time $t$ takes first place on the influence on the phase transition temperature $T_C$.

Thus, based on the plotted phase diagrams (Figs. 4 and 5) and $\Delta S_M(T)$ dependences (Fig. 6), it has been shown for the first time that there are powerful tools for manipulating magnetic phase transition temperatures and MCE of the Mn-containing perovskites as temperature $t_{ann}$, aging time $t$, and high hydrostatic pressure $P$, knowing of which is a crucial for both fundamental and practical purposes, as well as for creating new refrigerant and/or other electronic materials with desired functional parameters.

## 4. Conclusions

Based on the thorough and comprehensive analysis of the tremendous massive experimental data performed for the $La_{0.8-x}Cd_xNa_{0.2}MnO_3$ ($x = 0$ and $0.05$) NPs obtained under different $t_{ann} = 500$-$900$ °C and additionally for the $La_{0.7}A_{0.2}Mn_{1.1}O_3$ ($A = Na^+, Ag^+, K^+$) NPs obtained at the $t_{ann} = 900$ °C with using XRD, SEM, EDS, TEM, HRTEM, SAED, XPS, magnetic, and magnetocaloric methods, the following conclusions can be made with their extrapolation and generalization for the whole class of Mn-containing inorganic perovskites. The NPs with the smallest size have been the most structurally modified with increased size $D$, broadened bandwidth $W$, raised microstrains $\varepsilon$, and decreased dislocation density $\delta$ after $t = 3$ years. The $La_{0.8-x}Cd_xNa_{0.2}MnO_3$ samples show the presence of different valence $Mn^{3+}/Mn^{4+}$ manganese states. The NPs clearly demonstrate the "core-shell" structure with its reducing upon increasing $t_{ann}$. The structural-size effect has been detected, and its significant influence on magnetic and magnetocaloric properties has been observed. The smallest NPs have turned out to be the most magnetically "sensitive" ones to aging time $t$ with the increased phase transition temperatures: paramagnetic Curie temperature $\theta$, Curie temperature $T_C$, and blocking temperature $T_B$. The biggest NPs are the most stable ones with the slightest deviation in the phase transition temperatures. Time $t$ compared with an external hydrostatic pressure $P$ makes the Mn-containing magnetic system more homogeneous. The structural-size and particle's magnetic moment fluctuations have been observed, greatly impacting the phase transition temperatures and MCE of the smallest NPs. In contrast, their contribution to the bigger ones is reduced significantly. External hydrostatic pressure has the strongest influence on the $T_C$ of the biggest and $T_B$ of the smallest NPs, leading to their greatest increase and decrease, respectively. It has been found that the aging of the studied nanoparticles is not accompanied by degradation of their magnetic properties. The most significant improvement in MCE is for the biggest nanoparticles after $t = 3$ years. As it turned out, such underrated parameter as aging time $t$ has the greatest influence on the Curie temperature for the smallest NPs, achieving its maximum value $dT_C/dP \approx 100$ K/GPa for the LCNMO-500 ones. At the same time, the structural-size effect and external hydrostatic pressure $P$ have the highest effect on the biggest and more magnetically homogeneous NPs with the maximum value of $dT_C/dP \approx 91$ K/GPa for the LCNMO-900 and $dT_C/dP \approx 16$ K/GPa for the LNMO-900, respectively. The obtained results demonstrate different efficient ways to tune magnetic phase transition temperatures and MCE of the Mn-containing perovskites owing to internal irreversible (annealing temperature $t_{ann}$ and aging time $t$) and external reversible (hydrostatic $P$) pressures. This makes it possible to design new functional materials with specified parameters for various fields of science and technology, from medicine to space. In addition, this may prompt the creation of a new direction for studying the functional properties of perovskites, which, along with temperature and pressure, also depend on the aging time.

**Acknowledgments**
This work is supported by the European Union within the Project 101120397 — APPROACH.

**Declaration of Competing Interest**
The authors declare that they have no known competing financial interests or personal relationships that could have appeared to influence the work reported in this paper.

# Supplementary Material

# Structural-size effect, aging time, and pressure dependent functional properties of Mn-containing perovskite nanoparticles

Danyang Su[a], N.A. Liedienov[a,b,*], V.M. Kalita[c,d,e], I.V. Fesych[f], Wei Xu[g], A.V. Bodnaruk[e], Yu.I. Dzhezherya[c,d,e], Quanjun Li[a], Bingbing Liu[a], G.G. Levchenko[a,b,h,*]

[a]*State Key Laboratory of Superhard Material, Jilin University, 130012 Changchun, China*
[b]*Donetsk Institute for Physics and Engineering named after O.O. Galkin, NAS of Ukraine, 03028 Kyiv, Ukraine*
[c]*National Technical University of Ukraine "Igor Sikorsky Kyiv Polytechnic Institute", 03056 Kyiv, Ukraine*
[d]*Institute of Magnetism, NAS of Ukraine and MES of Ukraine, 03142 Kyiv, Ukraine*
[e]*Institute of Physics, NAS of Ukraine, 03028 Kyiv, Ukraine*
[f]*Taras Shevchenko National University of Kyiv, 01030 Kyiv, Ukraine*
[g]*State Key Laboratory of Inorganic Synthesis and Preparative Chemistry, College of Chemistry, Jilin University, 130012 Changchun, China*
[h]*International Center of Future Science, Jilin University, 130012 Changchun, China*

Corresponding author
*E-mail addresses*:   nikita.ledenev.ssp@gmail.com (N.A. Liedienov)
g-levch@ukr.net (G.G. Levchenko)

**Process of the formation of a perovskite structure**

The derivatogram of the powder mixture of La–Na–Mn–O nitrate-citrate gel after self-ignition at 200 °C can be conditionally divided into three temperature regions (see Fig. S1): (i) thermal process from room temperature to 210 °C associated with slight weight loss (-4.0 %) which is a result of evaporation of physically adsorbed water and surface hydroxyl groups removal from the combusted gel; (ii) the second significant weight loss (-13.0 %) occurs in the temperature range between 210-500 °C and can be attributed to the decomposition of lanthanum oxycarbonate $La_2O_2CO_3$ and sodium nitrite $NaNO_2$, the presence of which in the self-ignited powder composition is confirmed by the XRD data along with the presence of manganese oxide $Mn_3O_4$ (see Fig. S2); (iii) after 500 °C, the TG curve shows a slight mass loss (-3.3 %) and then an insignificant increase in the mass of the sample (+2.0 %), which may be associated with the implementation of redox processes in the oxide matrix with the formation of different oxidation states of manganese ions [1, 2]. Additionally, the changing mass of the sample is accompanied by a broad endo-effect. Considering the obtained thermal results, the L(C)NMO manganite synthesis was performed at 500, 700, and 900 °C to define the effect of heat treatment temperature on their functional properties.

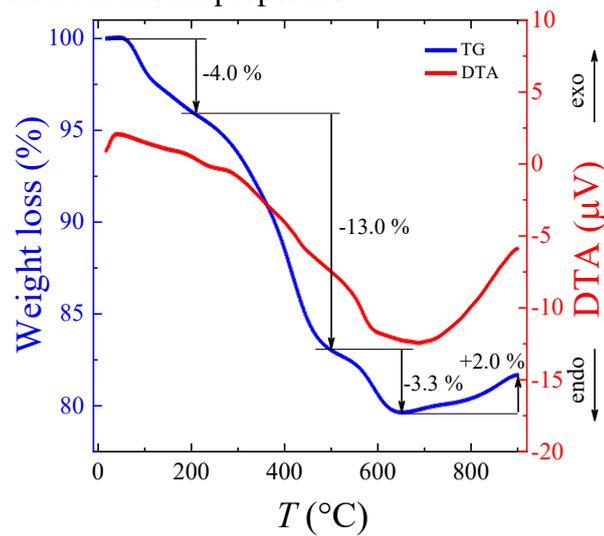

**Fig. S1.** TG/DTA curves of the La–Na–Mn–O self-ignited powder.

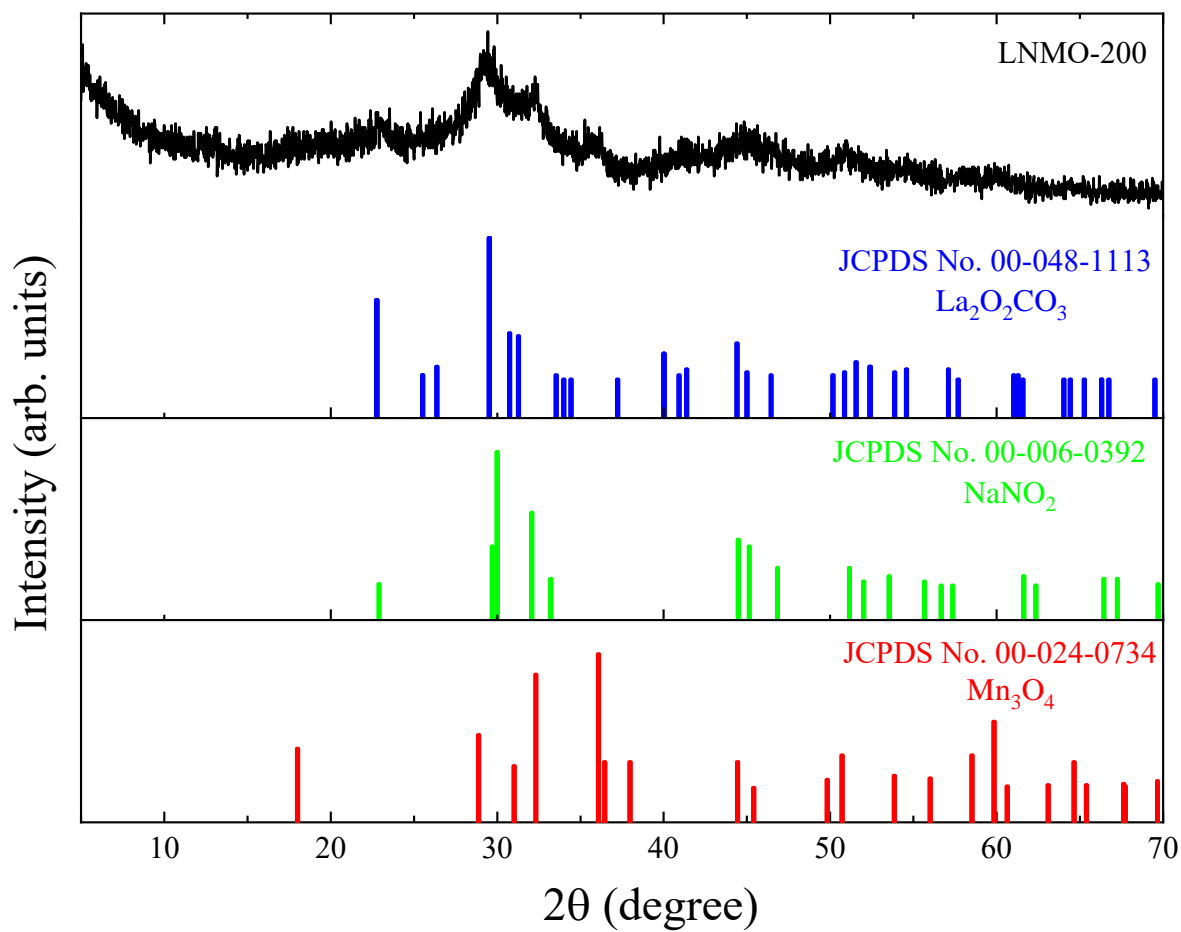

**Fig. S2.** XRD pattern for the as-prepared $La_{0.8}Na_{0.2}MnO_3$ powder obtained after self-ignition at 200 °C. The standard reflexes of monoclinic lanthanum carbonate $La_2O_2CO_3$ (JCPDS Card No. 00-048-1113), orthorhombic sodium nitrite $NaNO_2$ (JCPDS Card No. 00-006-0392), and tetragonal manganese oxide $Mn_3O_4$ (JCPDS Card No. 00-024-0734) are also noted.

## X-ray diffraction data analysis

The average crystallite size of the L(C)NMO samples was calculated by the X-ray line broadening method using the Scherrer formula [1]:

$$D_{012} = \frac{K \cdot \lambda}{\beta_{012} \cdot \cos\theta_{012}},$$

where $D_{012}$ (nm) is the average size of crystallites along the direction normal to the diffraction plane (012), $K$ is a constant related to crystallite shape, ordinarily equal to 0.9, $\lambda$ is the X-ray wavelength 0.15406 (nm) of Cu$K_\alpha$-radiation, $\beta_{012}$ is the integral breadth of the peak related to the diffraction plane (012) and $\theta_{012}$ is the Bragg angle in radians for the crystallographic plane (012). Considering that the integral width of the peak in the diffraction pattern is approximated by the pseudo-Voigt function with a large (up to 90% or more) contribution of the Lorentz function, the Lorentzian was used to describe the shape of the diffraction reflection at $2\theta \approx 22.8°$ (see Fig. S3). In order to exclude the instrumental broadening $\beta_{inst}$, standard silicon (Si) X-ray powder diffraction data is recorded under the same condition and is eliminated from the observed peak width. The true integrated peak width was calculated using the formula [2]:

$$\beta_{012} = \sqrt{\beta_{exp}^2 - \beta_{inst}^2},$$

where $\beta_{exp}$ is the experimental peak width of the sample at half maximum intensity, and $\beta_{inst}$ is an instrumental broadening of the diffraction line, which depends on the design features of the diffractometer or instrumental broadening (in radians). The crystallite size is assumed to be the size of a coherently diffracting domain, and it is not necessarily the same as particle size.

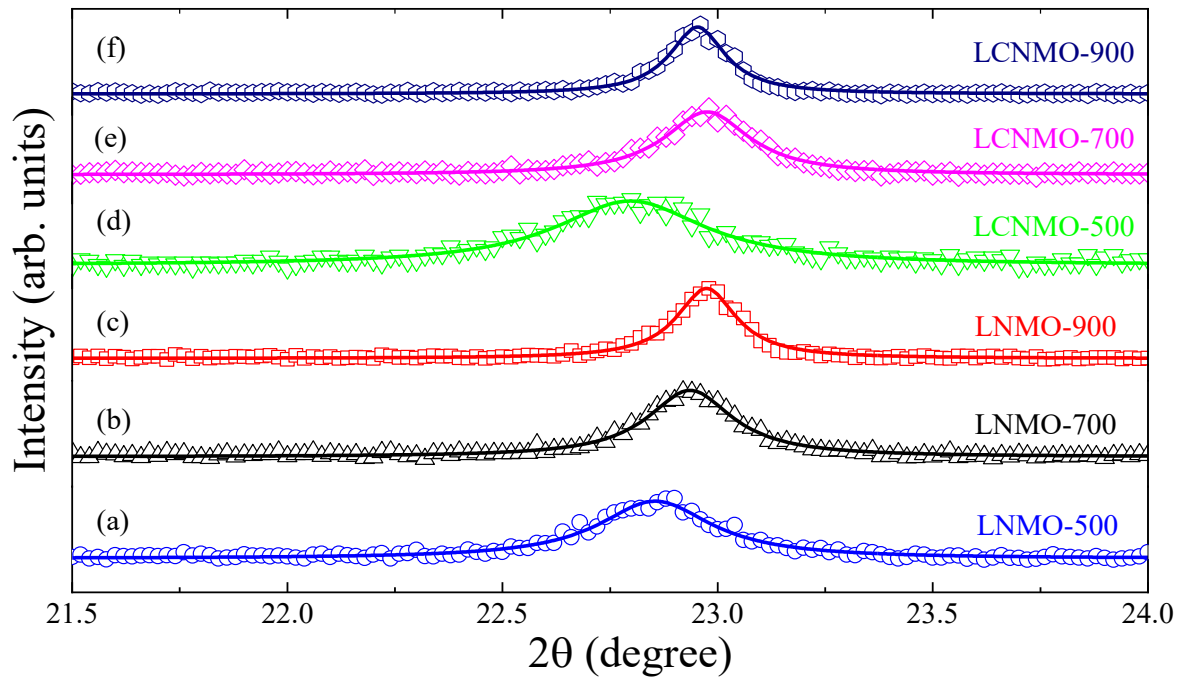

**Fig. S3**. Fitting a typical Bragg peak (012) with $2\theta \approx 22.8°$ of the LNMO (a, b, c) and LCNMO (d, e, f) perovskite phases using a Lorentz function.

Fig. S4 shows an annealing temperature dependence of the nanoparticle size $D_{XRD}(t_{ann})$ and dislocation density $\delta(t_{ann})$ for the L(C)NMO manganites. The dislocation density ($\delta$), which represents the number of defects in the sample, is defined as the length of dislocation lines per unit volume of the crystal and is calculated using the equation:

$$\delta = 1 / D^2_{XRD}.$$

The size of the coherent scattering regions $D_{XRD}$ increases proportionally to the annealing temperature $t_{ann}$, whereas the dislocation density $\delta$ decreases, indicating particle enlargement and agglomeration.

This leads to increasing the nanocrystallinity of the L(C)NMO samples and reducing the concentration of defects.

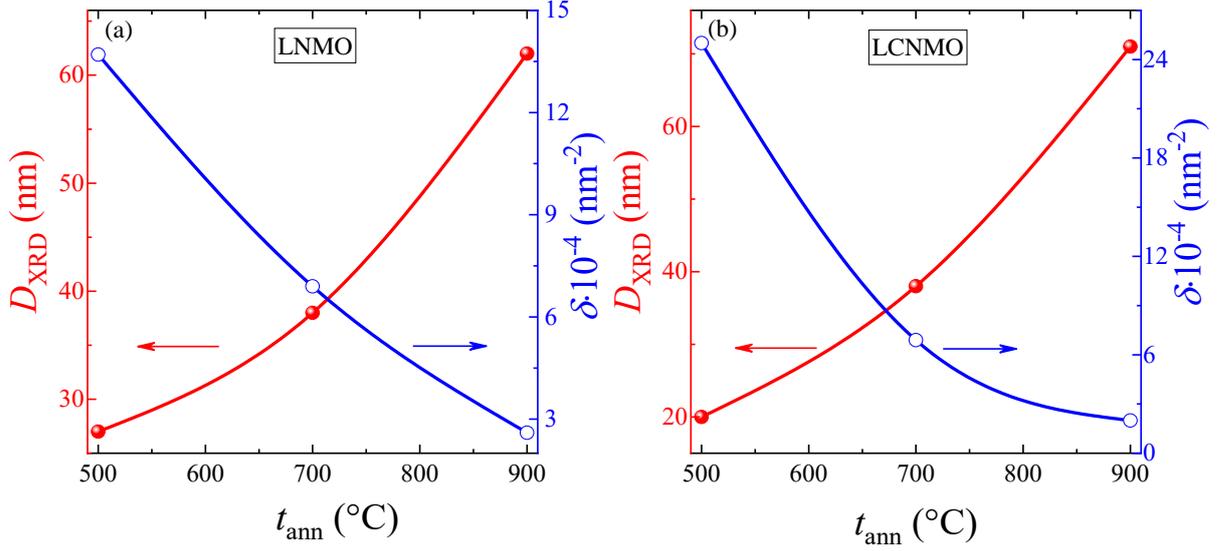

**Fig. S4**. Variation of average crystallite size ($D_{XRD}$) and dislocation density ($\delta$) for the LNMO (*a*) and LCNMO (*b*) samples obtained under different annealing temperatures $t_{ann}$.

**Table S1**

XRD broadening parameters of (012) and (024) reflexes of the L(C)NMO samples obtained under different annealing temperatures $t_{ann}$.

| Name | $2\theta_{012}$ (degree) | $2\theta_{024}$ (degree) | $\beta_{exp-012}$ (degree) | $\beta_{exp-024}$ (degree) | $\beta_{inst}$ (degree) | $\beta_{012}$ (degree) | $\beta_{024}$ (degree) | $\beta_{024}/\beta_{012}$ | $\cos\theta_{012}/\cos\theta_{024}$ | $\mathrm{tg}\theta_{024}/\mathrm{tg}\theta_{012}$ |
|---|---|---|---|---|---|---|---|---|---|---|
| LNMO-500 | 22.8548 | 46.8573 | 0.321 | 0.411 | 0.106 | 0.3030 | 0.3971 | 1.3 | 1.1 | 2.1 |
| LNMO-700 | 22.9358 | 46.9198 | 0.240 | 0.323 | 0.106 | 0.2155 | 0.3051 | 1.4 | 1.1 | 2.1 |
| LNMO-900 | 22.9739 | 46.9499 | 0.168 | 0.243 | 0.106 | 0.1305 | 0.2187 | 1.7 | 1.1 | 2.1 |
| LCNMO-500 | 22.7976 | 46.6714 | 0.425 | 0.833 | 0.106 | 0.4112 | 0.8262 | 2.0 | 1.1 | 2.1 |
| LCNMO-700 | 22.9753 | 46.9563 | 0.240 | 0.331 | 0.106 | 0.2150 | 0.3136 | 1.5 | 1.1 | 2.1 |
| LCNMO-900 | 22.9531 | 46.9473 | 0.156 | 0.249 | 0.106 | 0.1150 | 0.2253 | 2.0 | 1.1 | 2.1 |

It is well known that the ratio of the half-width of X-ray peaks, for example, (012) and (024) for a rhombohedral $R\bar{3}c$ perovskite structure, lies between $\cos\theta_{024}/\cos\theta_{012} < \beta_{024}/\beta_{012} < \mathrm{tg}\theta_{024}/\mathrm{tg}\theta_{012}$. Based on the coincidences of $\beta_{024}/\beta_{012} \approx \mathrm{tg}\theta_{024}/\mathrm{tg}\theta_{012}$ or $\beta_{024}/\beta_{012} \approx \cos\theta_{024}/\cos\theta_{012}$, the reason for the broadening of X-ray peaks results in the lattice microstrains or the crystallite size, respectively. If $\beta_{024}/\beta_{012}$ is in the middle of this interval, both factors affect the line broadening.

Strain-induced broadening arising from crystal defects and distortion is calculated employing the Stokes-Wilson equation [3]:

$$\varepsilon = \frac{\beta_\varepsilon}{4\tan\theta_{hkl}}.$$

The two equations presented below assume that the size ($\beta_D$) and strain ($\beta_\varepsilon$) broadening represent the total integral width ($\beta_{hkl}$) of a Bragg peak [4]. The angle dependences of the size and strain broadening in the Williamson-Hall analysis are described by the following equations:

$$\beta_{hkl} = \beta_D + \beta_\varepsilon = \frac{K\cdot\lambda}{D_{hkl}\cdot\cos\theta_{hkl}} + 4\varepsilon\cdot\tan\theta_{hkl}.$$

$$\beta_{hkl}\cdot\cos\theta_{hkl} = \frac{K\cdot\lambda}{D_{hkl}} + 4\varepsilon\cdot\sin\theta_{hkl}.$$

The structural constants $K$, $\lambda$, and $\beta_{hkl}$ were used the same as for the Scherrer formula (see above). During the linear fitting of the $\beta_{hkl} \cdot \cos\theta_{hkl}$ term against $4\sin\theta_{hkl}$ for the rhombohedrally distorted perovskite L(C)NMO samples, the lattice strain from the slope ($\varepsilon$) can be estimated. As it turned out, a positive slope of the calculated microstrains for the perovskite L(C)NMO samples indicates the action of tensile forces on the crystal lattice (see Fig. S5).

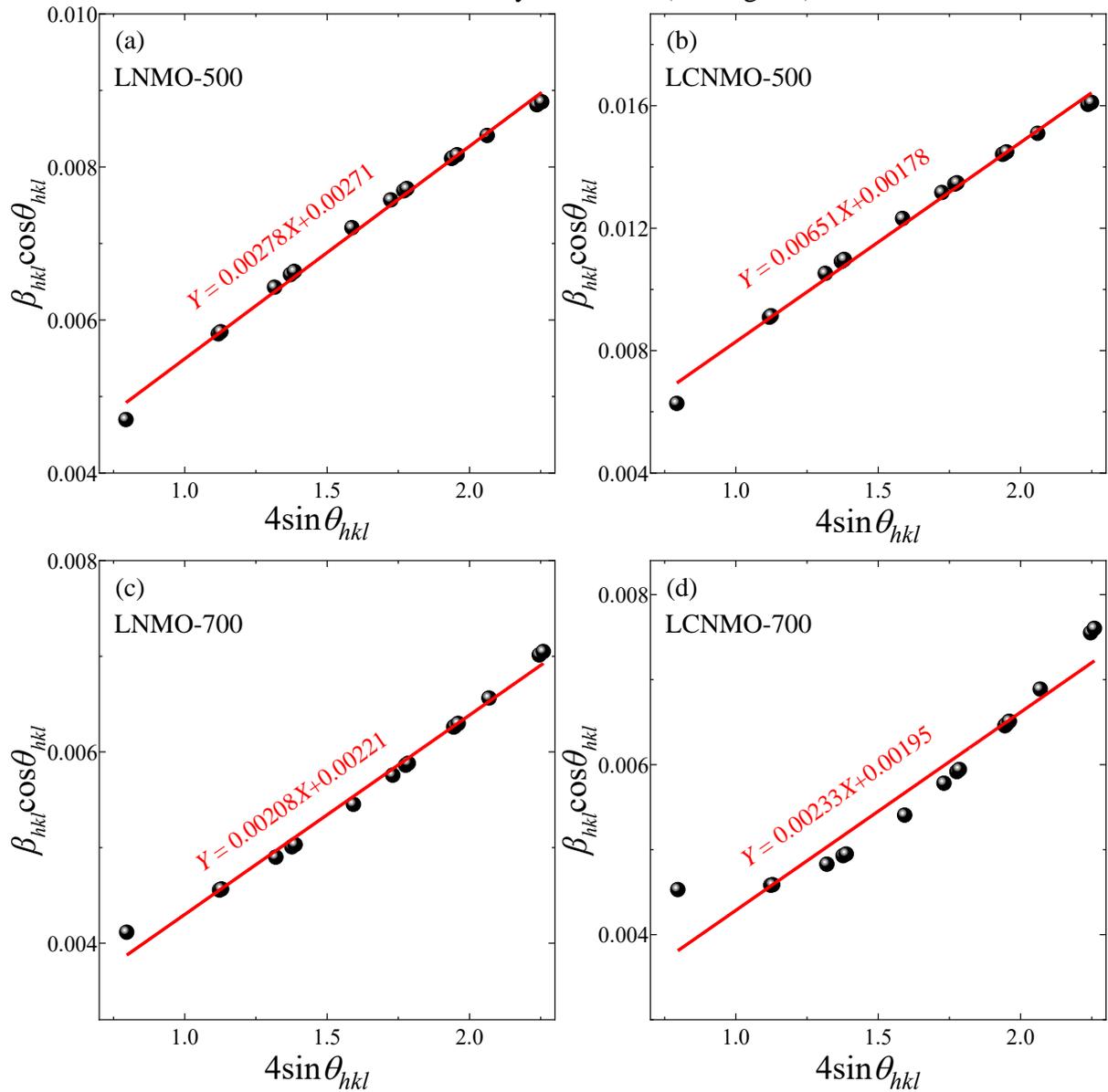

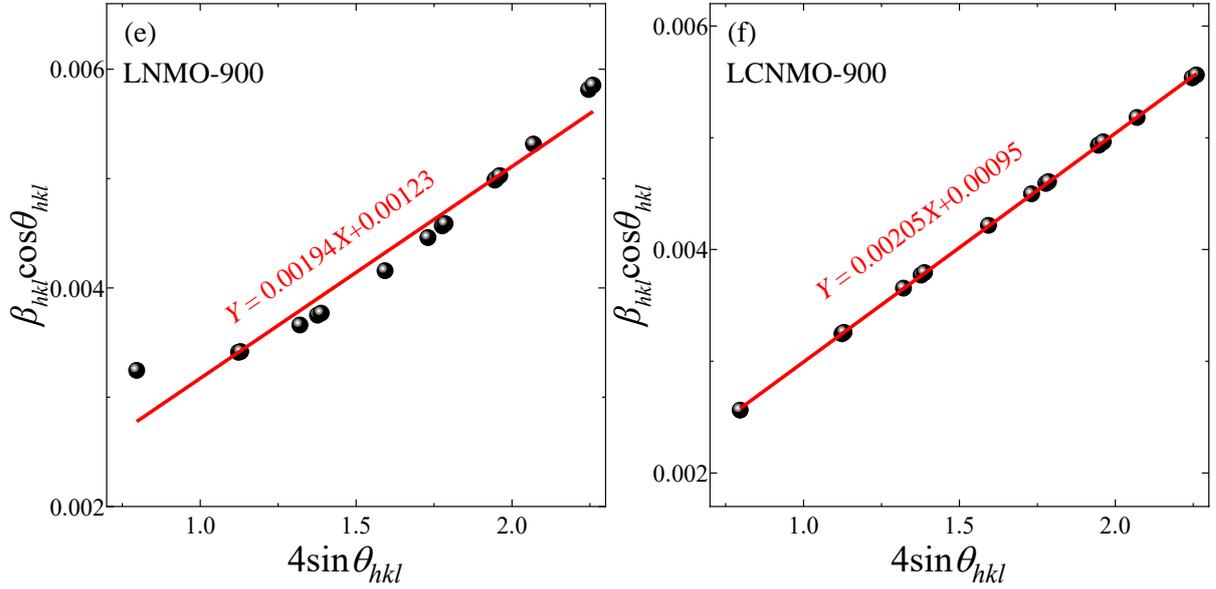

**Fig. S5**. The Williamson-Hall plots of the LNMO (a, c, e) and LCNMO (b, d, f) samples obtained under different annealing temperatures to determine microstrains $\varepsilon$.

The X-ray density $\rho_{XRD}$ of the L(C)NMO samples was calculated using powder X-ray diffraction data according to the formula:

$$\rho_{XRD} = \frac{Z \cdot M_r}{V \cdot N_A},$$

where $Z$ is the number of formula units in the unit cell, $M_r$ is the molecular weight of the substituted lanthanum manganite (g/mol) without considering nonstoichiometry in oxygen ($\Delta = 0$), $V$ is a unit cell volume (cm$^3$), and $N_A$ is Avogadro number ($6.022 \cdot 10^{23}$ mol$^{-1}$).

### SEM and TEM data

**Table S2**

Chemical composition (at. %) according to EDS data for the L(C)NMO samples for ideal and real cases.

| $x$ | Case | Element (at. %) | | | | |
|---|---|---|---|---|---|---|
| | | La | Cd | Na | Mn | O |
| LNMO-500 | Ideal | 16.00 | – | 4.00 | 20.00 | 60.00 |
| | Real | 18.28 | – | 5.85 | 22.76 | 53.11 |
| LNMO-700 | Ideal | 16.00 | – | 4.00 | 20.00 | 60.00 |
| | Real | 18.66 | – | 5.42 | 22.61 | 53.31 |
| LNMO-900 | Ideal | 16.00 | – | 4.00 | 20.00 | 60.00 |
| | Real | 19.25 | – | 4.62 | 22.48 | 53.66 |
| LCNMO-500 | Ideal | 15.00 | 1.00 | 4.00 | 20.00 | 60.00 |
| | Real | 13.33 | 0.48 | 3.81 | 15.81 | 66.57 |
| LCNMO-700 | Ideal | 15.00 | 1.00 | 4.00 | 20.00 | 60.00 |
| | Real | 9.92 | 0.42 | 4.6 | 13.24 | 71.82 |
| LCNMO-900 | Ideal | 15.00 | 1.00 | 4.00 | 20.00 | 60.00 |
| | Real | 13.45 | 0.66 | 2.21 | 16.56 | 67.22 |

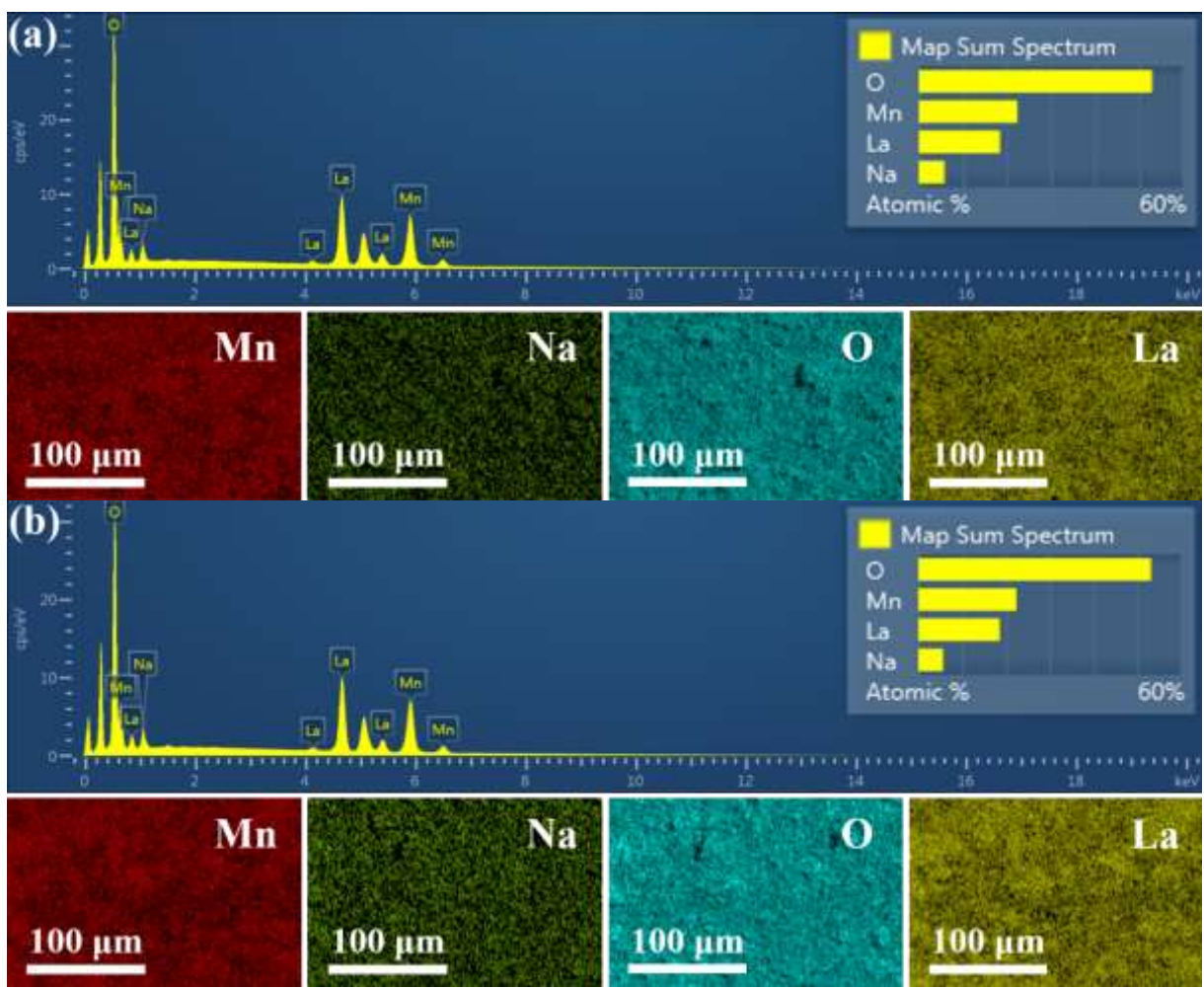

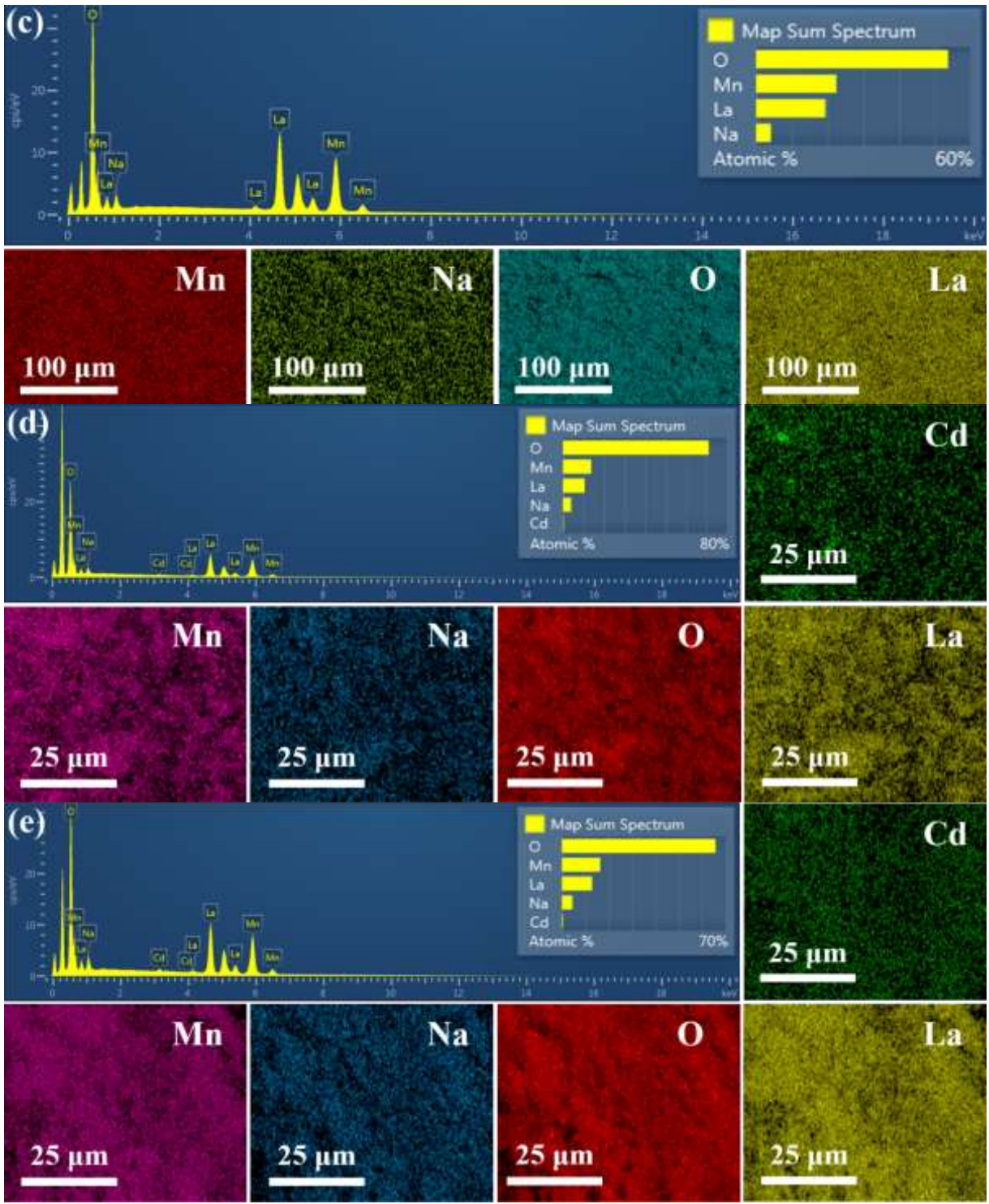

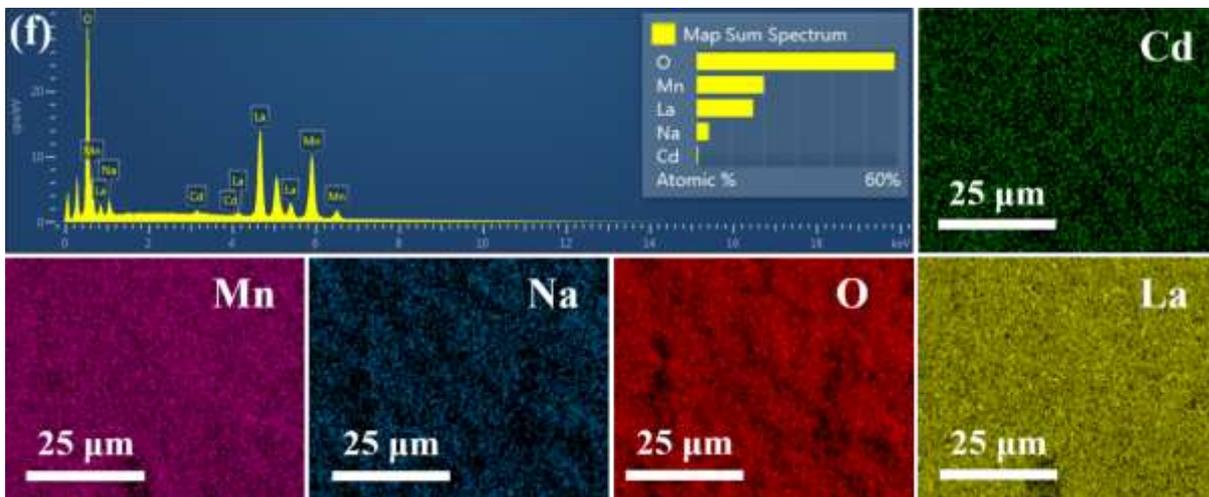

**Fig. S6**. EDS data for the LNMO (a, b, c) and LCNMO (d, e, f) samples obtained under different annealing temperatures $t_{ann}$.

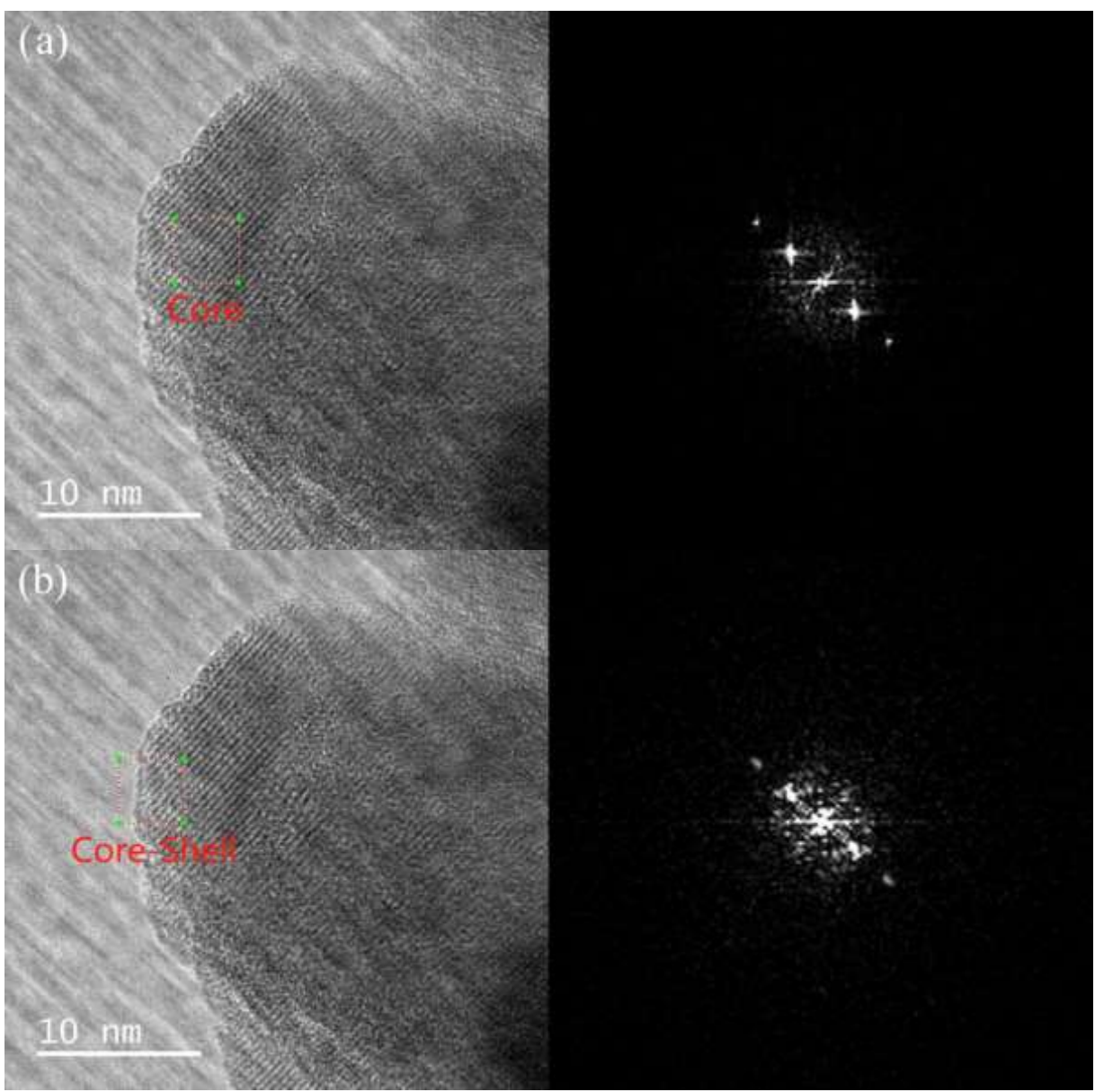

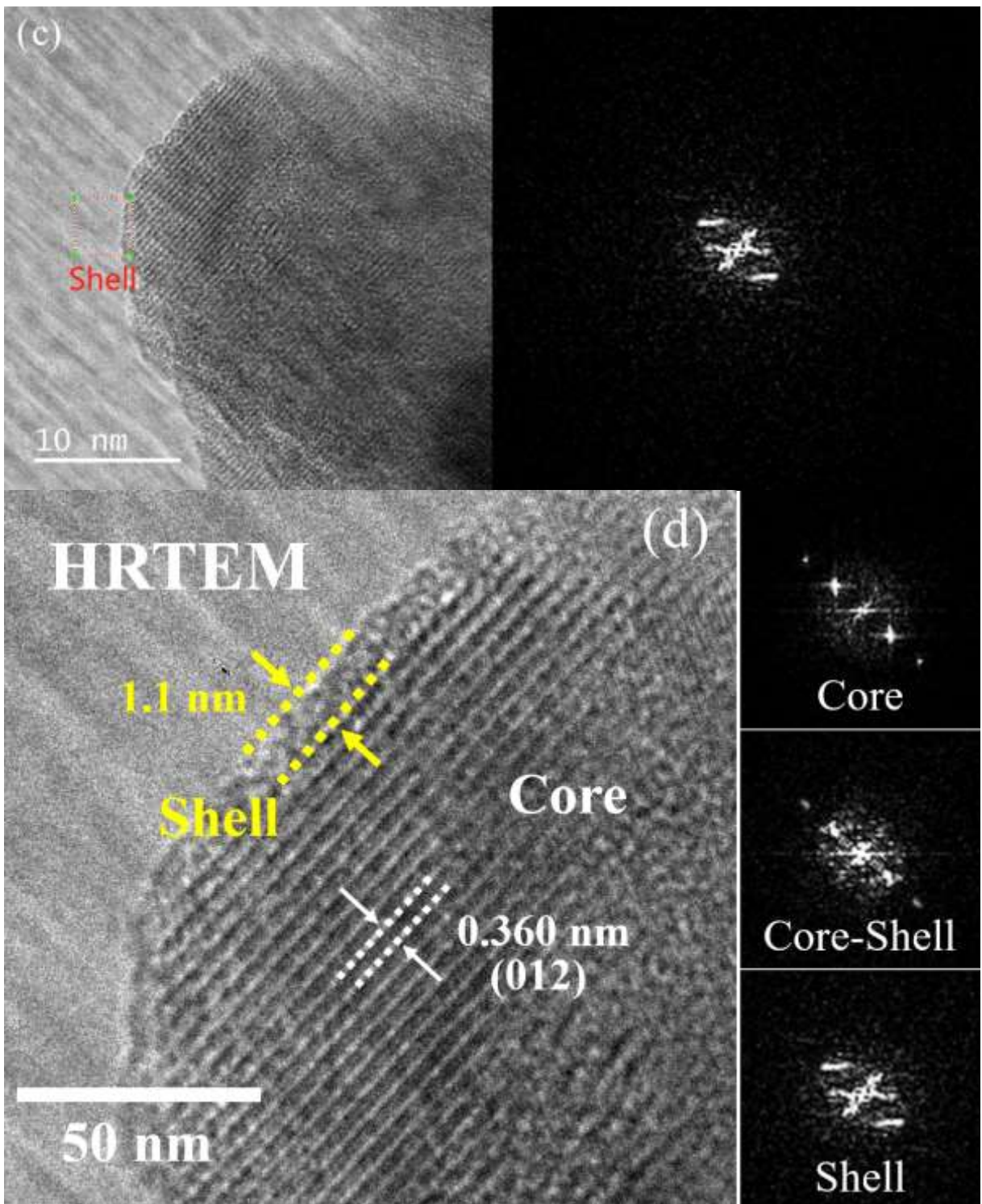

**Fig. S7**. The core-shell contributions for the LNMO-500 (a, b, c, d) sample based on the HRTEM data using the FFT approach.

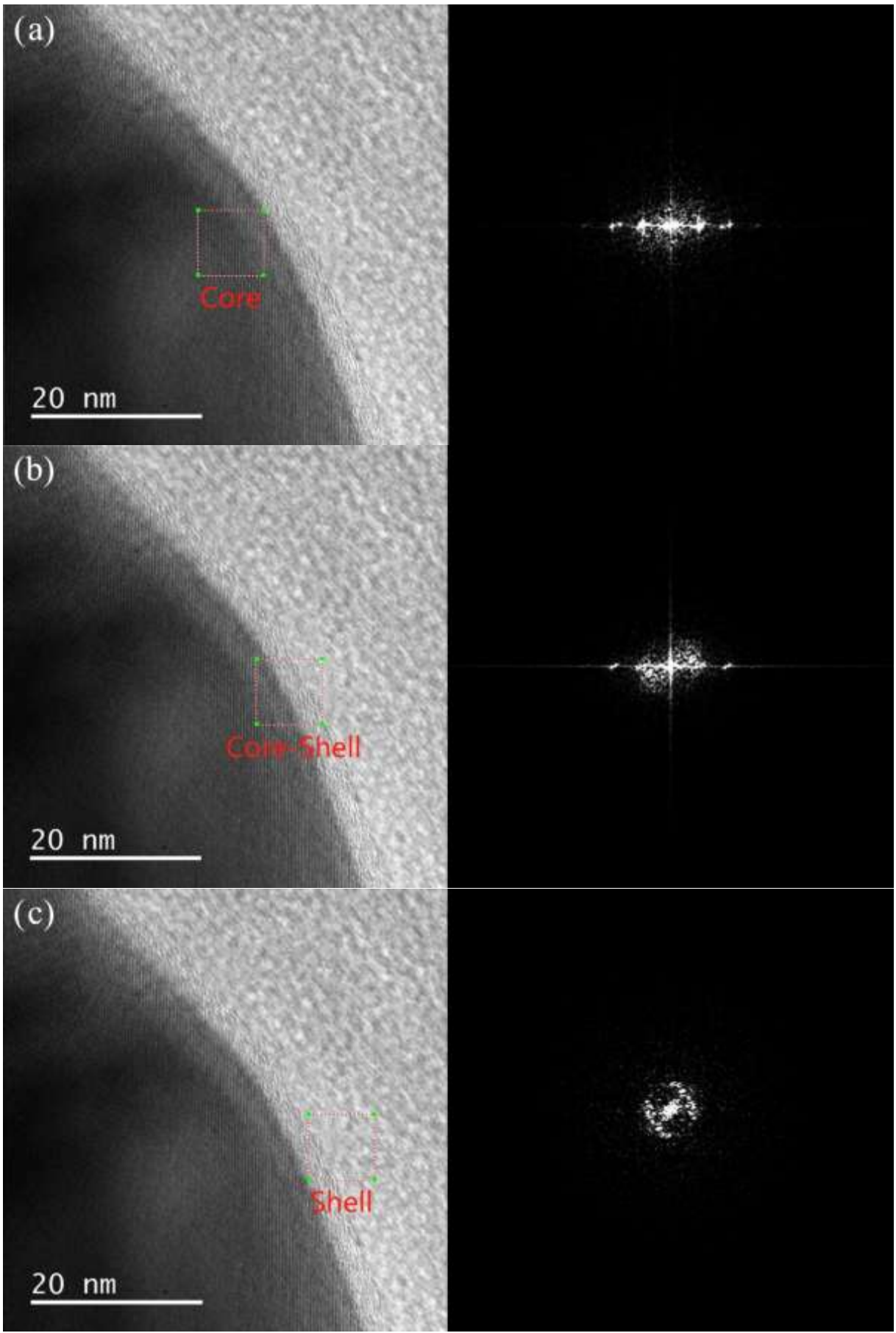

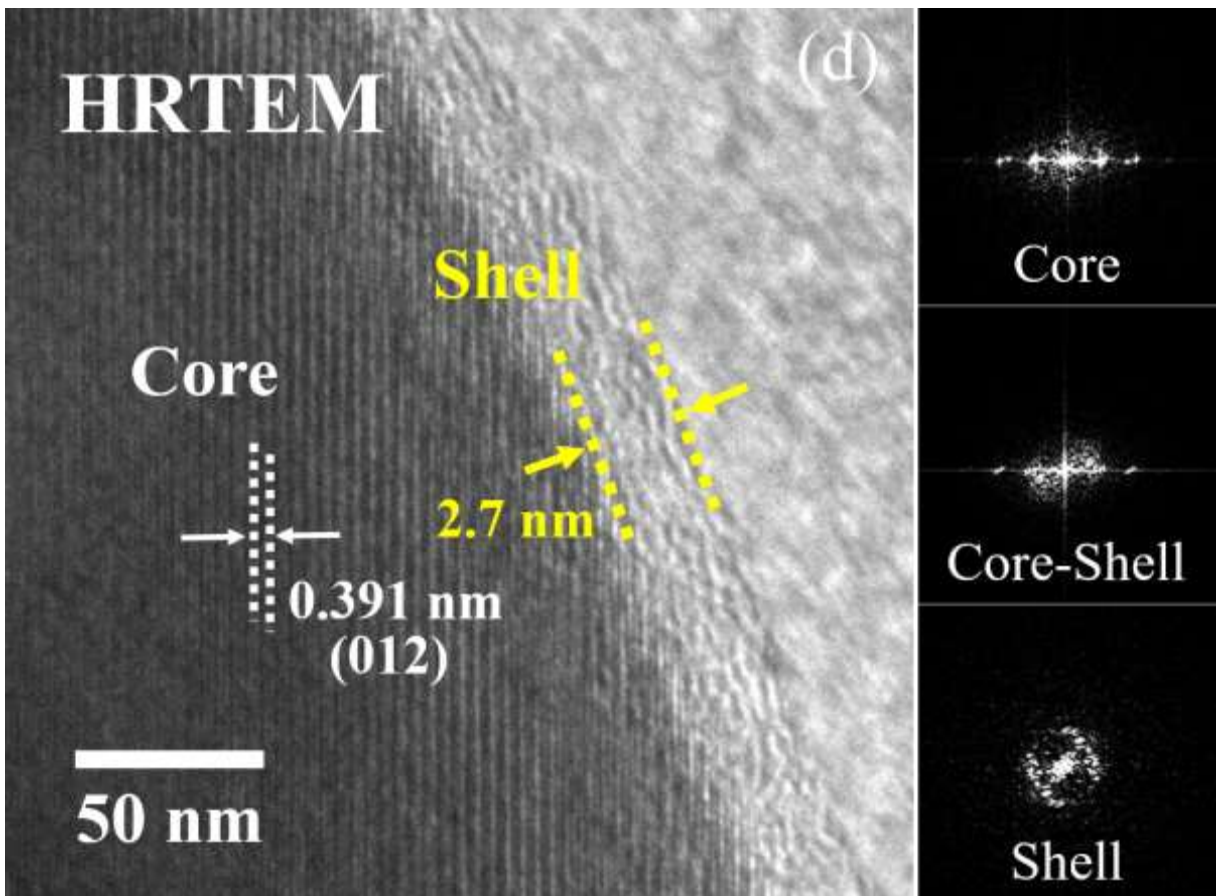

**Fig. S8**. The core-shell contributions for the LCNMO-500 (a, b, c, d) sample based on the HRTEM data using the FFT approach.

## Magnetic properties of the initial L(C)NMO samples

Fig. S9 shows the field dependences of the magnetization for the L(C)NMO samples under different temperatures. At $T = 100$ K, all L(C)NMO samples demonstrate a FM magnetization type with the saturation behavior. With increasing $t_{ann}$, the magnetization of the powders increases. For samples with smaller particle sizes annealed at $t_{ann} \leq 700$ °C, the magnetization of undoped LNMO samples is greater than that of doped LCNMO samples, whereas at $t_{ann} = 900$ °C, the magnetization behavior is opposite. This is most likely due to forming a more homogeneous "none-core-shell" structure [5].

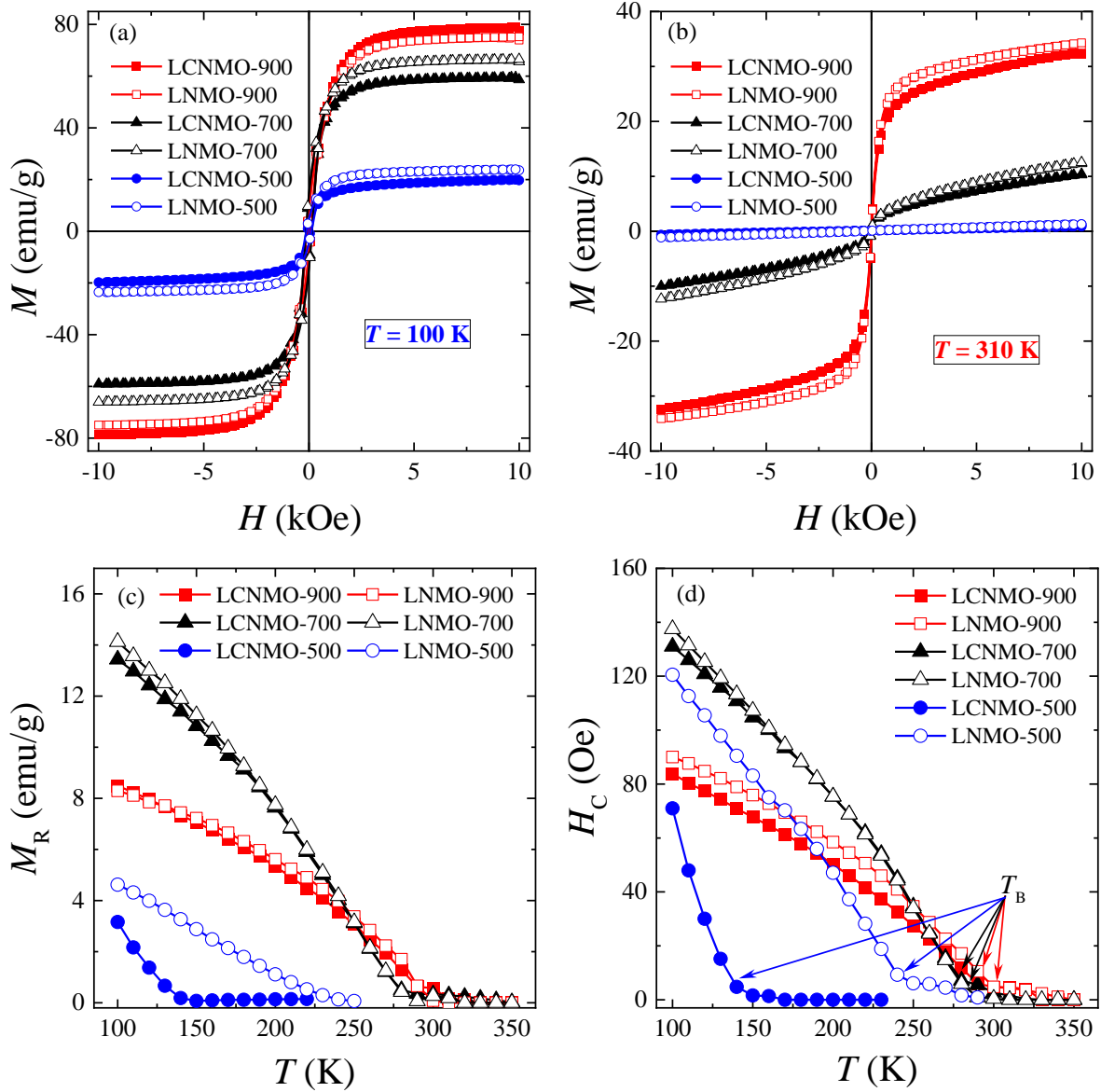

**Fig. S9**. The field dependences of the magnetization for the L(C)NMO samples at $T = 100$ (a) and 310 K (b), as well as their temperature dependences of the residual magnetization $M_R(T)$ (c) and coercivity $H_C(T)$ (d).

At $T = 310$ K, the L(C)NMO-900 samples retain the field dependence for FM-type magnetization, but with a significantly reduced saturation magnetization. At the same time, the magnetization of the LNMO-900 sample becomes greater than the magnetization of the LCNMO-900 one. The L(C)NMO-700 samples at higher temperatures do not saturate even upon 10 kOe. The L(C)NMO-500 samples are magnetized linearly and are in a PM state.

From the hysteresis loops, the values of residual magnetization ($M_R$) and coercivity ($H_C$) were found (see Fig. S9 (c, d)). It is clear that even at low temperatures, the values of the coercivity are small (see Fig. S9 (d)), so it can be assumed that the magnitude of the magnetic anisotropy field of the particles is also small and the particles can be considered weakly anisotropic. The decreasing coercivity with increasing temperature does not follow Brown's law with a root dependence on the temperature $H_C \neq T^{1/2}$ (see Fig. S9(d)). This is because the particles of the samples under study are not an ensemble of Stoner-Wohlfarth particles with a constant (not changing with temperature) magnetization modulus. In all L(C)NMO samples, the spontaneous magnetization of particles $M_S$ depends on temperature; unlike Stoner-Wohlfarth particles, they have $M_S(T) \neq$ const. This is the main reason why the magnetization of such ensembles of particles also does not satisfy the Langevin function. However, despite this, the particles under study behave SPM at high temperatures. Moreover, the temperature dependence of the coercivity changes its behavior at a blocking temperature $T_B$ (see Fig. S9(d)).

Additionally, the temperature dependence of the coercivity on the square of the temperature was plotted in Fig. S10. It can be seen that for almost all L(C)NMO samples, the magnitude of the coercivity is directly proportional to the square of the temperature

$$H_C(T) = \frac{H_{C, T=100\ K}}{(1-100^2/T_B^2)}\left(1 - \frac{T^2}{T_B^2}\right),$$

where $H_{C,\ T=100\ K}$ is the coercivity at $T = 100$ K and $T_B$ is the blocking temperature. The blocking temperature $T_B$ corresponds to a kink in the $H_C(T^2)$ dependence (see Fig. S10). For the L(C)NMO-900, the interval of superparamagnetic magnetization $T_C$–$T_B$ is larger (see below) than for particles with a smaller size at L(C)NMO-700, since these particles are more homogeneous. Their coercivity and remanent magnetization are almost half as much (see Fig. S9(c, d)). It is well known that particles with a size greater than 150 nm are non-uniformly magnetized, i.e. vortex type. In our case, the particle size of all L(C)NMO samples is smaller, and it can be assumed that they are all in a single-domain state.

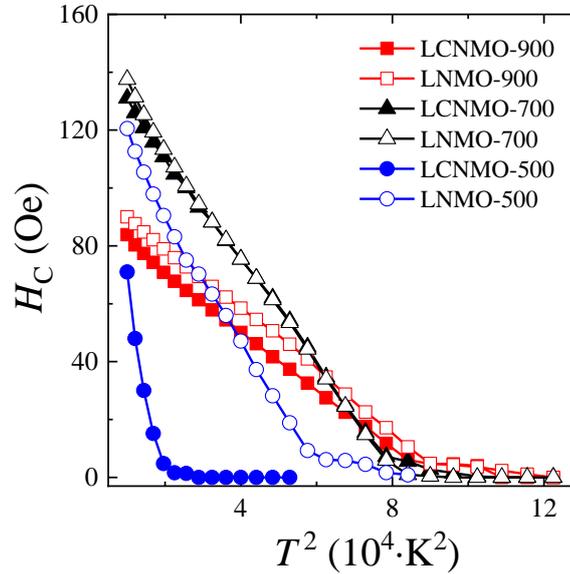

**Fig. S10**. The temperature dependences of the coercivity $H_C(T^2)$.

Fig. S11 (a) and (b) shows the temperature dependences of the magnetizations of all L(C)NMO samples obtained in a constant magnetic field of $H = 2$ and 10 kOe, respectively. The temperature dependences of magnetization $M(T)$ for the L(C)NMO-900 samples with large particle sizes decrease faster with increasing temperature than for other samples (see Fig. S11 (c, d)). This means that the L(C)NMO-900 samples are more magnetically homogeneous and have fewer defects (see Table 1). The LCNMO-900 sample at $T < 290$ K has greater magnetization than the LNMO-900. However, the situation totally changes at $T > 290$ K. For the samples with lower $t_{ann}$, doping with Cd is simultaneously accompanied by a decrease in the Curie temperature and saturation magnetization (see Fig. S11). For the L(C)NMO samples with $t_{ann} = 500$ and 700 °C and a higher concentration of defects, the curves have a flatter decline, and their derivatives $dM/dT(T)$ are very wide compared to the L(C)NMO-900 ones. This may mean that these samples are more inhomogeneous, and their critical index value for the temperature dependence of spontaneous magnetization varies greatly or is entirely absent.

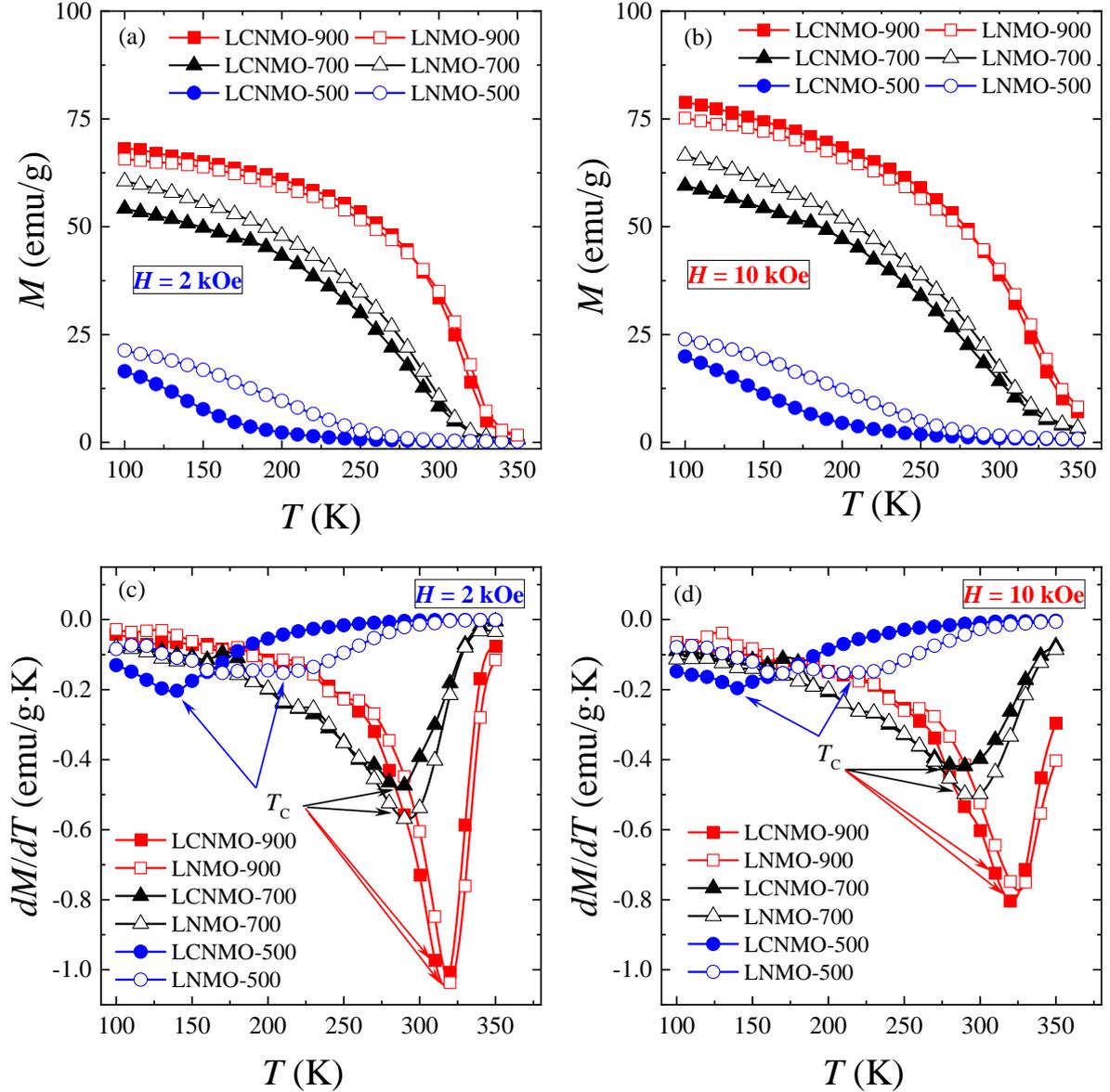

**Fig. S11**. The temperature dependences of magnetization $M(T)$ and derivative magnetization $dM/dT(T)$ for the L(C)NMO samples in the fields of $H = 2$ (a, c) and 10 kOe (b, d), respectively.

Fig. S12 shows the temperature dependences of the spontaneous magnetization $M_S(T)$ and its approximation by the power-law $M(T) \sim ((T-T_C)/T_C)^\beta$ with a critical exponent $\beta$. The $\beta$ is around 0.36-0.37 for the L(C)NMO-700, indicating the 3D-Heisenberg model with the theoretical value of $\beta = 0.365$ [6], whereas for the L(C)NMO-900, the $\beta$ is in the range of 0.28-0.31, showing the 3D-Ising model with the $\beta = 0.325$ [6]. The linear asymptotes of the $M_S(T)$ intersect the temperature axis at the $T_C$ points, and their values agree with the above-found $T_C$ (see Fig. S11). For the L(C)NMO-500, obtaining a linear approximation with the $T_C$ was impossible, indicating the absence of their critical behavior. For the L(C)NMO-900, the index $\beta$ value is lower than for the L(C)NMO-700, which explains the more pronounced critical behavior. Thus, the L(C)NMO critical index with different particle sizes and defects and their behavior at the critical point are significantly different.

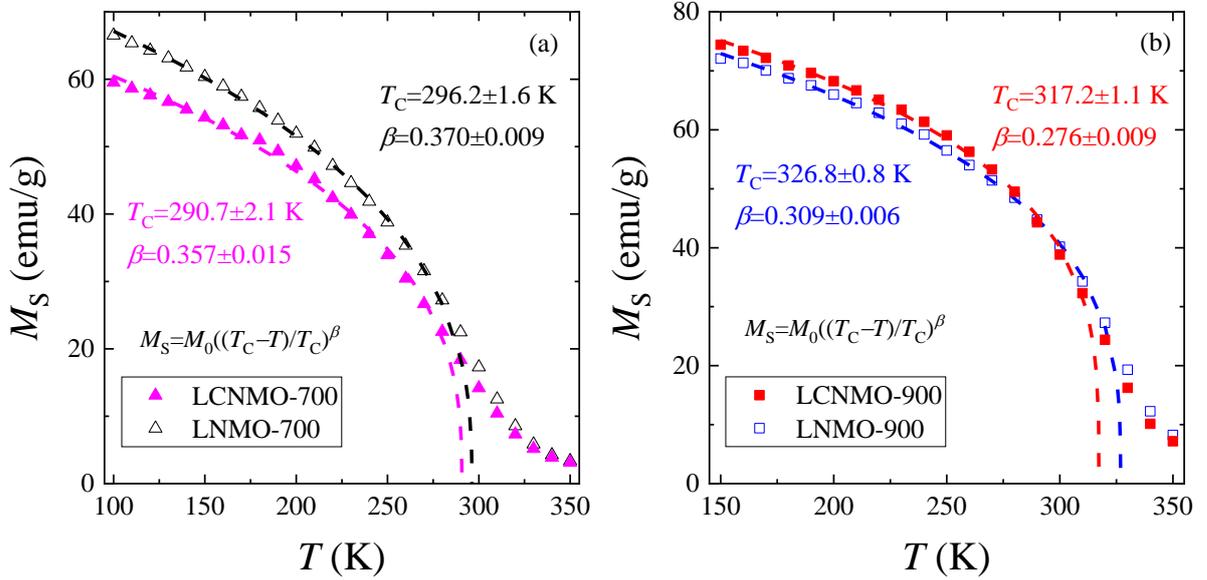

**Fig. S12**. The temperature dependences of the spontaneous magnetization $M_S(T)$ in the field $H = 10$ kOe for defining the Curie temperature $T_C$ and critical exponent $\beta$ of the L(C)NMO-700 (a) and L(C)NMO-900 (b) samples.

Fig. S13 shows the temperature dependences for the inverse magnetic susceptibility $\chi^{-1} = M/H$ in a field $H = 2$ kOe. For the LNMO-500, the behavior of magnetic susceptibility below 290 K is similar to the Griffiths phase separation (PS) [7], which is consistent with the fact that the particle structure of these samples has the largest number of defects. Therefore, in the PM phase near the Curie point, they behave like magnetically inhomogeneous particles with a scatter of Curie temperatures. For the LCNMO-500, the reverse magnetic susceptibility above 290 K behaves linearly with a PM Curie temperature $\theta = 52$ K. For other samples, the behavior of the magnetic susceptibility corresponds to their FM state.

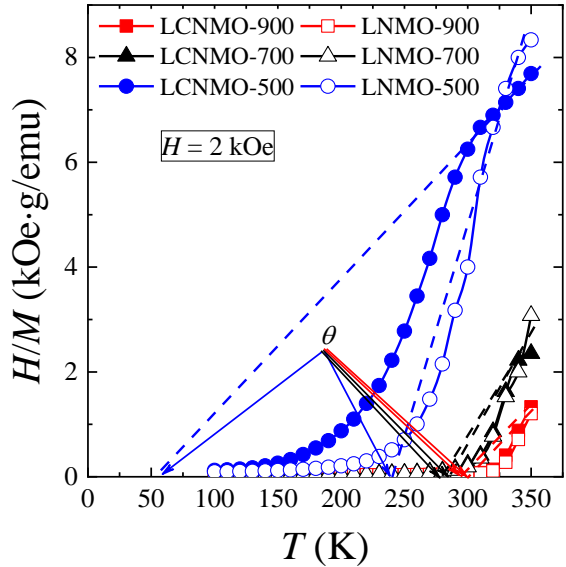

**Fig. S13**. The temperature dependences for the inverse magnetic susceptibility $\chi^{-1} = H/M(T)$ for the L(C)NMO samples in a field $H = 2$ kOe.

The surface layer thickness $t_{shell}^{M(H)}$ was obtained from the following assumption [8]:
$$t_{shell}^{M(H)} \approx (D/2)\{1-[M_S(D)/M_S(D_{900°C})]^{1/3}\},$$
where $D$ is the particle size according to the XRD data, $M_S(D_{900°C})$ is the spontaneous magnetization for the biggest particles annealed at $t_{ann} = 900$ °C, and $M_S(D)$ is particle size dependent spontaneous magnetization (see Table S3).

**Table S3**

Particle size $D$, spontaneous magnetization $M_S$, and surface layer thickness $t_{shell}^{M(H)}$ for the L(C)NMO nanopowders under different $t_{ann}$.

| Sample | $t_{ann}$ (°C) | $D$ (nm) | $M_S$ (emu/g) | $t_{shell}^{M(H)}$ (nm) |
|---|---|---|---|---|
| LNMO | 500 | 27 | 32 | 3.9 |
|  | 700 | 38 | 78 | 0.8 |
|  | 900 | 62 | 88 | – |
| LCNMO | 500 | 20 | 30 | 3.1 |
|  | 700 | 38 | 70 | 1.5 |
|  | 900 | 71 | 90 | – |

**Magnetic properties of the L(C)NMO samples after three years**

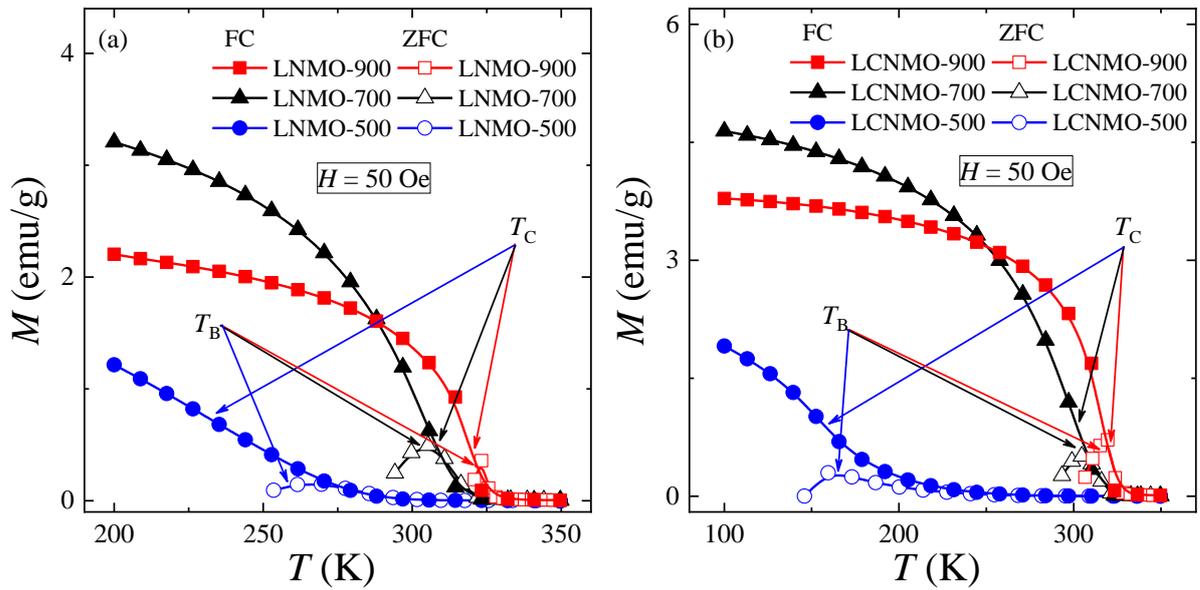

**Fig. S14**. The temperature dependences of magnetization $M_{ZFC-FC}(T)$ for the LNMO (a) and LCNMO (b) samples in the field of $H = 50$ Oe.

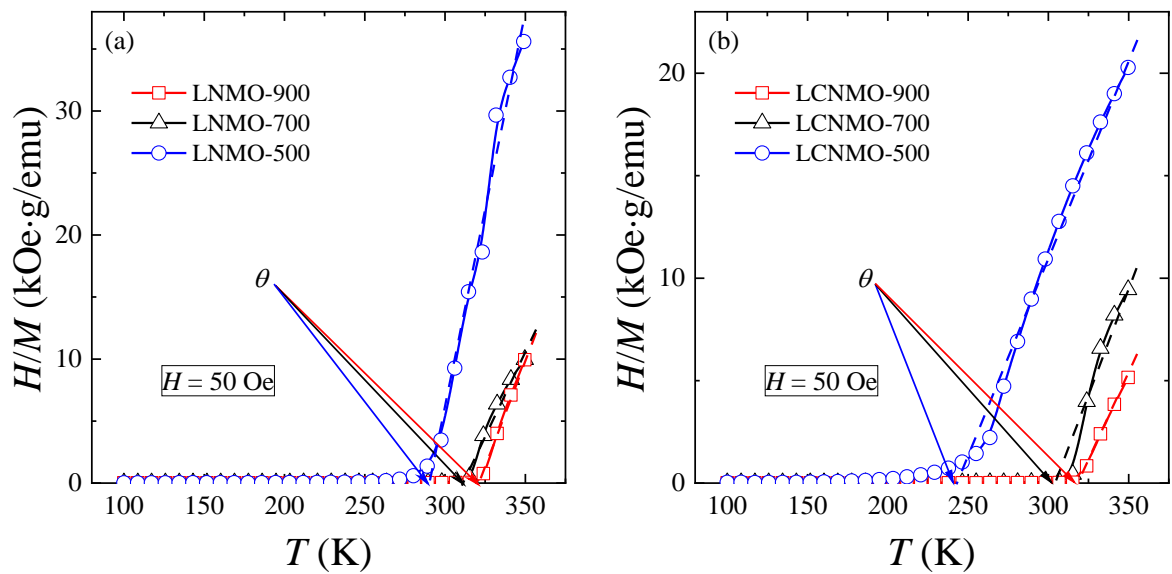

**Fig. S15**. The temperature dependences for the inverse magnetic susceptibility $\chi^{-1} = H/M(T)$ for the LNMO (a) and LCNMO (b) samples in the field of $H = 50$ Oe.

**Magnetic properties of the L(C)NMO samples after three years under high pressure**

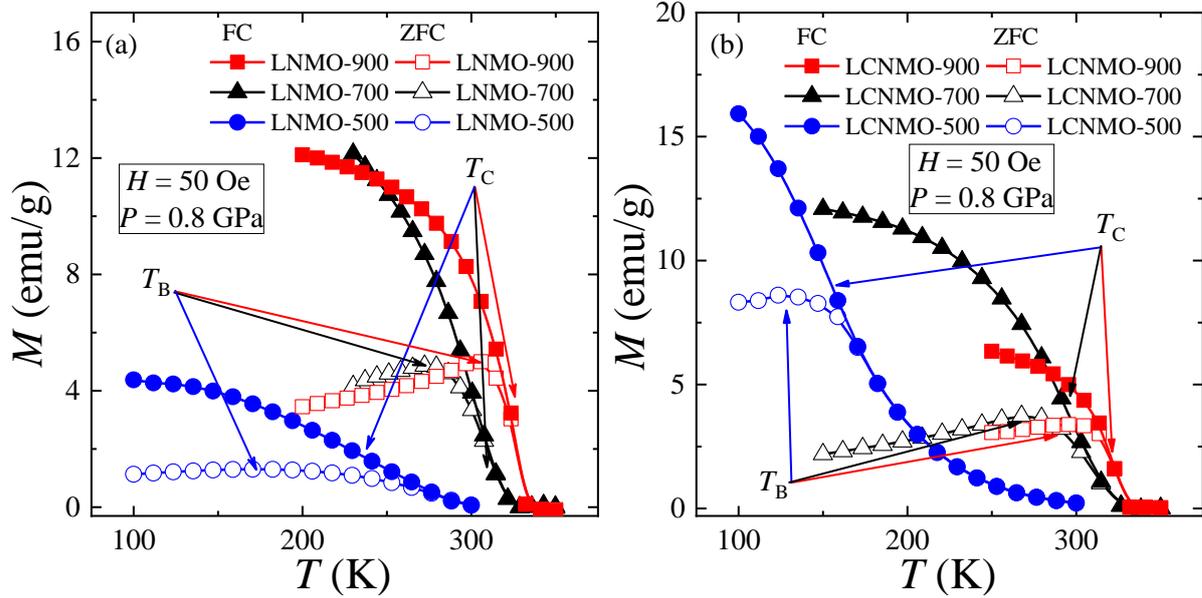

**Fig. S16**. The temperature dependences of magnetization $M_{ZFC-FC}(T)$ for the LNMO (a) and LCNMO (b) samples in the field of $H = 50$ Oe and under high-pressure $P = 0.8$ GPa.

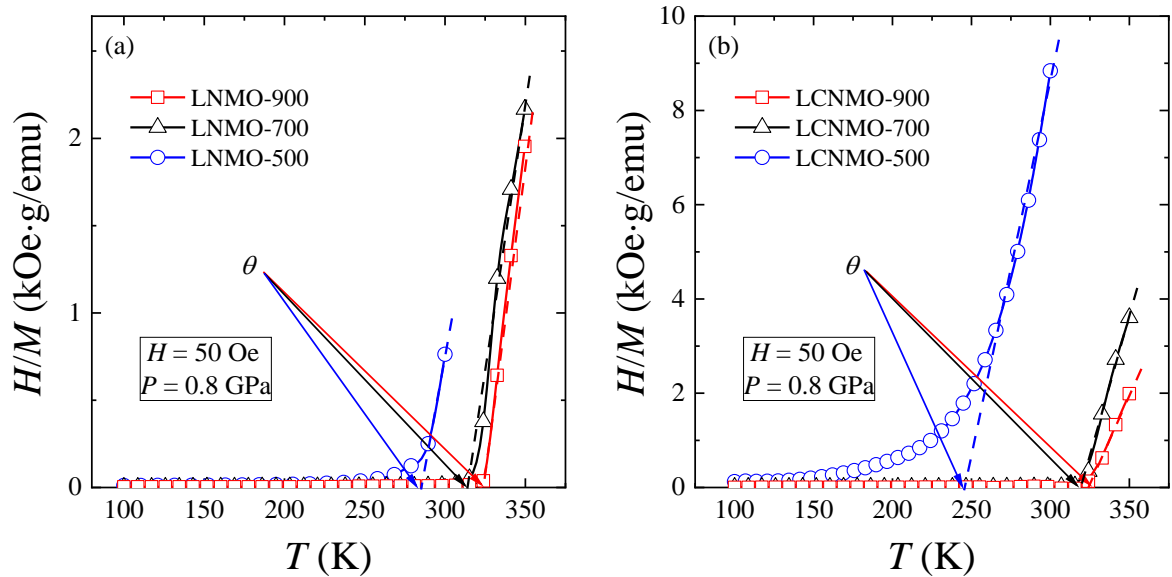

**Fig. S17**. The temperature dependences for the inverse magnetic susceptibility $\chi^{-1} = H/M(T)$ for the LNMO (a) and LCNMO (b) samples in the field of $H = 50$ Oe and under high-pressure $P = 0.8$ GPa.

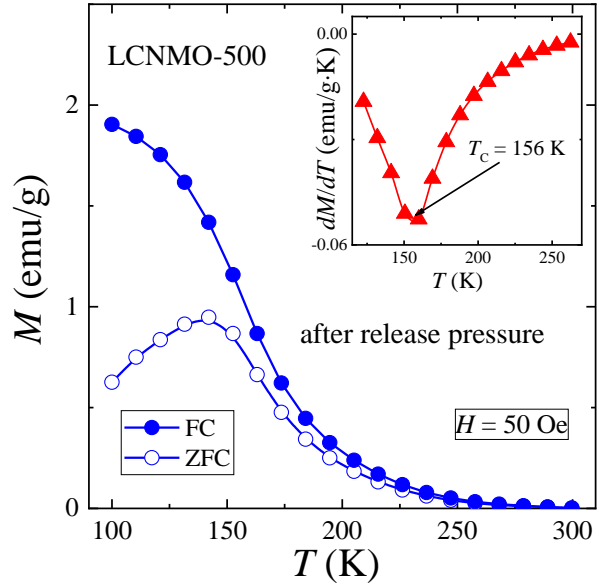

**Fig. S18**. The temperature dependences of magnetization $M_{ZFC-FC}(T)$ for the LCNMO-500 sample in the field of $H = 50$ Oe and after release pressure.

**Table S4**

Phase transition temperatures (PM Curie temperature $\theta$, Curie temperature $T_C$, and blocking temperature $T_B$) for the L(C)NMO nanopowders depending on the structural and size effects $t_{ann}$, time $t$, and pressure $P$ parameters.

| Sample | $t_{ann}$ (°C) | $t$ (year) | $P = 0$ GPa | | | $P = 0.8$ GPa | | |
|---|---|---|---|---|---|---|---|---|
| | | | $\theta$ (K) | $T_C$ (K) | $T_B$ (K) | $\theta$ (K) | $T_C$ (K) | $T_B$ (K) |
| LNMO | 500 | 0 | 241 | 210 | 242 | – | – | – |
| | | 3 | 290 | 231 | 265 | 298 | 236 | 181 |
| | 700 | 0 | 310 | 292 | 283 | – | – | – |
| | | 3 | 311 | 306 | 305 | 326 | 313 | 284 |
| | 900 | 0 | 325 | 320 | 298 | – | – | – |
| | | 3 | 321 | 321 | 323 | 338 | 334 | 309 |
| LCNMO | 500 | 0 | 52 | 137 | 132 | – | – | – |
| | | 3 | 243 | 156 | 161 | 248 | 161 | 124 |
| | 700 | 0 | 306 | 286 | 280 | – | – | – |
| | | 3 | 304 | 298 | 297 | 318 | 304 | 272 |
| | 900 | 0 | 321 | 316 | 290 | – | – | – |
| | | 3 | 319 | 319 | 319 | 325 | 325 | 297 |

**Structural properties of the L(C)NMO samples after three years**

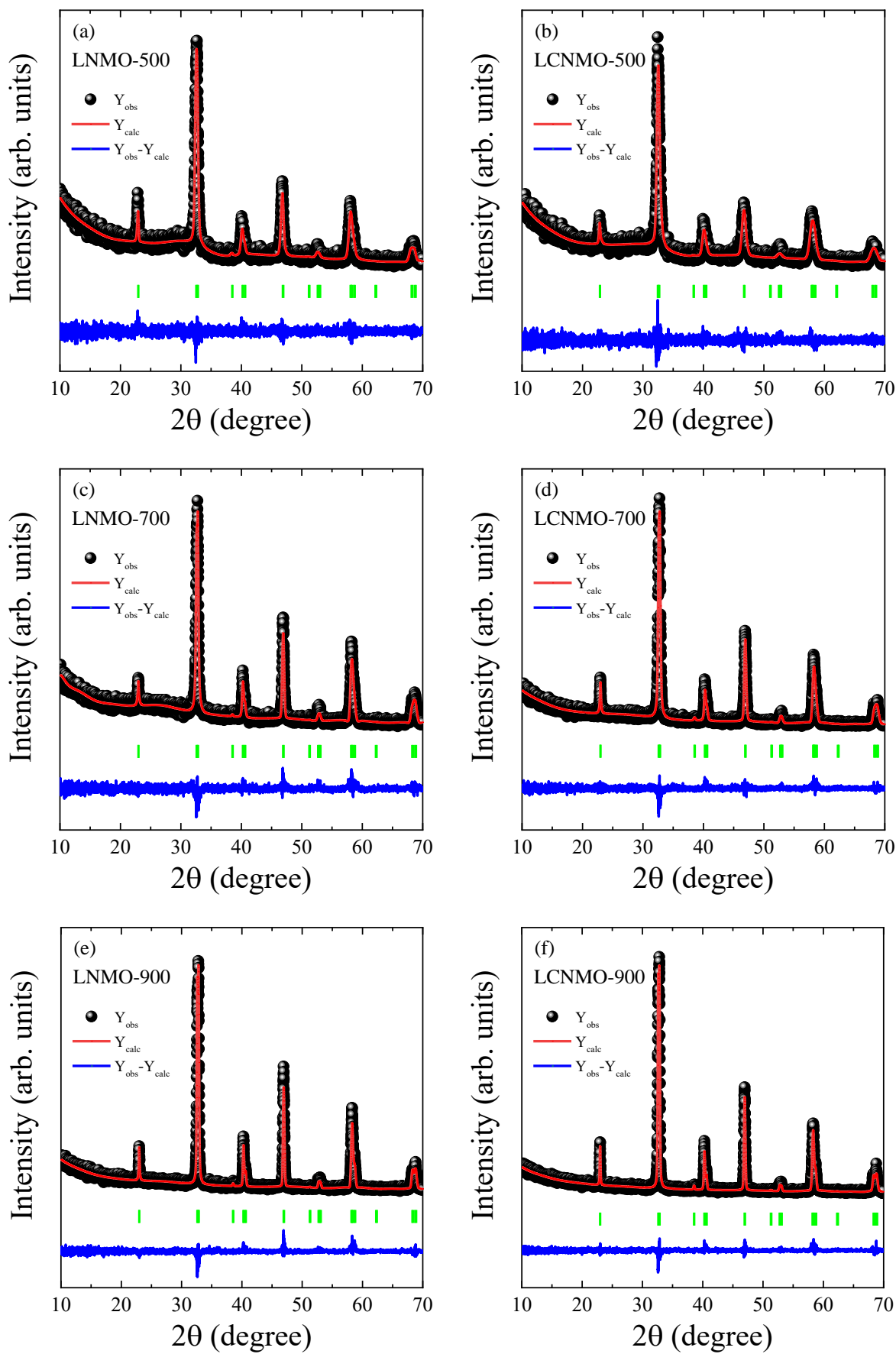

**Fig. S19.** Observed (solid black circles) and calculated (solid red line) X-ray powder diffraction patterns for the $La_{0.8}Na_{0.2}MnO_3$ (a, c, e) and $La_{0.75}Cd_{0.05}Na_{0.2}MnO_3$ (b, d, f) samples prepared at the different temperatures 500 °C (a, b), 700 °C (c, d), and 900 °C (e, f) after 3 years. The difference (blue line) between the observed and calculated spectra is plotted at the bottom. The ticks (green) indicate allowed Bragg peak positions.

**Table S5**

After three years, Rietveld refinement crystallographic parameters for the L(C)NMO nanopowders under different $t_{ann}$.

| Sample | LNMO | | | LCNMO | | |
|---|---|---|---|---|---|---|
| $t_{ann}$ (°C) | 500 | 700 | 900 | 500 | 700 | 900 |
| $D_{XRD}$ (nm) | 22±2 | 41±2 | 68±2 | 20±1 | 44±2 | 72±2 |
| $\delta \cdot 10^{-4}$ (nm$^{-2}$) | 20.0 | 6.0 | 2.2 | 26.1 | 5.1 | 1.9 |
| $\varepsilon$ (%) | 0.429(3) | 0.361(2) | 0.195(3) | 0.797(12) | 0.293(5) | 0.172(2) |
| Space group | $R\bar{3}c$ (No. 167) | $R\bar{3}c$ (No. 167) | $R\bar{3}c$ (No. 167) | $R\bar{3}c$ (No. 167) | $R\bar{3}c$ (No. 167) | $R\bar{3}c$ (No. 167) |
| $a$ (Å) | 5.50088(49) | 5.49326(20) | 5.49187(18) | 5.51475(85) | 5.49339(16) | 5.48878(9) |
| $b$ (Å) | 5.50088(49) | 5.49326(20) | 5.49187(18) | 5.51475(85) | 5.49339(16) | 5.48878(9) |
| $c$ (Å) | 13.33342(209) | 13.33079(80) | 13.33029(59) | 13.38622(447) | 13.33734(74) | 13.32800(37) |
| $V$ (Å$^3$) | 349.411(70) | 348.374(28) | 348.185(22) | 352.566(140) | 348.562(24) | 347.734(12) |
| Z | 6 | 6 | 6 | 6 | 6 | 6 |
| $<d_{Mn-O}>$ (Å) | 1.9599(39) | 1.9474(26) | 1.9493(22) | 1.9509(52) | 1.9557(22) | 1.9552(16) |
| $\theta_{<Mn-O-Mn>}$ (°) | 162.9(5) | 167.7(3) | 166.5(5) | 170.3(6) | 163.9(3) | 163.5(2) |
| W | 0.0938 | 0.0965 | 0.0960 | 0.0961 | 0.0947 | 0.0947 |
| $\rho$ (g/cm$^3$) | 6.235 | 6.254 | 6.257 | 6.142 | 6.212 | 6.227 |
| $R_p$ (%) | 8.1 | 10.5 | 8.9 | 8.3 | 8.4 | 8.2 |
| $R_{wp}$ (%) | 10.8 | 14.1 | 12.1 | 11.1 | 11.4 | 11.4 |
| $R_{exp}$ (%) | 11.1 | 13.6 | 11.4 | 11.3 | 11.2 | 11.4 |
| $\chi^2$ (%) | 1.0 | 1.1 | 1.1 | 1.0 | 1.0 | 1.0 |

**Table S6**

XRD broadening parameters of (012) and (024) reflexes of the L(C)NMO samples obtained under different annealing temperatures $t_{ann}$ after three years.

| Name | $2\theta_{012}$ (degree) | $2\theta_{024}$ (degree) | $\beta_{exp-012}$ (degree) | $\beta_{exp-024}$ (degree) | $\beta_{inst}$ (degree) | $\beta_{012}$ (degree) | $\beta_{024}$ (degree) | $\beta_{024}/\beta_{012}$ | $\cos\theta_{012}/\cos\theta_{024}$ | $tg\theta_{024}/tg\theta_{012}$ |
|---|---|---|---|---|---|---|---|---|---|---|
| LNMO-500 | 22.8507 | 46.7728 | 0.368 | 0.432 | 0.060 | 0.3631 | 0.4278 | 1.2 | 1.1 | 2.1 |
| LNMO-700 | 22.9433 | 46.9086 | 0.207 | 0.318 | 0.060 | 0.1981 | 0.3123 | 1.6 | 1.1 | 2.1 |
| LNMO-900 | 22.9504 | 46.9285 | 0.134 | 0.233 | 0.060 | 0.1198 | 0.2251 | 1.9 | 1.1 | 2.1 |
| LCNMO-500 | 22.8104 | 46.6652 | 0.419 | 0.837 | 0.060 | 0.4147 | 0.8348 | 2.0 | 1.1 | 2.1 |
| LCNMO-700 | 22.9657 | 46.9326 | 0.193 | 0.323 | 0.060 | 0.1834 | 0.3174 | 1.7 | 1.1 | 2.1 |
| LCNMO-900 | 22.9932 | 46.9816 | 0.127 | 0.228 | 0.060 | 0.1119 | 0.2200 | 2.0 | 1.1 | 2.1 |

**Structural and magnetic properties of the L$A$MO samples ($A$ = Na$^+$, Ag$^+$, K$^+$) before and after 3 years**

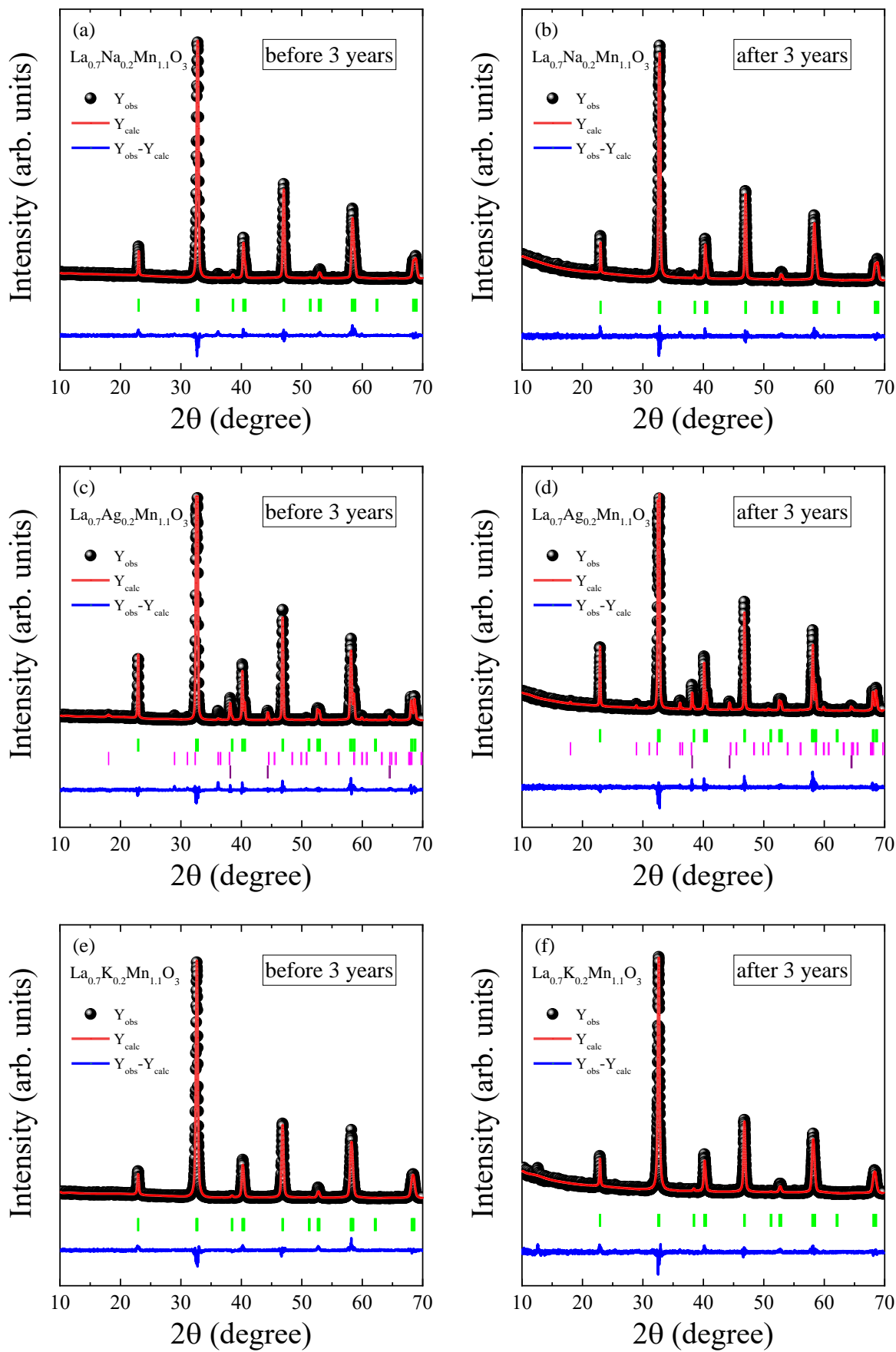

**Fig. S20.** Observed (solid black circles) and calculated (solid red line) X-ray powder diffraction patterns for the $La_{0.7}Na_{0.2}Mn_{1.1}O_3$ (a, b), $La_{0.7}Ag_{0.2}Mn_{1.1}O_3$ (c, d), and $La_{0.7}K_{0.2}Mn_{1.1}O_3$ (e, f) samples measured before (a, c, e) and after 3 years (b, d, f). The difference (blue line) between the observed and calculated spectra is plotted at the bottom. The ticks (green) indicate the main allowed Bragg peak positions. Additional magenta and navy bars are from hausmannite $Mn_3O_4$ and metallic Ag for the $La_{0.7}Ag_{0.2}Mn_{1.1}O_3$ (c, d) only, respectively.

**Table S7**

Rietveld refinement crystallographic parameters for the L$A$MO nanopowders ($A$ = Na$^+$, Ag$^+$, K$^+$) before and after three years.

| $A$ | Na$^+$ | | Ag$^+$ | | K$^+$ | |
|---|---|---|---|---|---|---|
| $t$ (year) | 0 | 3 | 0 | 3 | 0 | 3 |
| $D_{XRD}$ (nm) | 52±2 | 64±2 | 75±3 | 76±2 | 32±1 | 39±1 |
| $\delta \cdot 10^{-4}$ (nm$^{-2}$) | 3.7 | 2.4 | 1.8 | 1.8 | 9.9 | 6.5 |
| $\varepsilon$ (%) | 0.167(7) | 0.180(6) | 0.060(3) | 0.049(1) | 0.172(3) | 0.266(9) |
| Space group | $R\bar{3}c$ (No. 167) | $R\bar{3}c$ (No. 167) | $R\bar{3}c$ (No. 167) | $R\bar{3}c$ (No. 167) | $R\bar{3}c$ (No. 167) | $R\bar{3}c$ (No. 167) |
| $a$ (Å) | 5.4813(1) | 5.4844(1) | 5.5044(1) | 5.5078(1) | 5.4979(3) | 5.5016(1) |
| $b$ (Å) | 5.4813(1) | 5.4844(1) | 5.5044(1) | 5.5078(1) | 5.4979(3) | 5.5016(1) |
| $c$ (Å) | 13.3238(5) | 13.3326(3) | 13.3466(4) | 13.3555(2) | 13.3941(13) | 13.4015(6) |
| $V$ (Å$^3$) | 346.68(2) | 347.31(1) | 350.21(1) | 350.88(1) | 350.62(4) | 351.29(2) |
| $Z$ | 6 | 6 | 6 | 6 | 6 | 6 |
| $<d_{Mn-O}>$ (Å) | 1.9488(17) | 1.9514(13) | 1.9588(21) | 1.9604(14) | 1.9524(18) | 1.9566(16) |
| $\theta_{<Mn-O-Mn>}$ (°) | 165.4(4) | 164.8(3) | 163.9(3) | 163.8(4) | 167.2(4) | 165.7(3) |
| $W$ | 0.0960 | 0.0955 | 0.0941 | 0.0939 | 0.0956 | 0.0947 |
| $\rho$ (g/cm$^3$) | 6.043 | 6.032 | 6.465 | 6.453 | 6.066 | 6.055 |
| $R_p$ (%) | 9.5 | 6.7 | 10.1 | 7.1 | 9.0 | 7.2 |
| $R_{wp}$ (%) | 13.3 | 8.9 | 15.3 | 9.2 | 12.3 | 9.6 |
| $R_{exp}$ (%) | 9.6 | 4.3 | 10.1 | 4.4 | 9.5 | 4.6 |
| $\chi^2$ (%) | 1.9 | 4.3 | 2.3 | 4.4 | 1.7 | 4.2 |

**Table S8**

XRD broadening parameters of (012) and (024) reflexes of the L$A$MO samples ($A$ = Na$^+$, Ag$^+$, K$^+$) before and after three years.

| Name | $t$ (year) | $2\theta_{012}$ (degree) | $2\theta_{024}$ (degree) | $\beta_{exp-012}$ (degree) | $\beta_{exp-024}$ (degree) | $\beta_{inst}$ (degree) | $\beta_{012}$ (degree) | $\beta_{024}$ (degree) | $\beta_{024}/\beta_{012}$ | $\cos\theta_{012}/\cos\theta_{024}$ | $tg\theta_{024}/tg\theta_{012}$ |
|---|---|---|---|---|---|---|---|---|---|---|---|
| $La_{0.7}Na_{0.2}Mn_{1.1}O_3$ | 0 | 22.9903 | 47.0102 | 0.189 | 0.273 | 0.106 | 0.1565 | 0.2516 | 1.6 | 1.1 | 2.1 |
| | 3 | 23.0447 | 47.0469 | 0.140 | 0.251 | 0.060 | 0.1265 | 0.2437 | 1.9 | 1.1 | 2.1 |
| $La_{0.7}Ag_{0.2}Mn_{1.1}O_3$ | 0 | 22.8565 | 46.7675 | 0.151 | 0.215 | 0.106 | 0.1075 | 0.1871 | 1.7 | 1.1 | 2.1 |
| | 3 | 23.0818 | 46.9740 | 0.123 | 0.191 | 0.060 | 0.1074 | 0.1813 | 1.7 | 1.1 | 2.1 |
| $La_{0.7}K_{0.2}Mn_{1.1}O_3$ | 0 | 22.9014 | 46.8148 | 0.276 | 0.348 | 0.106 | 0.2548 | 0.3315 | 1.3 | 1.1 | 2.1 |
| | 3 | 23.0568 | 46.9562 | 0.216 | 0.312 | 0.060 | 0.2075 | 0.3062 | 1.5 | 1.1 | 2.1 |

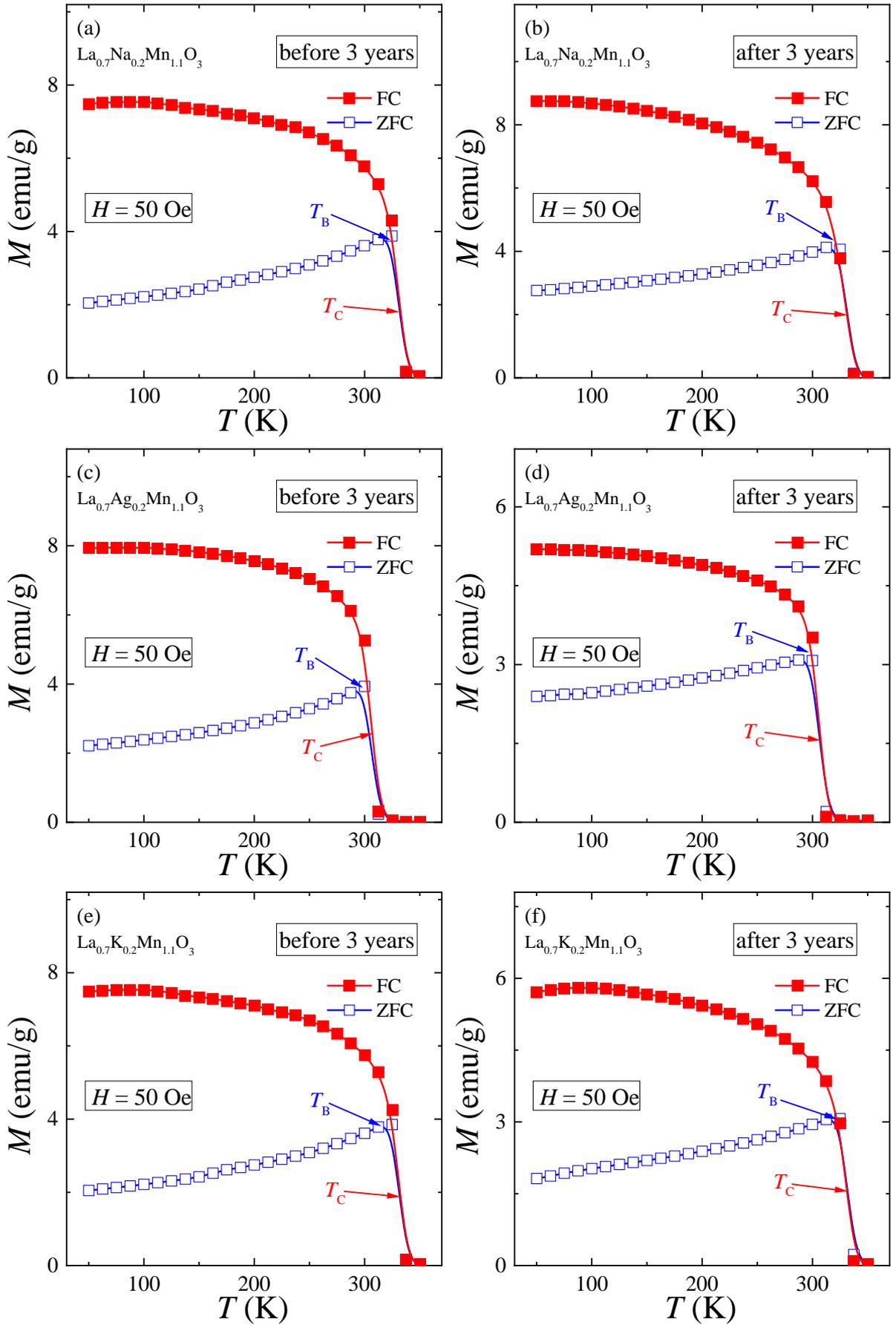

**Fig. S21**. The temperature dependences of magnetization $M_{ZFC-FC}(T)$ for the La$_{0.7}$Na$_{0.2}$Mn$_{1.1}$O$_3$ (a, b), La$_{0.7}$Ag$_{0.2}$Mn$_{1.1}$O$_3$ (c, d), and La$_{0.7}$K$_{0.2}$Mn$_{1.1}$O$_3$ (e, f) samples measured before (a, c, e) and after 3 years (b, d, f).

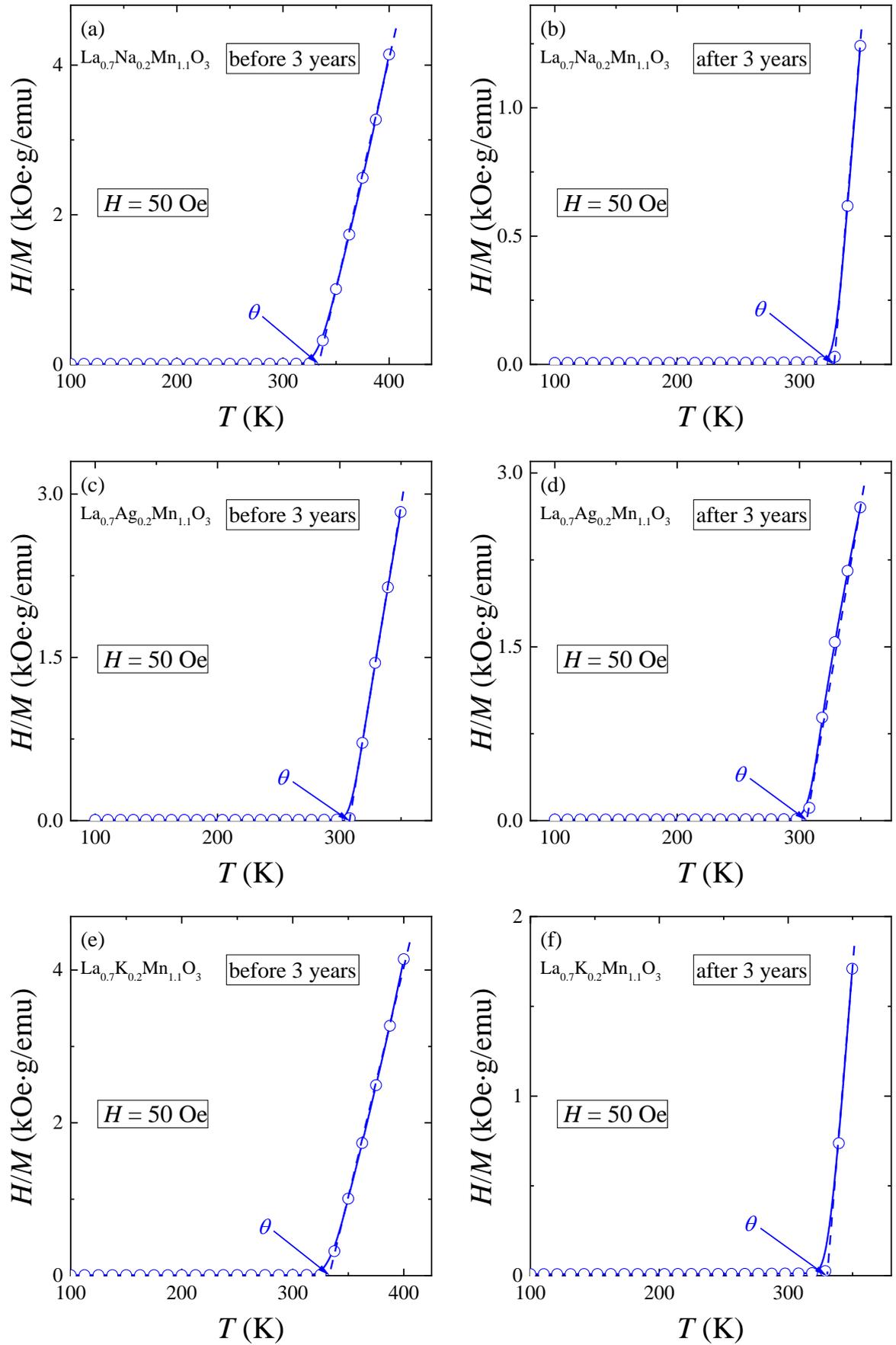

**Fig. S22**. The temperature dependences for the inverse magnetic susceptibility $\chi^{-1} = H/M(T)$ for the $La_{0.7}Na_{0.2}Mn_{1.1}O_3$ (a, b), $La_{0.7}Ag_{0.2}Mn_{1.1}O_3$ (c, d), and $La_{0.7}K_{0.2}Mn_{1.1}O_3$ (e, f) samples measured before (a, c, e) and after 3 years (b, d, f).

Phase transition temperatures (PM Curie temperature $\theta$, Curie temperature $T_C$, and blocking temperature $T_B$) for the L$A$MO nanopowders ($A$ = Na$^+$, Ag$^+$, K$^+$) depending on the time $t$.

| $A$ | $t$ (year) | $\theta$ (K) | $T_C$ (K) | $T_B$ (K) |
|---|---|---|---|---|
| Na$^+$ | 0 | 334 | 331 | 322 |
|  | 3 | 329 | 330 | 320 |
| Ag$^+$ | 0 | 307 | 308 | 301 |
|  | 3 | 304 | 307 | 294 |
| K$^+$ | 0 | 335 | 332 | 318 |
|  | 3 | 331 | 331 | 320 |

**SM9**
# Recalculation of the influence of structural-size effect and aging time on the change of Curie temperature in terms of pressure

The XRD patterns and refined structural parameters for the L(C)NMO-900 samples obtained under different pressures are presented in Fig. 23 and Tables S10 and S11.

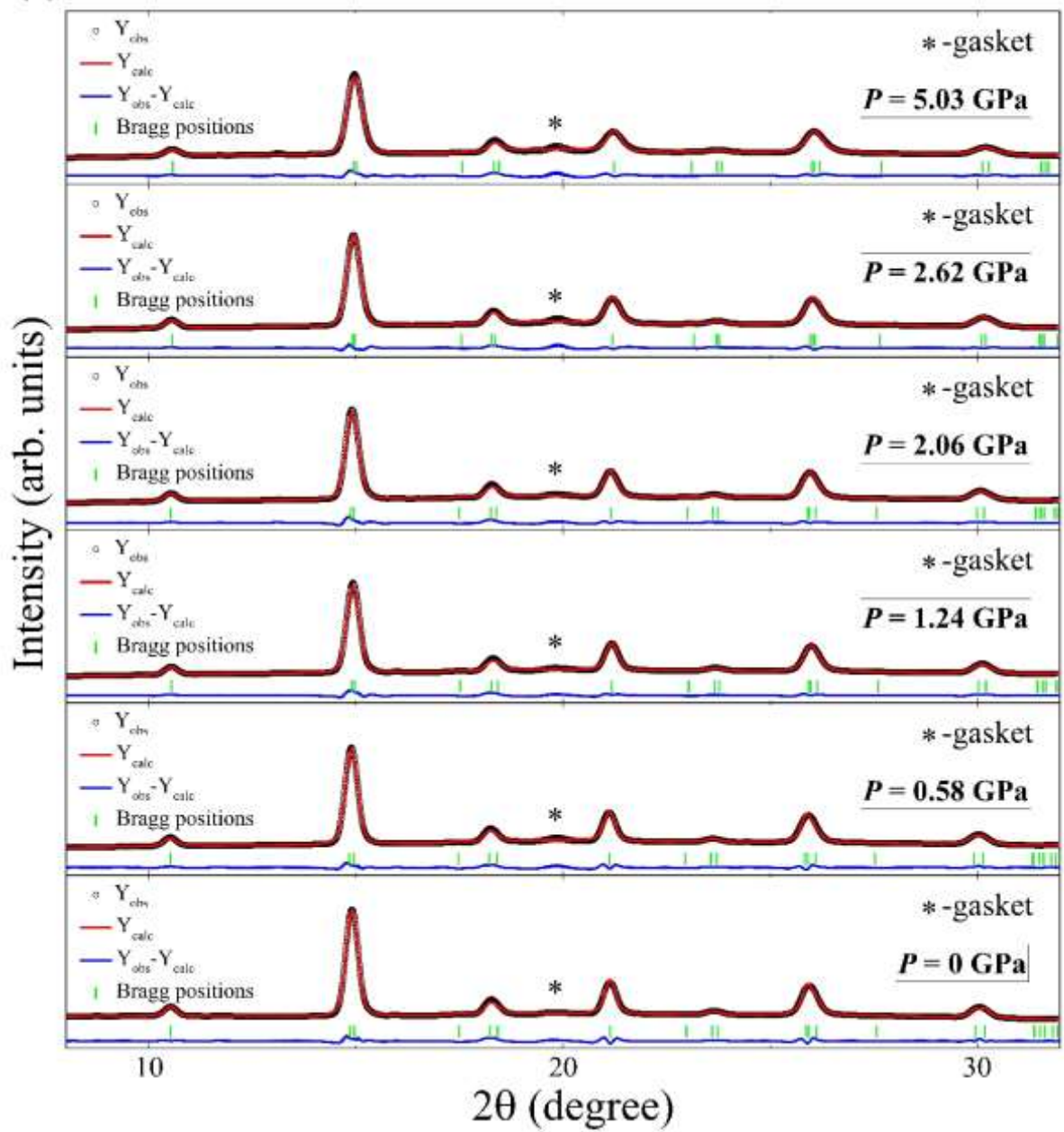

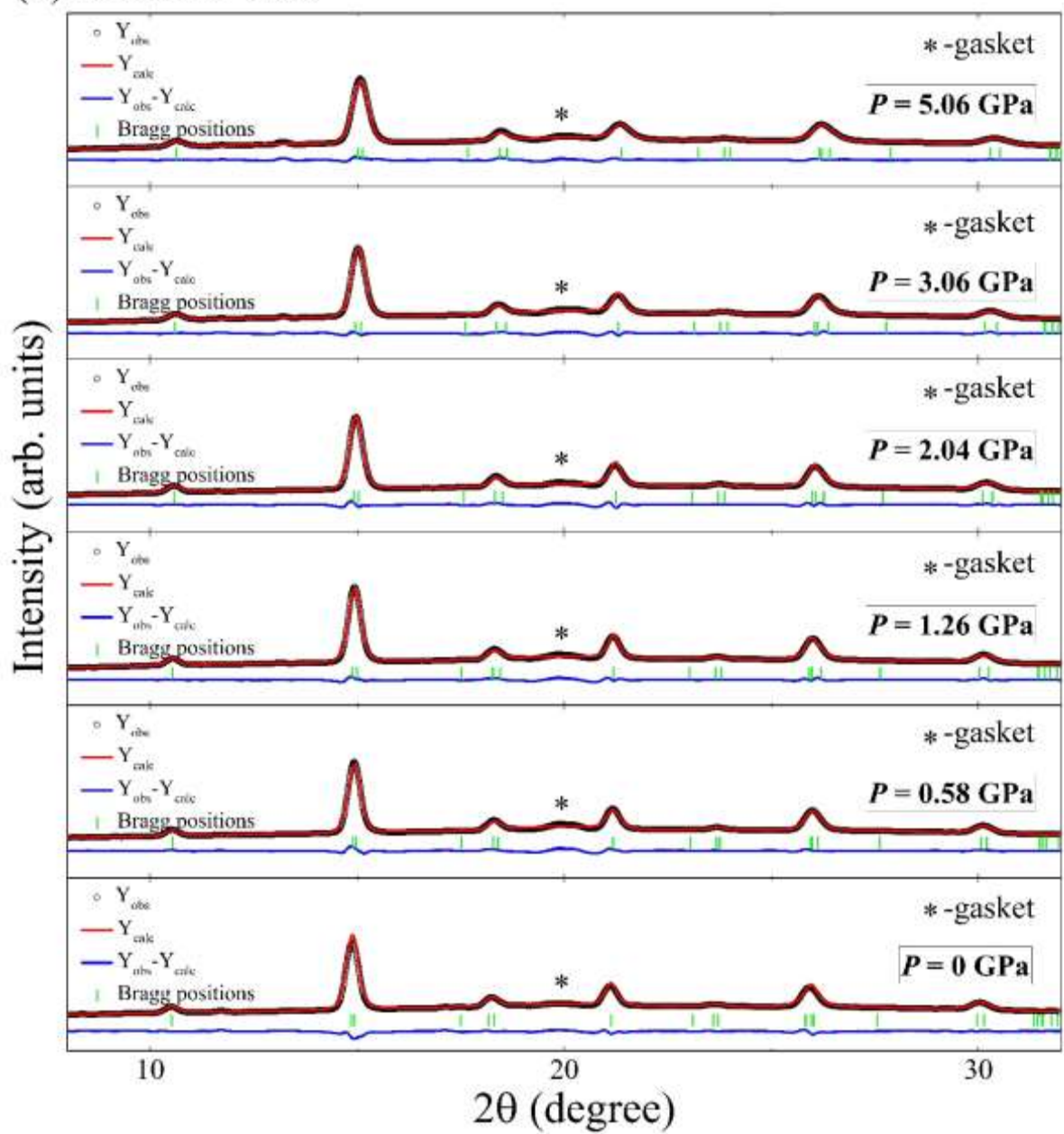

**Fig. 23**. Rietveld refinement XRD patterns of the LNMO-900 (a) and LCNMO-900 (b) samples obtained under different pressures *P*. The experimental (black circles) and calculated (red line) values with a difference (blue line) are plotted. The vertical bars indicate the angular positions of the allowed Bragg reflections (JCPDS Card No. 96-152-5859). The * symbol indicates the peak from the gasket.

**Table S10**

Rietveld refinement crystallographic parameters data for the LNMO-900 nanopowder under different pressures $P$.

| Sample | LNMO | | | | | |
|---|---|---|---|---|---|---|
| Pressure $P$ (GPa) | 0 | 0.58 | 1.24 | 2.06 | 2.62 | 5.03 |
| Crystal system | rhombohedral | | | | | |
| Space group | $R\bar{3}c$ (No. 167) | | | | | |
| Lattice parameters $a$, $b$, $c$ (Å) | $a = b =$ 5.488(9) $c =$ 13.296(45) $\alpha = \beta = 90°$; $\gamma = 120°$ | $a = b =$ 5.494(10) $c =$ 13.309(50) $\alpha = \beta = 90°$; $\gamma = 120°$ | $a = b =$ 5.474(15) $c =$ 13.294(75) $\alpha = \beta = 90°$; $\gamma = 120°$ | $a = b =$ 5.481(17) $c =$ 13.316(83) $\alpha = \beta = 90°$; $\gamma = 120°$ | $a = b =$ 5.445(30) $c =$ 13.398(150) $\alpha = \beta = 90°$; $\gamma = 120°$ | $a = b =$ 5.458(25) $c =$ 13.271(118) $\alpha = \beta = 90°$; $\gamma = 120°$ |
| Unit cell volume $V$ (Å³) | 346.9±1.4 | 348.0±1.6 | 345.1±2.4 | 346.4±2.6 | 344.0±4.7 | 342.5±3.7 |
| Z | 6 | | | | | |

**Table S11**

Rietveld refinement crystallographic parameters data for the LCNMO-900 nanopowder under different pressures $P$.

| Sample | LCNMO | | | | | |
|---|---|---|---|---|---|---|
| Pressure $P$ (GPa) | 0 | 0.58 | 1.26 | 2.04 | 3.06 | 5.06 |
| Crystal system | rhombohedral | | | | | |
| Space group | $R\bar{3}c$ (No. 167) | | | | | |
| Lattice parameters $a$, $b$, $c$ (Å) | $a = b =$ 5.457(22) $c =$ 13.486(112) $\alpha = \beta = 90°$; $\gamma = 120°$ | $a = b =$ 5.471(40) $c =$ 13.313(183) $\alpha = \beta = 90°$; $\gamma = 120°$ | $a = b =$ 5.477(25) $c =$ 13.266(113) $\alpha = \beta = 90°$; $\gamma = 120°$ | $a = b =$ 5.465(29) $c =$ 13.227(130) $\alpha = \beta = 90°$; $\gamma = 120°$ | $a = b =$ 5.451(35) $c =$ 13.174(141) $\alpha = \beta = 90°$; $\gamma = 120°$ | $a = b =$ 5.429(71) $c =$ 13.155(286) $\alpha = \beta = 90°$; $\gamma = 120°$ |
| Unit cell volume $V$ (Å³) | 347.8±3.5 | 345.1±5.9 | 344.7±3.7 | 342.1±4.2 | 339.1±4.7 | 335.7±9.6 |
| Z | 6 | | | | | |

To determine what pressure creates the structural-size effect and aging time for the L(C)NMO samples, the slope angle, i.e. $dV/dP$, was defined from the $V(P)$ dependences (Fig. 24). In the first approximation, it is assumed that the $dV/dP$ is the same for all LNMO and LCNMO samples obtained under different $t_{ann}$, respectively. Knowing the change in the volume $dV$ depending on the aging time $t$ and annealing temperature $t_{ann}$ using Tables 1 and S5, the $dT_C/dP$ values for all L(C)NMO samples using Table S4 can be recalculated (see Table S11).

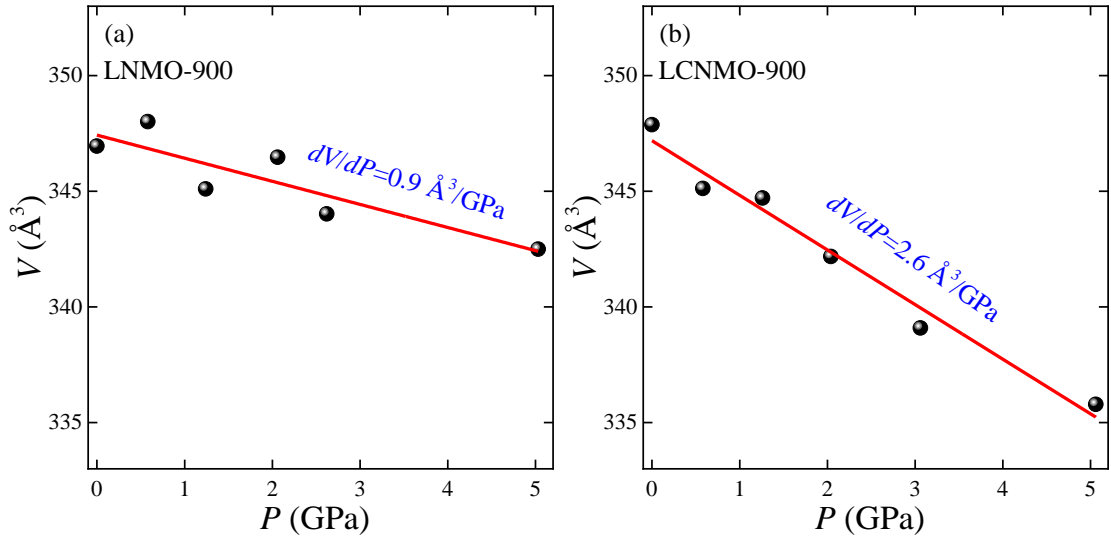

**Fig. S24**. The pressure dependences of the volume $V(P)$ for the LNMO-900 (a) and LCNMO-900 (b) samples.

**Table S12**

The change in the phase transition temperature $dT_C/dP$ depending on the internal (structural-size and aging time) and external (hydrostatic) pressures for all L(C)NMO nanopowders.

| Sample | $t_{ann}$ (°C) | External pressure | | Internal pressure | |
|---|---|---|---|---|---|
| | | | | Aging time | Structural-size effect |
| | | $\|dV/dP\|$ (Å³/GPa) | $dT_C/dP$ (K/GPa) | $dT_C/dP$ (K/GPa) | $dT_C/dP$ (K/GPa) |
| LNMO | 500 | | 6.3 | 11.9 | – |
| | 700 | 0.9 | 8.8 | 16.3 | 21.7 |
| | 900 | | 16.3 | 1.1 | 27.5 |
| LCNMO | 500 | | 6.3 | 99.6 | – |
| | 700 | 2.6 | 7.5 | 27.6 | 83.5 |
| | 900 | | 7.5 | 10.3 | 91.4 |

**Magnetocaloric properties of the L(C)NMO samples before and after three years**

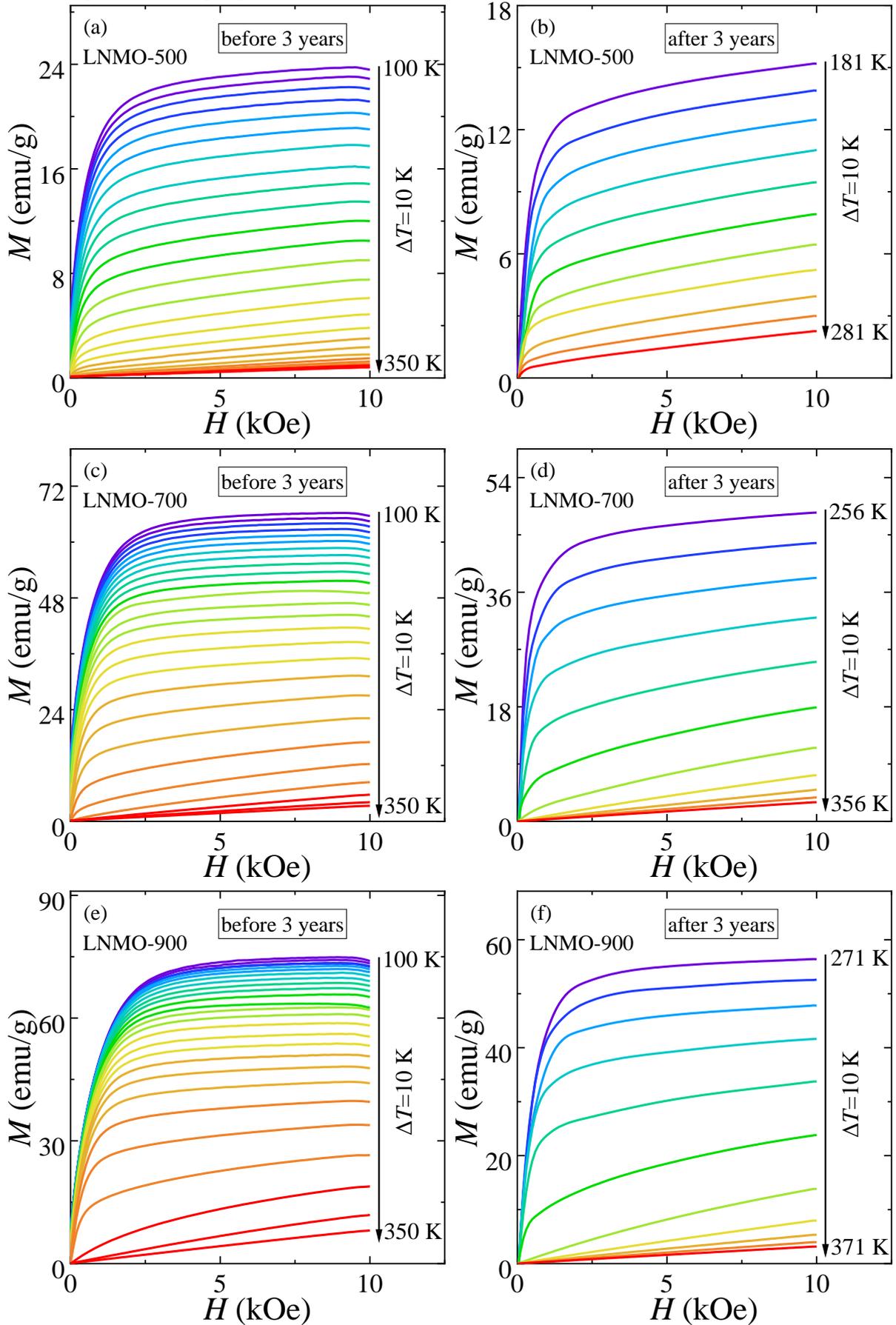

**Fig. S25**. The field dependences of magnetization $M(H)$ for the LNMO-500 (a, b), LNMO-700 (c, d), and LNMO-900 (e, f) samples measured before (a, c, e) and after 3 years (b, d, f).

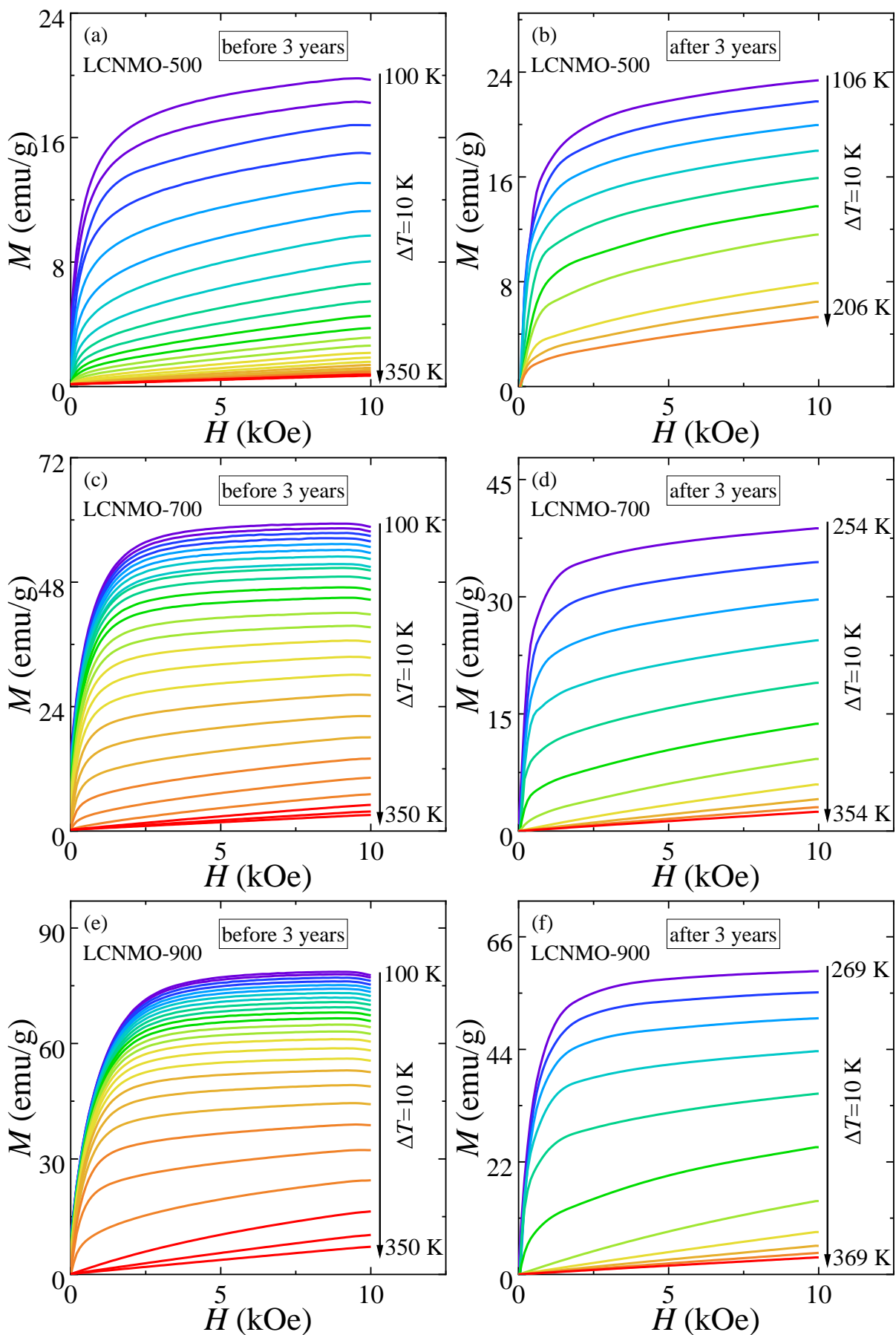

**Fig. S26**. The field dependences of magnetization *M*(*H*) for the LCNMO-500 (a, b), LCNMO-700 (c, d), and LCNMO-900 (e, f) samples measured before (a, c, e) and after 3 years (b, d, f).

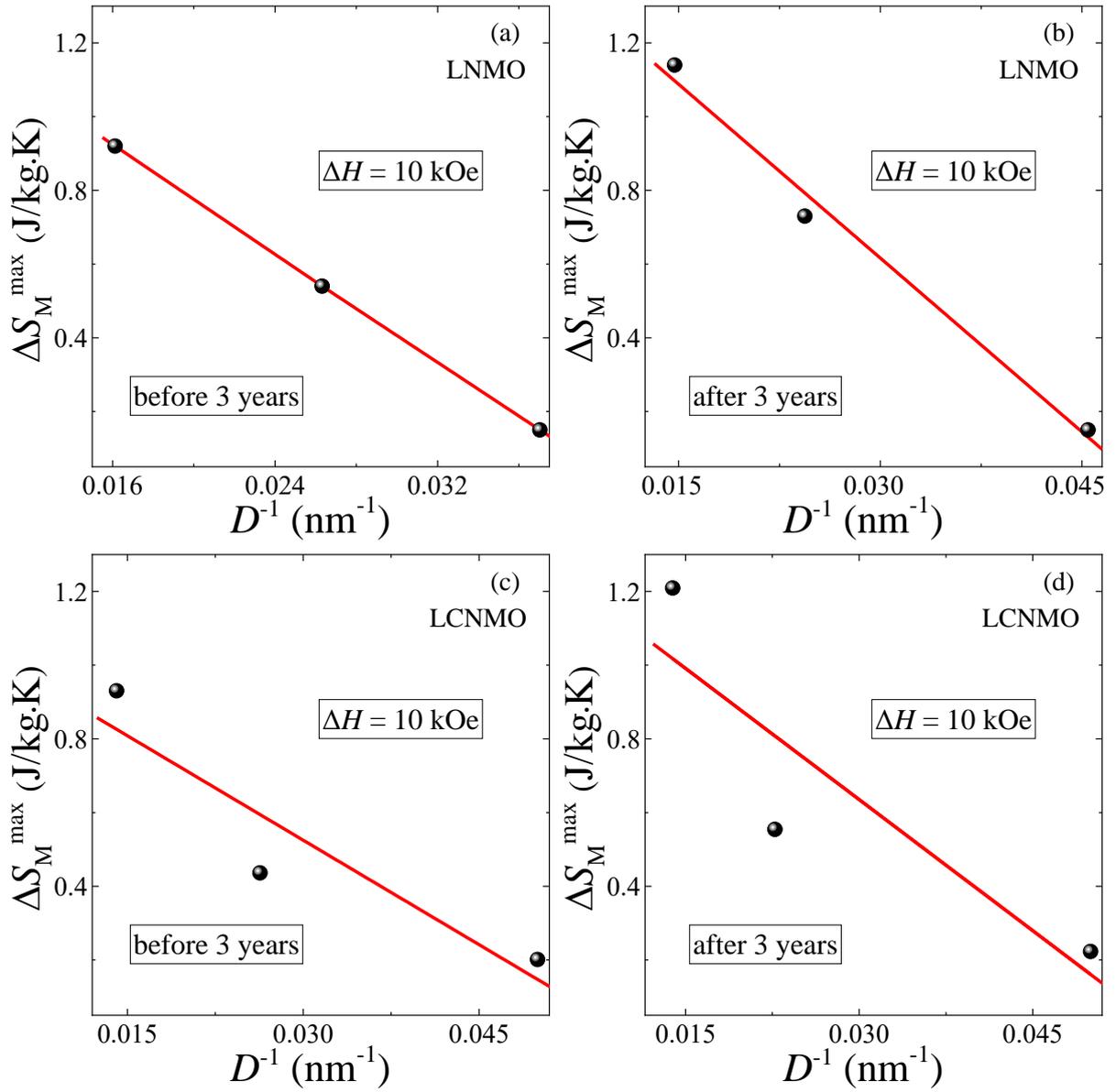

**Fig. S27**. The linear particle size dependences of magnetic entropy change $\Delta S_M^{max}(D^{-1})$ for the LNMO (a, b) and LCNMO (c, d) samples measured before (a, c) and after 3 years (b, d).